\documentclass[%
 amsmath,amssymb,
]{revtex4-2}

\usepackage{amsmath,amssymb,xcolor,mathrsfs}

\usepackage{graphicx}% Include figure files
\usepackage{dcolumn}% Align table columns on decimal point
\usepackage{bm}% bold math
\usepackage[hidelinks]{hyperref}
\begin{document}

\title{Frank Elastic Constants of Semiflexible Polymer Solutions}% Force line breaks with \\

\author{Quinn MacPherson}
 \affiliation{Science Division, College of the Sequoias}%Lines break automatically or can be forced with \\
 \altaffiliation[Previously at ]{Department of Physics, Stanford University.}

\date{\today}

\begin{abstract}
We derive the Frank elastic constants for nematic solutions of semiflexible polymers. We plot these results as a function of the coarse-grained Maier-Saupe quadrupole aligning strength and polymer stiffness ranging from rigid to highly flexible. The derivation uses the random phase approximation and combines the exact results for the statistics of a worm-like-chain with polymer field theory using a spherical harmonic basis. The results are evaluated using a numerical inverse Laplace transform.  We present the results in terms of microscopic features such as hairpins and polymer ends so the trends can be understood independently from the derivation.  Key findings are that for rigid polymers $K_{bend}>K_{splay}>K_{twist}$ while for flexible polymers $K_{splay}>K_{bend}>K_{twist}$.  For rigid polymers, the Frank elastic constants grow with the polymer length.  For flexible polymers the elastic constants grow with the persistence length, which becomes the characteristic length scale, with the exception of $K_{splay}$ at high alignment strengths which grows with polymer lengths due to the elimination of hairpins.
\end{abstract}

%Use showkeys class option if keyword
\maketitle

%\tableofcontents

\section{Introduction}

When oblong molecules are sufficiently concentrated they tend to align.  This liquid crystalline alignment results from the molecules being much easier to pack together when the molecules are oriented in roughly the same direction as their neighbors.  That is, the alignment is an entropic effect, though an enthalpic contribution is also possible.  Polymers, which are characterised by a length that is much larger than their diameter, are oblong in the extreme.  Solutions and melts of semiflexible and rigid polymers are particularly prone to alignment because their structural rigidity tends to align monomers along the polymer backbone, compounding the aligning tendency from packing.

\begin{figure}
\begin{centering}
\includegraphics[width=0.5\linewidth]{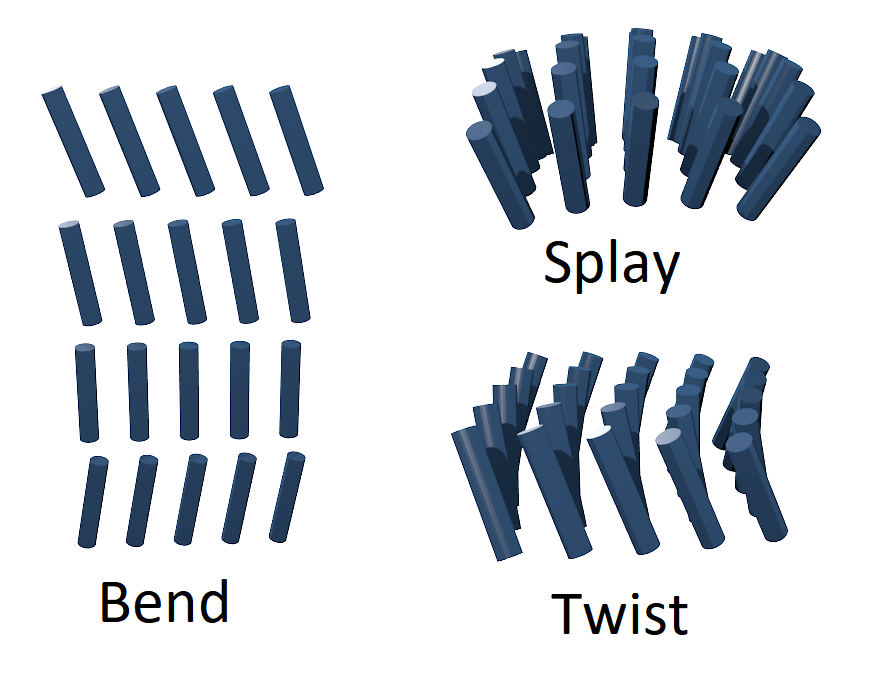}
\par\end{centering}
\caption[Bend twist and splay pictures]{\label{fig:Frank_Elastic_Pic_1}Pictorial representation of bend, twist, and splay.  The axes of the cylinders represent the direction of nematic ordering, $\vec{n}$.  We cite~\cite{selingerInterpretationSaddlesplayOseenFrank2018} for the format of this figure.}
\end{figure}

Even for highly packed solutions of rigid polymers, imperfections in the packing will cause the direction of alignment to wander so that distant positions in the solution will have different local directions of alignment.  The orientation of alignment is critical to the behavior of polymer solutions ranging from conducting polymers~\cite{rudnickiImpactLiquidCrystallineChain2019} to the packaging of DNA withing a viral capsid~\cite{svensekConfinedNematicPolymers2010}.  The deviations of the local direction of liquid crystal alignment over large distances can be classified into bend, twist, and splay which are depicted in figure~\ref{fig:Frank_Elastic_Pic_1}.

We formally define bend as $\vec{n}\times\nabla\times\vec{n}$, twist as $\vec{n}\cdot\nabla\times\vec{n}$, and splay as $\nabla\cdot\vec{n}$ where the unit vector $\vec{n}$ denotes the direction of alignment (we will formally define $\vec{n}$ in equation~\ref{eq:DefinitionOfn}).  If deviations in the direction of alignment occur slowly -- i.e. over large spatial distances -- then bend, twist, and splay will be small due the presence of the spatial derivative $\nabla$.  Because these deformations are small, we can write the energy of these deformation modes as a quadratic in bend, twist, and splay

\begin{eqnarray}
E_{Frank}&=&\frac{1}{2}K_{bend}\left\langle \left(\vec{n}\times\nabla\times\vec{n}\right)^{2}\right\rangle +\frac{1}{2}K_{twist}\left\langle \left(\vec{n}\cdot\nabla\times\vec{n}\right)^{2}\right\rangle+\frac{1}{2}K_{splay}\left\langle \left(\nabla\cdot\vec{n}\right)^{2}\right\rangle \label{eq:FrankElasticEnergy}
\end{eqnarray}

where $K_{bend}$, $K_{twist}$, and $K_{splay}$ are the Frank elastic constants for the (achiral) solution. In writing Eq.~\ref{eq:FrankElasticEnergy}, we have included only long distance (small k) terms and hence ignore higher order derivative terms such as the saddle-splay interaction.

The central theme of material science is to predict macroscopic material properties from microstructure, which in turn is generated by the molecular properties of the constituents. Here we consider the latter connection between the molecular properties (in this case, the polymer length, concentration, Kuhn length and monomer cross sectional area) and the microstructure (bend, twist, and splay).  We are particularly interested in the large effects that the flexibility or rigidity of the polymer backbone can have on the Frank elastic constants.

The first step to building a theoretic model of the elastic response of a solution is to model the thermodynamics of the solution.  In other words, when should we expect the aligned phase to even exist? The phase behavior of solutions of semiflexible polymers have been investigated by a number of authors~\cite{khokhlovLiquidcrystallineOrderingSolution1981, khokhlovLiquidcrystallineOrderingSolution1982, khokhlovInfluenceExternalField1982, khokhlovTheoryLiquidcrystallineOrdering1985, khokhlovTheoryNematicOrdering1986, semenovStatisticalPhysicsLiquidcrystalline1988, liuFreeEnergyFunctionals1993, spakowitzSemiflexiblePolymerSolutions2003}.  The interested reader is referred to these references for the resulting phase diagrams. These dictate the minimum possible degree of alignment where a nematic phase can be found.

Another important step is deriving the elastic constants for solutions in the rigid rod limit~\cite{straleyFrankElasticConstants1973, leeComputationsPhaseEquilibrium1986, marrucciElasticConstantsMaierSaupe1991}. This is the limiting case for a polymer of infinite Kuhn length, to which we will compare our results.
There has also been considerable work on deriving the Frank elastic constants of semiflexible polymer solutions under various approximations and conditions~\cite{odijkElasticConstantsNematic1986, shimadaCoefficiensGradientTerms1988, semenovStatisticalPhysicsLiquidcrystalline1988, doussalStatisticalMechanicsDirected1991, petschekMolecularstatisticalTheoryCurvature1992, satoFrankElasticConstants1996}.  Shimada~\cite{shimadaCoefficiensGradientTerms1988} makes the approximation of a weak aligning field.  Petschek and Terentjev~\cite{petschekMolecularstatisticalTheoryCurvature1992} estimated Frank elastic constants of a chain of monomers connected by stiff joints using a ground-state-dominance approximation\footnote{It appears that this approximation roughly corresponds to including only the leading pole in the complex integral we perform in section~\ref{sec:Numerical-Laplace-inversion}}. Santo and Teramoto~\cite{satoFrankElasticConstants1996} found the Frank constants for a freely jointed chain and tabulated results of previous papers. 

Exact results for the propagator of the wormlike chain in an aligning field were found by Spakowitz and Wong~\cite{spakowitzExactResultsSemiflexible2004, spakowitzEndtoendDistanceVector2005} and enable a better determination of polymer statistics. We use this propagator in conjunction with polymer field theory to more accurately determine the Frank elastic constants. While a connection between spherical harmonics and the Frank elastic modes has long been recognized~\cite{priestTheoryFrankElastic1973}, we introduce spherical tensor fields and compact summation notation which simplify and clarify the field theory derivation (see equations~\ref{eq:density} and~\ref{eq:SummationNotation}). The derivation of the Frank elastic constants we present remains analytical up to an inverse Laplace transform, which we perform numerically.  The results and their interpretation, which are intended to be intelligible without reference to the derivation, are presented in section~\ref{sec:Frank_Results}.

\section{Model and Assumptions}

\subsection{Solution of WLC's\label{sec:FrankAssumptions}}

We wish to study deformations in the alignment of liquid crystal polymer
solutions and melts. To this end we propose the following mathematical
model and approximations. The point(s) at which an approximation
is used in the subsequent derivation will be noted for those wishing
to relax these assumptions.
\begin{enumerate}
\item \label{enu:No_Polydispersity} We will assume all polymers have the same length, $L$, i.e. no polydispersity.
\item \label{enu:ContinuousPolymers}We describe the $j^{th}$ polymer as
a continuous path in space $\vec{r}_{j}\left(s\right)$, given as a function
of path length $s\in[0,L]$.
We define the orientation direction of the polymer at point $s$ as
the unit vector $\vec{u}_{j}\left(s\right)=\frac{\partial\vec{r}_{j}\left(s\right)}{\partial s}$.
\item \label{enu:Semi-flexable}Each polymer will be assumed to obey the
statistics of an inextensible Worm Like Chain (WLC). That is, the
mechanical energy of $n_{p}$ polymers is given by 
\begin{equation}
E_{poly}\equiv\frac{1}{2}\ell_{p}\sum_{j=1}^{n_{p}}\int_{0}^{L}ds\left(\frac{\partial\vec{u}_{j}}{\partial s}\right)^{2}\label{eq:WLC_energy}
\end{equation}
where $\ell_{p}$ is the persistence length of the polymer, which sets
the polymer stiffness. In other words, the energy is proportional
to the square of the curvature of the polymers. Along with all other energies
in this paper, $E_{poly}$ is assumed to be in units of $k_{b}T$.
\item \label{enu:cross_section_A} The cross sectional area of the polymer
is sufficiently small that its diameter is smaller than the other
length scales of interest in the problem.
\item The polymers are achiral.  That is, they do not exhibit a natural twist direction that affects the solution properties.  As an example of how this could be violated, imagine a naturally twisted polymer and further imagine that the alignment interactions between adjacent chains depend on this twist, such that a natural twist is conferred to the polymer solution.
\item \label{enu:ReversalSymmetry}The system is unchanged by reversing
any polymer by swapping its $s=0$ and $s=L$ ends.  An example of how this could be violated would be a solution of polymers with a negative charge on one polymer end and a positive charge on the other.  Such polymers would have dipole-dipole interactions and/or experience the aligning effect of an external field.
\item \label{enu:TranslationalInvariance}The system is translationally
invariant. With no boundary conditions the material can be thought
of as being infinite in extent.
\item \label{enu:GoodSolvent}We will assume the solvent is good enough
that the solution does not segregate into a polymer rich and polymer
lean phase. The range of applicability of this assumption in the presence
of an alignment interaction is studied in~\cite{spakowitzSemiflexiblePolymerSolutions2003}.
\item \label{enu:TwoBody}Polymer interactions are assumed to be two body
interactions, that is, the energy is quadratic in polymer density.
\item \label{enu:ShortRange}Polymer interactions are local, meaning that
their range is well below the length scales of relevance to the liquid
crystal alignment fluctuation under study.
\item \label{enu:Only_x_ and_a}There are only two polymer interactions.
The first of these is the Flory-Huggins interaction which governs
the energy of mixing of polymer and solvent. The second is a Maier-Saupe
interaction which captures the tendency of adjacent polymer segments
to align with each other by assigning an energy proportional to $\left(\vec{u}_{1}\cdot\vec{u}_{2}\right)^{2}$
for two polymer segments in close spatial proximity.
\item \label{enu:Small_Perterbation}The preferential direction of alignment
is approximately in the $z$ direction. The density and the preferred
direction of alignment of the polymers will only make small deviations
from their mean field values. Note that the orientation of particular
polymers can differ substantially from the $z$ direction.
\item \label{enu:RPA}The effective potential experienced by a single polymer,
created by the others, undergoes only small fluctuations from its
mean value. This allows the Random Phase Approximation (RPA).
\item \label{enu:AzimuthallySymmetric}The average alignment field will
be azimuthally symmetric with respect to the aligning axis. In other
words, there is not a second axis of alignment that could result from
ribbon-like polymers with an asymmetric cross-section.
\item \label{enu:small_k}We are primarily interested in coordinated deviations
of the alignment over large distances (small $\vec{k}$). 
\end{enumerate}
\subsection{Fuzzball and Rigid Rod}

The properties of the WLC are made more evident by comparison to the
rigid rod and an object I will refer to as a fuzzball. The rigid rod
is a straight rod of length $L$ and cross sectional area $A$. The
rigid rod is a WLC in the limit as $\ell_{p}\to\infty$. Because it
has no flexibility, the rigid rod has no relevant internal energy
$E_{poly}$. The fuzzball, like the rigid rod, has a single direction
and no internal energy. However, the density of the fuzzball is spread
out in a symmetric Gaussian about a point. The fuzzball can be thought
of as a crude model of a small molecule or an electric quadruple whose
field falls off as a Gaussian.

\section{Summary of Derivation}

The results of this paper are intended to be intelligible without
consulting the derivation. However, the derivation contains a number
of interesting tricks and insights that the reader may find useful
for similar problems. 
%Most of the mathematics used here can be found
%elsewhere but here it is combined together with efficient notation.
In particular, we describe a real spherical harmonic tensor distribution, extended summation notation for describing orientation over space,
mean field polymer field theory, fluctuations away from this mean field,
and WLC propagators derived via stone fence diagrams.

In this summary we provide a brief description of the approach we
use to calculate the Frank elastic constants for a solution of semiflexible
polymers. A similar procedure is followed for the rigid rod and fuzz-ball
examples. We begin by describing
the polymer solution with the partition function 

\begin{equation}
Z\propto\int\mathscr{D}\left[\vec{\mathrm{r}}_{j}\left(s\right)\right]
\exp\left(
-E_{poly}-E_{FH}-E_{MS}
\right)\label{eq:Partiation_Funtion_long}
\end{equation}

where the polymer bending energy $E_{poly}$ is given in equation~\ref{eq:WLC_energy}, the $E_{FH}$ is the Flory-Huggins interaction and $E_{MS}$ is the Maier-Saupe energy representing the preference for nearby polymers to align.  The latter two are given by the sum of pairwise interactions

\begin{equation}
    E_{FH}=\frac{A^{2}}{2}\sum_{i=1}^{n_{p}}\int_{0}^{L}ds_{1}\sum_{j=1}^{n_{p}}\int_{0}^{L}ds_{2}\chi\delta\left(\mathrm{\vec{r}_{i}}\left(s_{1}\right)-\mathrm{\vec{r}_{j}}\left(s_{2}\right)\right)\label{eq:Fory_energy_long}
\end{equation}

and 

\begin{eqnarray}
    E_{MS}=&\frac{A^{2}}{3}\sum_{i=1}^{n_{p}}\int_{0}^{L}ds_{1}\sum_{j=1}^{n_{p}}\int_{0}^{L}ds_{2}a\left(\vec{u}_{i}(s_{1})\cdot\vec{u}_{j}\left(s_{2}\right)\right)^{2}\delta\left(\mathrm{\vec{r}_{i}}\left(s_{1}\right)-\mathrm{\vec{r}_{j}}\left(s_{2}\right)\right). \label{eq:Maier-Saupe_energy_long}
\end{eqnarray}

where $A$ is the cross sectional area of the polymers. A factor of $1/2$ is included to account for double counting~\footnote{without the factor of $1/2$ the $E_{MS}$ would have a $2/3$ rather than a $1/3$ per the customary definition of $a$}. 

Equations~\ref{eq:Partiation_Funtion_long}-\ref{eq:Maier-Saupe_energy_long} are then simplified significantly by using
the local density and orientation distribution,
$\phi$, which we define in equation \ref{eq:density} and
use throughout the remainder of the derivation. In this notation $E_{FH}$ and $E_{MS}$ are
replaced with $\phi V\phi$ where $V$ is the interaction matrix defined
in equation \ref{eq:V_potential}.

To make the system mathematically tractable we first solve the partition
function for the mean field of $\phi$ which we write as $\left\langle \phi\right\rangle ^{MF}$.
The self consistent equation \ref{eq:SelfConsistantEquation}
requires that $\left\langle \phi\right\rangle ^{MF}$ is the density
that would be expected in a field $\left\langle \phi\right\rangle ^{MF}V$
which $\left\langle \phi\right\rangle ^{MF}$ generates and applies
to itself. The Frank elastic constants are defined in terms of fluctuations
about this mean field solution, as will be discussed shortly.

In order to calculate $\phi$ for a WLC in the presence of an aligning
mean field we follow~\cite{yamakawaHelicalWormlikeChains1997, spakowitzSemiflexiblePolymerSolutions2003, spakowitzExactResultsSemiflexible2004} in using stone fence diagrams. These
diagrams allow us to calculate the Laplace transform (from chain length
$N$ to Laplace variable $p$) of $\left\langle \phi\right\rangle $
as well as the Laplace transforms of products, e.g. $\mathscr{\mathcal{L}}_{N\to p}\left\langle \phi_{1}\phi_{2}\right\rangle $.
We are able to numerically invert the Laplace transforms via path
integration in the complex plane. The construction of expectation
values is described in section \ref{sec:Evaluating-expectations} and
 the description of the diagrams is given in section \ref{sec:Stone-Fence-diagrams}.

In section \ref{subsec:Fluctuations} we investigate fluctuations
about the mean field solution. The fluctuations are divided into Fourier
modes via a Fourier transform from position $\vec{r}$ to wave vector
$\vec{k}$. To first approximation, the energy of each mode $\tilde{\phi}\left(k\right)$
is quadratic, 
\begin{equation}
E\approx\tilde{\phi}\left(\vec{k}\right)\Gamma\left(\vec{k}\right)\tilde{\phi}\left(-\vec{k}\right)
\end{equation}
where the ``spring constants'' for each of these modes are combined
in the matrix $\Gamma$.

In section \ref{subsec:Frank-Elastic-Energies1} we define
the nematic director and the Frank elastic constants and manipulate
them to be written in terms of particular long-wavelength elements
of $\Gamma$. This allows us to write expressions for the Frank elastic
constants in terms of numerical inverse Laplace transforms in section
\ref{sec:Calculate_FEC}. The numerical evaluation of the inverse transform is described in section \ref{sec:Numerical-Laplace-inversion}.

\section{Summation Notation \& System Description}

The material properties at a point $\vec{r}$ in space are determined
by the position and orientation of polymers in the vicinity of $\vec{r}$.
For a particular configuration $\left\{ \vec{\mathrm{r}}_{j}\left(s\right)\right\} $
of the system, the local density and orientation distributions can
be mathematically described by
\begin{equation}
\hat{\phi}_{l,m}\left(\vec{r}\right)\equiv\sqrt{\frac{4\pi}{2\ell+1}}\sum_{j=1}^{n_{p}}A\int_{0}^{L}dsY_{l,m}\left(\vec{\mathrm{u}}_{j}\left(s\right)\right)\delta\left(\vec{r}-\vec{\mathrm{r}}_{j}\left(s\right)\right)\label{eq:density}
\end{equation}
where $Y_{\ell,m}$ is the real
spherical harmonic function described in appendix \ref{sec:Appendix:-Real-spherical}.
Note that we will use the roman $\vec{\mathrm{r}}$ and $\vec{\mathrm{u}}$
for position and orientation of points along a polymer and italicized
$\vec{r}$ and $\vec{u}$ for generic position in space. These continuous
polymers (assumption \ref{enu:ContinuousPolymers}) could be replaced
with discrete beads by replacing the integral in equation \ref{eq:density} with
a sum. While $\hat{\phi}$ is formally defined as sum of infinitely
thin space curve delta distributions, it is easier to interpret after integrating over a coarse grained volume $\Delta V$ in the vicinity
of the position $\vec{r}$ in the material. For example, the scalar
component of the density $\hat{\phi}_{0,0}$ is interpreted as the
local volume fraction of polymer; integrating $\hat{\phi}_{0,0}$
over the region of space $\Delta V$ gives the volume of polymer within
that region. The values of $\hat{\phi}_{l,m}\left(\vec{r}\right)$
for $\ell>0$ describe the degree and direction of alignment of the
polymers. In equation~\ref{eq:density} the units of volume from $Ads$ cancel the volume units from the delta function so that $\hat{\phi}$ is unitless, as would be expected from a volume fraction.

We write the partition function describing the set of configurations
that $\hat{\phi}$ can take on as
\begin{equation}
Z=\frac{1}{n_{p}!}\int\mathscr{D}\left[\vec{\mathrm{r}}_{j}\left(s\right)\right]\exp\left(-E_{poly}-\frac{1}{2}\hat{\phi}_{1}V_{12}\hat{\phi}_{2}\right)\label{eq:PartitionFunction}
\end{equation}
The $1/n_{p}!$ prefactor accounts for the indistinguishably of the
polymers but is not consequential the following discussion. The functional
integral $\int\mathscr{D}\left[\vec{\mathrm{r}}_{j}\left(s\right)\right]$
refers to an integral over all possible configurations of the system,
that is, it integrates the position of each point of each polymer
over all space 
\begin{equation}
\int\mathscr{D}\left[\vec{\mathrm{r}}_{j}\left(s\right)\right]=\lim_{\Delta s\to0}\prod_{j=1}^{n_{p}}\int_{-\infty}^{\infty}d\vec{\mathrm{r}}_{j,s=0}\int_{-\infty}^{\infty}d\vec{\mathrm{r}}_{j,s+\Delta s}....\int_{-\infty}^{\infty}d\vec{\mathrm{r}}_{j,L}\label{eq:functional_intigral}
\end{equation}
The WLC polymer energy $E_{poly}$ defined in equation~\ref{eq:WLC_energy} effectively
reduces the region of integration to continuous smooth polymers.
The term $\frac{1}{2}\hat{\phi}_{1}V_{12}\hat{\phi}_{2}$ in equation~\ref{eq:PartitionFunction} describes the interaction potential between
polymers. The interaction is quadratic in $\hat{\phi}$ in keeping
with assumption~\ref{enu:TwoBody}. The subscripts in $\hat{\phi}_{1}V_{12}\hat{\phi}_{2}$
refer to an extended summation notation that sums over $\ell$ and
$m$ indices and integrates over space such that
\begin{align}
\hat{\phi}_{1}V_{12}\hat{\phi}_{2}\equiv&
\sum_{\ell_1=0}^{\infty}\sum_{m_1=-\ell_1}^{\ell_1}
\sum_{\ell_2=0}^{\infty}\sum_{m_2=-\ell_2}^{\ell_2}
\int d\vec{r_1}d\vec{r_2} \hat{\phi}_{\ell_1,m_1}(\vec{r}_1)V_{\ell_1,\ell_2}^{m_1,m_2}(\vec{r}_1,\vec{r}_2)\hat{\phi}_{\ell_2,m_2}(\vec{r}_2) \label{eq:SummationNotation}
\end{align}
%and
%\begin{align}
%A_{1}B_{1}\equiv\sum_{\ell=0}^{\infty}\sum_{m=-\ell}^{\ell}\int d\vec{r}A_{l,m}\left(\vec{r}\right)B_{l,m}\left(\vec{r}\right)
%\end{align}

Combining the concise summation notation with the spherical harmonic
indices for orientation allows us to manipulate the tensor field $\hat{\phi}$ in
much the same way as scalar fields in previous publications~\cite{leiblerTheoryMicrophaseSeparation1980, maoImpactConformationalChemical2016}.
Note that when $\phi$ has a single subscript this is understood to be the subscript from the summation notation and when it has two as in $\phi_{\ell,m}$ these refer to $\ell$ and $m$ as should also be clear from context.
The notation $\hat{\phi}_{1}V_{12}\hat{\phi}_{2}$ is analogous to
the vector notation $\vec{\phi}^{T}V\vec{\phi}$ for matrix $V$ and
vector $\vec{\phi}$ which have an infinite number of indices in order
to capture all $\ell,m,$ and $\vec{r}$.
While this notation allows
for a wide variety of orientationally dependent and ranged potentials,
we will restrict our discussion to local potentials (assumption \ref{enu:ShortRange})
with $V\propto\delta\left(\vec{r}_{1}-\vec{r}_{2}\right)$. We will
also restrict $V$ to be non-zero only for a Flory-Huggins scalar potential ($\ell=0$)
and a Maier-Saupe tensor potential $(\ell=2)$ in keeping with assumption
\ref{enu:Only_x_ and_a}. Interactions of order $\ell=1$ violate
the assumed reversal symmetry (assumption \ref{enu:ReversalSymmetry}).
Including interaction terms of order $\ell>2$ would relax assumption \ref{enu:Only_x_ and_a} to systematically
refine the rotational detail of inter-chain interactions with the
interaction constants depending on the chemistry and sterics of the
polymer in question. 

Under the above assumptions we write
\begin{equation}
V_{12}=\left(\chi\delta_{\ell_{1},0}-\frac{2}{3}a\delta_{\ell_{1},2}\right)\delta_{m_{1},m_{2}}\delta_{\ell_{1},\ell_{2}}\delta\left(\vec{r}_{1}-\vec{r}_{2}\right)\label{eq:V_potential}
\end{equation}
where $\chi$ is the widely used Flory-Huggins mixing parameter. The second term in the parentheses
provides an energetic benefit to polymer segments that align $\vec{u}_{1}\approx\vec{u}_{2}$
or antialign $\vec{u}_{1}\approx-\vec{u}_{2}$. The
factor of $2/3$ maintains the meaning of the Maier-Saupe parameter,
$a$, from~\cite{liuFreeEnergyFunctionals1993} and~\cite{spakowitzSemiflexiblePolymerSolutions2003}. The negative sign on the $\ell=2$
term in equation \ref{eq:V_potential} is chosen so that a positive
$a$ corresponds to a preferential alignment. The equivalence to the
Maier-Saupe formulation~\cite{liuFreeEnergyFunctionals1993, spakowitzSemiflexiblePolymerSolutions2003} in terms of $\vec{u}\otimes\vec{u}-\frac{1}{3}I$
rather than $Y_{2}^{m}\left(\vec{u}\right)$ results from the relation
\begin{align}
&\sum_{i,j=1}^{3}\left(\vec{u}_{1}\otimes\vec{u}_{1}-\frac{1}{3}I\right)_{ij}\left(\vec{u}_{2}\otimes\vec{u}_{2}-\frac{1}{3}I\right)_{ij}  =\left(\vec{u}_{1}\cdot\vec{u}_{2}\right)^{2}-\frac{1}{3}=\frac{2}{3}P_{2}\left(\vec{u}_{1}\cdot\vec{u}_{2}\right)
=\frac{2}{3}\sum_{m=-2}^{2}Y_{2,m}\left(\vec{u}_{1}\right)Y_{2,m}\left(\vec{u}_{2}\right)\label{eq:MaierSaupe_YY_relation}
\end{align}
for all unit vectors $\vec{u}_{1}$ and $\vec{u}_{2}$. 

\subsection*{Fuzzball and Rigid Rod}

In contrast to equation \ref{eq:density}, the density distribution
for solution of fuzzballs is given as 
\begin{equation}
\hat{\phi}_{\ell,m}^{FB}=\sqrt{\frac{4\pi}{2\ell+1}}\sum_{j=1}^{n_{p}}\frac{v_{p}}{\left(2\pi\sigma^{2}\right)^{3/2}}\exp\left(-\frac{\left(\vec{r}-\vec{\mathrm{r}}_{j}\right)^{2}}{2\sigma^{2}}\right)Y_{\ell,m}\left(\vec{u}_{j}\right)\label{eq:density_FB}
\end{equation}
where $\sigma$ is the standard deviation of the field falloff and
defines the size of the fuzzball. $v_{p}$ is the effective volume
(or amplitude if you prefer) of the fuzzball. For rigid rods the density
distribution is
\begin{equation}
\hat{\phi}_{\ell,m}^{RR}\left(\vec{r}\right)=\sqrt{\frac{4\pi}{2\ell+1}}\sum_{j=1}^{n_{p}}Y_{\ell,m}\left(\vec{\mathrm{u}}_{j}\right)A\int_{0}^{L}ds\delta\left(\vec{\mathrm{r}}_{j}+s\vec{\mathrm{u}}_{j}-\vec{r}\right)
\end{equation}
where $\vec{r}_{j}$ is the location of one end of the rod and $\vec{u}_{j}$
is the orientation of the rod. For the rigid rod, the integration in the above equation simplifies to \begin{equation*}
    \hat{\phi}_{\ell,m}^{RR}\left(\vec{r}\right)=\sqrt{\frac{4\pi}{2\ell+1}}\sum_{j=1}^{n_{p}}Y_{\ell,m}\left(\vec{\mathrm{u}}_{j}\right)AL\delta\left(\vec{\mathrm{r}}_{j}+L\vec{\mathrm{u}}_{j}-\vec{r}\right)\label{eq:density_RR}
\end{equation*} The partition functions for these two
solutions are similar
\begin{align}
Z^{RR/FB}=\int d\vec{\mathrm{r}}_{j}\int d\vec{\mathrm{u}}_{j}\exp\left(\frac{1}{2}\chi\left(\phi_{0,0}^{RR/FB}\left(\vec{r}\right)\right)^{2} \right.
\left.+\frac{1}{3}a\int d\vec{r}\sum_{m=-2}^{2}\left(\phi_{2,m}^{RR/FB}\left(\vec{r}\right)\right)^{2}\right)
\end{align}
and, in contrast to the WLC, simply require integrating over $n_{p}$
positions $\vec{r}_{j}$ and $n_{p}$ orientations $\vec{u}_{j}$.

\section{Field Manipulations}

We introduce a delta function at every point in space and orientation
$\delta\left(\phi-\hat{\phi}\right)$. Extending the Fourier representation
of a delta function $\delta\left(a-b\right)=\frac{1}{2\pi}\int_{-\infty}^{\infty}dp\left(ip\left(a-b\right)\right)$
to a delta function for a tensor distribution $\phi$ we have
\begin{equation}
\delta\left(\phi-\hat{\phi}\right)\propto\int\mathscr{D}W\exp\left(iW_{1}\left(\phi_{1}-\hat{\phi}_{1}\right)\right)\label{eq:functional_delta_function}
\end{equation}
where the functional integral $\int\mathscr{D}W$ integrates $W_{\ell}^{m}\left(\vec{r}\right)$
for each $\vec{r}$, $\ell$, and $m$ combination. Like $\phi$,
the Fourier variable\textbf{ }$W$\textbf{ }is also a tensor distribution.\textbf{
}Inserting this delta function into the partition function (equation
\ref{eq:PartitionFunction}) in the fashion of $f(x)=\int dyf(y)\delta\left(x-y\right)$
we have
\begin{equation}
Z\propto\int\mathscr{D}W\mathscr{D}\phi\mathscr{D}\left[\vec{\mathrm{r}}^{\left(1\right)}\left(s\right)\right]e^{-E_{poly}-\frac{1}{2}\phi_{1}V_{12}\phi_{2}+iW_{1}\left(\phi_{1}-\hat{\phi}_{1}\right)}
\end{equation}

Note that the $E_{poly}$ and $\hat{\phi}$ only depend on $\vec{r}$
while $\phi_{1}V_{12}\phi_{2}$ and $W_{1}\phi_{1}$ only depend on
$\phi$ so these terms can be separated
\begin{align}
Z\propto\int\mathscr{D}W\mathscr{D}\phi e^{-\frac{1}{2}\phi_{1}V_{12}\phi_{2}+iW_{1}\phi_{1}}
\left[\int\mathscr{D}\vec{\mathrm{r}}^{\left(1\right)}e^{-E_{poly}^{\left(1\right)}-iW_{1}\hat{\phi}_{1}^{\left(1\right)}}\right]^{n_{p}}\label{eq:sperated_Z}
\end{align}
This separation is quite useful, and is why we introduced the delta function~\ref{eq:functional_delta_function} and the $W$ field.

The second integral in~\ref{eq:sperated_Z} contains no inter-chain interactions other than
through the $W$ field and all chains are identical (assumption~\ref{enu:No_Polydispersity}) so it has been factored into a product of $n_{p}$
terms. The superscript $\ ^{\left(1\right)}$ on $\vec{\mathrm{r}},$
$E_{poly}$, and $\hat{\phi}$ indicates a single chain. Defining the contents of the square bracket
in equation~\ref{eq:sperated_Z} as $z_{p}$ we have
\begin{equation}
Z\propto\int\mathscr{D}W\mathscr{D}\phi e^{-\frac{1}{2}\phi_{1}V_{12}\phi_{2}+iW_{1}\phi_{1}+n_{p}\ln\left(z_{p}\right)}\label{eq:SimplePartition}
\end{equation}
\begin{equation}
z_{p}=\int\mathscr{D}\left[\vec{\mathrm{r}}^{\left(1\right)}\left(s\right)\right]e^{-E_{poly}^{\left(1\right)}-iW_{1}\hat{\phi}_{1}^{\left(1\right)}}\label{eq:Single_polymer_partition}
\end{equation}

While strictly speaking $W$ is simply an integration variable, we
can assign a physical meaning to it as a chemical potential. The term
$iW_{1}$ in equation \ref{eq:Single_polymer_partition} is the effective
potential experienced by a single polymer. Meanwhile, the term $iW_{1}$
in equation \ref{eq:SimplePartition} represents the potential that
the single polymers impose to deform the density field $\phi$. In
this way single polymers don't communicate directly with each other,
but communicate though the local chemical potential $W$.  
We will later integrate out $W$ between equations~\ref{Zlong} and~\ref{eq:QuadraticPartision}.

\section{Mean Field Solution\label{subsec:Mean-Field-Solution}}

\subsection{Mean Field Solution for Polymer Solutions}

In this section we will solve for the homogeneous mean field solution
$\hat{\phi}$ of partition function \ref{eq:PartitionFunction} so
that we can expand about $\hat{\phi}$ in section \ref{subsec:Fluctuations}.
The essence of the mean field solution is that rather than calculating
all interactions inside and between polymers, we calculate the statistics
of a single polymer with $E_{FH}$ and $E_{MS}$ replaced by the average field
the polymer experiences. The self consistency requirement sets the
mean field strength to that generated by $n_{p}$ such single polymers.
The field experienced by a single polymer can be found by differentiating
the potential $\frac{1}{2}\hat{\phi}_{1}V_{12}\hat{\phi}_{2}$ with
respect to the field generated by a single polymer, $\hat{\phi}^{\left(1\right)}$,
to get $\left<\phi_{1}\right>^{MF}V_{12}$. The self consistent equation sets the
expectation value of $n_{p}\left\langle \hat{\phi}^{\left(1\right)}\right\rangle $
to be the mean field strength $\left<\phi\right>^{MF}$.
\begin{equation}
\left<\phi\right>^{MF}=\frac{\int\mathscr{D}\left[\vec{\mathrm{r}}^{\left(1\right)}\right]n_{p}\hat{\phi}^{(1)}\exp\left(-E_{poly}-\left<\phi\right>_{1}^{MF}V_{12}\hat{\phi}_{2}^{(1)}\right)}{\int\mathscr{D}\left[\vec{\mathrm{r}}^{\left(1\right)}\right]\exp\left(-E_{poly}-\left<\phi\right>_{1}^{MF}V_{12}\hat{\phi}_{2}^{(1)}\right)}\label{eq:SelfConsistantEquation}
\end{equation}

Equation \ref{eq:SelfConsistantEquation} can alternatively be derived
by introducing $W$ as was done in equations \ref{eq:SimplePartition}
and \ref{eq:Single_polymer_partition} and then minimizing the integrand
of equation \ref{eq:SimplePartition} with respect to a constant $\phi$
and $W$ which gives $\left<\phi\right>^{MF}$ and $iW_{1}^{MF}=V_{12}\left<\phi\right>_{2}^{MF}$. 

For values of the Maier-Saupe parameter $a$ below its critical value
$a^{\star}$, the homogeneous mean field solution will simply be a
constant density with no net alignment $\left<\phi\right>_{\ell,m}^{MF}=\left<\phi\right>_{0,0}^{MF}\delta_{\ell,0}$.
Above $a^{\star}$ the tendency for polymers to align will overpower
entropy and break rotational symmetry creating a direction of preferred
orientation. We will align this direction with the $\hat{z}$ axis.
Because the mean field solution is constant throughout all space (assumptions
\ref{enu:TranslationalInvariance} and \ref{enu:GoodSolvent}) the
$\ell=0$ component of $\left<\phi\right>_{\ell,m}^{MF}=\left<\phi\right>_{0,0}^{MF}\delta_{\ell,0}+\left<\phi\right>_{2,0}^{MF}\delta_{\ell,0}\delta_{m,0}$.
Because $\left<\phi\right>_{0,0}^{MF}$ is a constant scalar offset it can be canceled
from equation \ref{eq:SelfConsistantEquation} with no consequence.
However, there will be a non-zero $\phi_{2,0}$ which will need to
be solved for. Under the potential \ref{eq:V_potential}, the aligning
field energy for a single polymer is $\left<\phi\right>_{1}^{MF}V_{12}\hat{\phi}_{2}^{(1)}=-\gamma\int_{0}^{N}dsY_{2,0}\left(\vec{\mathrm{u}}\left(s\right)\right)$
where $N=L/2\ell_{p}$ is chain length nondimensionalized by the Kuhn
length\footnote{For WLCs the Kuhn length is twice the persistence length.}
and 
\begin{equation}
\gamma=\frac{2}{3}\sqrt{\frac{4\pi}{5}}2\ell_{p}Aa\left<\phi\right>_{2,0}^{MF}\label{eq:Def_gamma}
\end{equation}
 is the strength of the aligning field which follows from equations
\ref{eq:V_potential} and \ref{eq:density}. The value of gamma can
roughly be thought of as the aligning energy in $kT$'s per persistence
length of polymer. 

\begin{figure}
\begin{centering}
\includegraphics[width=1.0\linewidth]{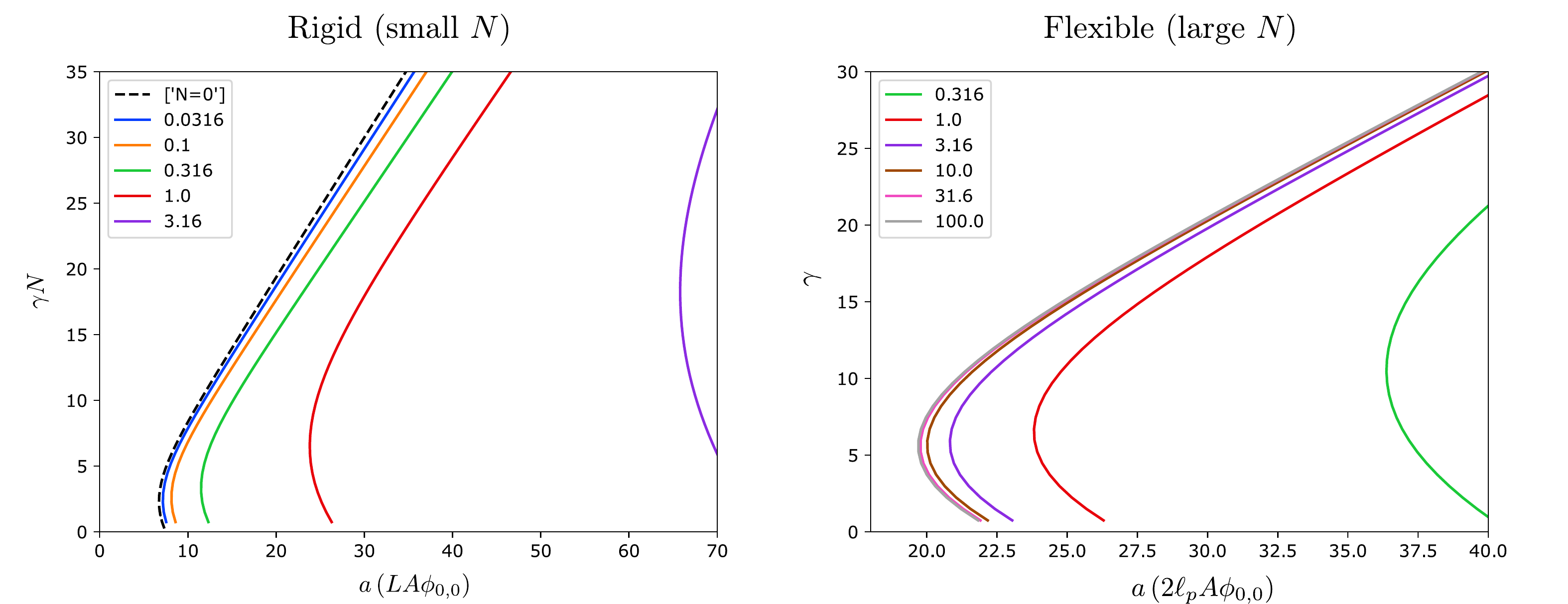}
\par\end{centering}
\caption[Field strength vs. Maier-Saupe parameter]{\label{fig:Mean-field-strength}Mean field strength $\gamma$ as a
function of Maier-Saupe $a$. The lower branch is nonphysical; the
nose corresponds to the limit of meta-stability of the aligned phase.
Curves are colored by the length of the polymer in Kuhn lengths, $N$.  The left plot is nondimensionalized by polymer length, which is convenient for rigid polymers.  The right plot is nondimensionalized by $2\ell_p$, which is convenient for flexible polymers
The dashed curve represents the rigid rod. }

\end{figure}

The procedure for calculating the value of $\left<\phi\right>_{2,0}^{MF}$ as a
function of $\gamma$ will be derived below, resulting in formula \ref{eq:phi_MF_matrix_basis}
with the inverse Laplace transform being performed numerically as
described in section \ref{sec:Numerical-Laplace-inversion}. In figure
\ref{fig:Mean-field-strength} we plot the mean field strength $\gamma$
in terms of $a$. These are plotted by solving~\ref{eq:Def_gamma} for $a$ as a function
of $\gamma$.
\begin{equation}
a=\frac{\gamma}{\frac{2}{3}\sqrt{\frac{4\pi}{5}}2\ell_{p}A\left<\phi\right>_{2,0}^{MF}\left(\gamma\right)}
\end{equation}
When working with relatively stiff polymers $L<2\ell_{p}$ it is natural
to nondimensionalize $a$ by the polymer volume $LA$. When working
with flexible polymers it is natural to nondimensionalize $a$ by
the volume of a Kuhn length $2\ell_{p}A$. Because a stronger $a$
is needed in dilute solutions, it is also convenient to multiply $a$
by the volume fraction of polymer $\phi_{0,0}$.

\subsection{Mean Field Solution for Rigid Rod and Fuzzball}

As the fuzzball and rigid rod have no Kuhn length, we instead define
their field strengths $\gamma$ in terms of their respective volumes
\begin{equation}
\gamma^{FB}=\frac{2}{3}a\left\langle \phi_{2,0}\right\rangle ^{MF}\sqrt{\frac{4\pi}{5}}v_{p}
\end{equation}
\begin{equation}
\gamma^{RR}=\frac{2}{3}a\left\langle \phi_{2,0}\right\rangle ^{MF}\sqrt{\frac{4\pi}{5}}LA
\end{equation}
so that the energy associated with each object is $E^{\left(1\right)}=-\gamma Y_{2,0}\left(\vec{u}_{i}\right)$.
The partition function for the orientation of a single fuzzball or
rigid rod is given by
\begin{equation}
z^{FB/RR}=\int d\vec{u}_{i}\exp\left(\gamma Y_{2,0}\left(\vec{u}_{i}\right)\right)
\end{equation}
To calculate expectation values for this partition function it is
useful to define the matrix
\begin{equation}
\mathbf{M}_{\ell_{1},\ell_{2}}^{m}=\int d\vec{u}_{i}Y_{\ell_{1},m}\left(\vec{u}_{i}\right)\exp\left(\gamma Y_{2,0}\left(\vec{u}_{i}\right)\right)Y_{\ell_{2},m}\left(\vec{u}_{i}\right)\label{eq:Def_of_M}
\end{equation}
The numerical values of $\mathbf{M}$ can be obtained
by matrix exponentiation 
\begin{equation}
\mathbf{M}^{m}=\exp\left(\gamma\mathbf{J}_{\left(2\right)}^{m,0,m}\right)
\end{equation}
where 
\begin{equation}
\left(\mathbf{J}_{\left(2\right)}^{m,0,m}\right)_{\ell_1,\ell_2}=\int d\vec{u}Y_{\ell_{1},m}\left(\vec{u}\right)Y_{2,0}\left(\vec{u}\right)Y_{\ell_{2},m}\left(\vec{u}\right)
\end{equation}
which is given in appendix \ref{sec:Appendix:-Real-spherical}.
As an example of how $\mathbf{M}$ can be used to calculate
an expectation, consider
\begin{align}
\left\langle \phi_{\ell,m}^{RR/FB}\right\rangle ^{MF} & =n_{p}\frac{\int d\vec{\mathrm{r}}\int d\vec{\mathrm{u}}\exp\left(\gamma Y_{2,0}\left(\vec{\mathrm{u}}_{j}\right)\right)\phi_{\ell,m}^{\left(1\right)RR/FB}}{\int d\vec{\mathrm{r}}\int d\vec{\mathrm{u}}\exp\left(\gamma Y_{2,0}\left(\vec{\mathrm{u}}_{j}\right)\right)}
\end{align}
which, with the help of $Y_{0,0}=\frac{1}{\sqrt{4\pi}}$, simplifies
to
\begin{equation}
\left\langle \phi_{\ell,m}^{RR}\right\rangle ^{MF}=\frac{n_{p}AL}{\mathscr{V}}\sqrt{\frac{1}{2\ell+1}}\frac{\mathbf{M}_{0,\ell_{2}}^{0}}{\mathbf{M}_{0,0}^{0}}\delta_{m,0}
\end{equation}
\begin{equation}
\left\langle \phi_{\ell,m}^{FB}\right\rangle ^{MF}=\frac{n_{p}v_{p}}{\mathscr{V}}\frac{1}{\sqrt{2\ell+1}}\frac{\mathbf{M}_{0,\ell}^{0}}{\mathbf{M}_{0,0}^{0}}\delta_{m,0}
.\end{equation}
 From the above we can verify the densities, $\left\langle \phi_{0,0}^{RR}\right\rangle ^{MF}=\frac{n_{p}AL}{\mathscr{V}}$
and $\left\langle \phi_{0,0}^{FB}\right\rangle ^{MF}=\frac{n_{p}v_{p}}{\mathscr{V}}$,
and get the mean field alignment $\left\langle \phi_{0,2}^{RR/FB}\right\rangle ^{MF}$
and then solve for the Maier-Saupe parameter
\begin{equation}
a^{RR}LA\phi_{00}=\frac{15\gamma^{RR}\boldsymbol{M}_{0,0}^{0}\left(\gamma^{RR}\right)}{4\sqrt{\pi}\mathbf{M}_{0,2}^{0}\left(\gamma^{RR}\right)}
\end{equation}
\begin{equation}
a^{FB}v_{p}\phi_{00}=\frac{15\gamma^{FB}\mathbf{M}_{0,0}^{0}\left(\gamma^{FB}\right)}{4\sqrt{\pi}\mathbf{M}_{0,2}^{0}\left(\gamma^{FB}\right)}
\end{equation}
plotted in figure \ref{fig:Mean-field-strength}.

\section{Stone Fence Diagrams\label{sec:Stone-Fence-diagrams}}

In this section we derive a propagator \textendash{} equations \ref{eq:Gmll}
and \ref{eq:LaplaceProp} \textendash{} which we will use to calculate
$\left\langle \phi\right\rangle ^{MF}$ in the next section.

To obtain a numerical value of $\left<\phi\right>^{MF}$ we follow the approach
of~\cite{spakowitzSemiflexiblePolymerSolutions2003, spakowitzExactResultsSemiflexible2004, spakowitzEndtoendDistanceVector2005} beginning with the propagator for
a semi-flexible WLC (assumption \ref{enu:Semi-flexable}) without
an aligning field~\cite{spakowitzExactResultsSemiflexible2004} (i.e. as if $\left<\phi\right>_{2}^{MF}=0$):
\begin{equation}
G_{o}\left(\vec{\mathrm{u}}|\vec{\mathrm{u}}_{0};L\right)=\sum_{\ell=0}^{\infty}\sum_{m=-\ell}^{\ell}Y_{\ell,m}\left(\vec{\mathrm{u}}\right)Y_{\ell,m}\left(\vec{\mathrm{u}}_{0}\right)e^{-\ell\left(\ell+1\right)\frac{L}{2\ell_{p}}}\label{eq:FreeChainPropagator}
\end{equation}
which gives the probability distribution for the orientation $\vec{\mathrm{u}}$
at one end of a chain of length $L$, given that the other end is
orientated in the $\vec{u}_{o}$ direction. (Note that in~\cite{spakowitzExactResultsSemiflexible2004} complex spherical harmonics were used. Because $\sum_{m=-\ell}^{\ell}Y_{\ell}^{m}\left(\vec{\mathrm{u}}\right)Y_{\ell}^{m*}\left(\vec{\mathrm{u}}_{0}\right)=P_{\ell}\left(\vec{u}\cdot\vec{u}_{0}\right)=\sum_{m=-\ell}^{\ell}Y_{\ell,m}\left(\vec{\mathrm{u}}\right)Y_{\ell,m}\left(\vec{\mathrm{u}}_{0}\right)$
equation \ref{eq:FreeChainPropagator} also holds for real spherical
harmonics.) $G_{o}$ ranges between the rigid rod $\lim_{L/2\ell_{p}\to0}G_{o}\left(\vec{\mathrm{u}}|\vec{\mathrm{u}}_{0};L\right)=\delta\left(\vec{\mathrm{u}}-\vec{\mathrm{u}}_{0}\right)$
and flexible chain $\lim_{L/2\ell_{p}\to\infty}G_{o}\left(\vec{\mathrm{u}}|\vec{\mathrm{u}}_{0};L\right)=\frac{1}{4\pi}$
and has the property $G_{o}\left(\vec{\mathrm{u}}|\vec{\mathrm{u}}_{0};L\right)=\int d\vec{\mathrm{u}}_1 G_{o}\left(\vec{\mathrm{u}}_{1}|\vec{\mathrm{u}}_{0};L_{1}\right)G_{o}\left(\vec{\mathrm{u}}|\vec{\mathrm{u}}_{1};L-L_{1}\right)$
where $\int d\vec{\mathrm{u}}_{1}$ is meant as integration over the
unit sphere. Going forward we will nondimensionalize the polymer length
by the Kuhn length $2\ell_{p}$ to give $N=L/2\ell_{p}$. 

The propagator $G_{o}$ allows us to calculate statistics about any point along a free chain. For example, suppose we have a WLC constrained
to start from orientation $\vec{u}_{0}$ and end at orientation $\vec{u}_{f}$
and there is some function $f$ that depends on the orientation $\vec{u}_{1}$
at a point $N_{1}=L_{1}/2\ell_{p}$ along a polymer. We can calculate
$f$'s expectation value via 
\begin{equation}
\left\langle f\left(\vec{u}_{1}\right)\right\rangle _{o}=\frac{\int d\vec{u}_{1}G_{o}\left(\vec{u}_{1}|\vec{\mathrm{u}}_{0};N_{1}\right)f\left(\vec{u}_{1}\right)G_{o}\left(\vec{\mathrm{u}}_{f}|\vec{u}_{1};N-N_{1}\right)}{G_{o}\left(\vec{\mathrm{u}}_{f}|\vec{\mathrm{u}}_{0};N\right)}\label{eq:IntermediateOrientation}
\end{equation}
This expectation value is precisely the type of quantity that we need
to calculate to obtain the propagator $G$ for a WLC which is in an
aligning field $\phi_{1}^{MF}V_{12}$ from equation \ref{eq:SelfConsistantEquation}.
In particular, if we use the expectation value in equation \ref{eq:IntermediateOrientation}
to include the aligning field $\gamma$ as a re-weighting of $G_{o}$
then the partition function for the end orientation is
\begin{equation}
G\left(\vec{\mathrm{u}}_{f}|\vec{\mathrm{u}}_{0},N\right)=G_{o}\left(\vec{\mathrm{u}}_{f}|\vec{\mathrm{u}}_{0};N\right)\left\langle \exp\left[\gamma\int_{0}^{N}dsY_{2,0}\left(\vec{\mathrm{u}}\left(s\right)\right)\right]\right\rangle _{o}\label{eq:orientation_partition}
\end{equation}
where the prefactor $G_{o}\left(\vec{\mathrm{u}}_{f}|\vec{\mathrm{u}}_{0};N\right)$
cancels the denominator of \ref{eq:IntermediateOrientation}. Please
note that $G$ is an unnormalized distribution that is proportional
to the probability distribution of $\vec{\mathrm{u}}_{f}$ for a given
$\vec{\mathrm{u}}_{0}$,
\begin{equation}
P\left(\vec{\mathrm{u}}_{f}|\vec{\mathrm{u}}_{0};N\right)=\frac{G\left(\vec{\mathrm{u}}_{f},\vec{\mathrm{u}}_{0},N\right)}{\int d\vec{\mathrm{u}}'_{f}G\left(\vec{\mathrm{u}}'_{f},\vec{\mathrm{u}}_{0},N\right)}\label{eq:rewightedPropagator}.
\end{equation}
$P$ and $G$ differ in that $\frac{1}{4\pi}\int d\vec{\mathrm{u}}_{0}P\left(\vec{\mathrm{u}}_{f}|\vec{\mathrm{u}}_{0};N\right)$
gives the probability distribution of $\vec{\mathrm{u}}_{f}$ assuming
$\vec{\mathrm{u}}_{0}$ is uniform over the unit sphere, while $\frac{1}{4\pi}\int d\vec{\mathrm{u}}_{0}G\left(\vec{\mathrm{u}}_{f}|\vec{\mathrm{u}}_{0},N\right)$
is proportional to the probability distribution of $\vec{\mathrm{u}}_{f}$
for a polymer of length $N$ in an aligning field whose ends are free
to rotate. Like $G_{0}$, the propagator $G$ has the property $G\left(\vec{\mathrm{u}}|\vec{\mathrm{u}}_{0};L\right)=\int d\vec{\mathrm{u}}G\left(\vec{\mathrm{u}}_{1}|\vec{\mathrm{u}}_{0};L_{1}\right)G\left(\vec{\mathrm{u}}|\vec{\mathrm{u}}_{1};L-L_{1}\right)$.
We can think of $G\left(\vec{\mathrm{u}}|\vec{\mathrm{u}}_{0};L\right)$
as the partition function for $\vec{u}$ and $\vec{u}_{0}$. The quantity
$G\left(\vec{\mathrm{u}}|\vec{\mathrm{u}}_{0};L\right)$ is useful
because it allows us to calculate chain statistics in the presence
of the $\gamma$ field. For example, the expectation of $f\left(\vec{\mathrm{u}}_{1}\right)$
for a chain in field of strength $\gamma$ and freely rotating ends
is given by
\begin{equation}
\left\langle f\left(\vec{\mathrm{u}}_{1}\right)\right\rangle =\frac{\int d\vec{\mathrm{u}}_{0}\int d\vec{\mathrm{u}}_{1}\int d\vec{\mathrm{u}}_{f}G\left(\vec{\mathrm{u}}_{1}|\vec{\mathrm{u}}_{0};L_{1}\right)f\left(\vec{\mathrm{u}}_{1}\right)G\left(\vec{\mathrm{u}}_{f}|\vec{\mathrm{u}}_{1};L-L_{1}\right)}{\int d\vec{\mathrm{u}}_{0}\int d\vec{\mathrm{u}}_{1}\int d\vec{\mathrm{u}}_{f}G\left(\vec{\mathrm{u}}_{1}|\vec{\mathrm{u}}_{0};L_{1}\right)G\left(\vec{\mathrm{u}}_{f}|\vec{\mathrm{u}}_{1};L-L_{1}\right)}\\
\end{equation}
where the denominator can be simplified to $\int d\vec{\mathrm{u}}_{0}\int d\vec{\mathrm{u}}_{f}G\left(\vec{\mathrm{u}}_{f}|\vec{\mathrm{u}}_{0};L\right)$.

We now turn to calculating $G\left(\vec{\mathrm{u}}|\vec{\mathrm{u}}_{0};L\right).$
Whereas equation \ref{eq:IntermediateOrientation} depended only on
the orientation at $s_{1}$, the expectation value in equation~\ref{eq:orientation_partition} depends
on the orientation at every point along the chain. In order to expand
$G$ in terms of $G_{o}$, we Taylor expand the exponential and reorder
the integrals 
\begin{widetext}
\begin{align}
\exp\left[\gamma\int_{0}^{N}dsY_{2,0}\left(\vec{\mathrm{u}}\left(s\right)\right)\right] & =\sum_{n=0}^{\infty}\frac{1}{n!}\left(\gamma\int_{0}^{N}dsY_{2,0}\left(\vec{\mathrm{u}}\left(s\right)\right)\right)^{n}\label{eq:gamma_expansion}\\
 & =\sum_{n=0}^{\infty}\left(\gamma^{n}\int_{0}^{N}ds_{n}...\int_{0}^{s_{3}}ds_{2}\int_{0}^{s_{2}}ds_{1}\prod_{i}^{n}Y_{2,0}\left(\vec{\mathrm{u}}_{i}\right)\right)\label{eq:TimeOrdering}
\end{align}
where we have introduced the short hand $\vec{\mathrm{u}}_{i}\equiv\vec{\mathrm{u}}\left(s_{i}\right)$.
The factor of $\frac{1}{n!}$ canceled the $n!$ from ordering the
integrals. To calculate the the $n=2$ term, for example, we have
\begin{align}
&\left\langle \gamma^{2}\int_{0}^{N}ds_{2}\int_{0}^{s_{2}}s_{1}Y_{2,0}\left(\vec{\mathrm{u}}_{2}\right)Y_{2,0}\left(\vec{\mathrm{u}}_{1}\right)\right\rangle _{o} \nonumber\\
&\hspace{1cm}=\gamma^{2}\int_{0}^{N}ds_{2}\int_{0}^{s_{2}}ds_{1}\int d\vec{\mathrm{u}}_{2}\int d\vec{\mathrm{u}}_{1}
G_{o}\left(\vec{\mathrm{u}}_{3}|\vec{\mathrm{u}}_{2};N-s_{2}\right)Y_{2,0}\left(\vec{\mathrm{u}}_{2}\right)
G_{o}\left(\vec{\mathrm{u}}_{2}|\vec{\mathrm{u}}_{1};s_{2}-s_{1}\right)Y_{2,0}\left(\vec{\mathrm{u}}_{1}\right)
G_{o}\left(\vec{\mathrm{u}}_{1}|\vec{\mathrm{u}}_{0};s_{1}\right) \label{eq:n2term}
\end{align}
\end{widetext}

\begin{figure}
\begin{centering}
\includegraphics[width=0.5\linewidth]{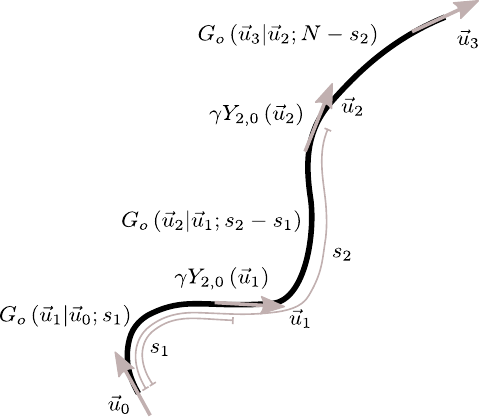}
\par\end{centering}
\caption[Propagator schematic]{\label{fig:propagator} Schematic illustrating the right hand side of equation~\ref{eq:n2term}.  The thick black curve represents a polymer $N$ Kuhn lengths long.  Arrows represent unit vectors $\vec{u_0}$, $\vec{u_1}$, $\vec{u_2}$, $\vec{u_3}$ at path lengths $0$, $s_1$, $s_2$, and $N$ along the polymer respectively.}
\end{figure}

To simplify equation \ref{eq:n2term}, it is convenient to perform
a Laplace transform from $N\to p$ . Defining the Laplace transform
of equation \ref{eq:FreeChainPropagator} as $\breve{G}_{o}\left(p\right)\equiv\mathcal{L}\left[G\right]=\int_{0}^{\infty}dNG_{o}\left(N\right)e^{-pN}$
gives
\begin{equation}
\breve{G}_{o}\left(\vec{\mathrm{u}}|\vec{\mathrm{u}}_{0};p\right)=\sum_{\ell=0}^{\infty}\sum_{m=-\ell}^{\ell}Y_{\ell,m}\left(\vec{\mathrm{u}}_{f}\right)Y_{\ell,m}\left(\vec{\mathrm{u}}_{0}\right)\frac{1}{p+\ell\left(\ell+1\right)}\label{eq:laplace_of_G0}
\end{equation}

The convolution rule for Laplace transforms says that $\mathcal{L}\left(f\right)\mathcal{L}\left(g\right)=\mathcal{L}\left[f*g\right]=\mathcal{L}\left[\int_{0}^{t}d\tau f\left(\tau\right)g\left(t-\tau\right)\right]$.
In equation \ref{eq:n2term} we have something of the form $\mathcal{L}\left[f*g*h\right]$.
By plugging the convolution rule into itself recursively we arrive
at the general relation $\mathcal{L}\left[f_{1}*f_{2}*...f_{n+1}\right]=\prod_{j=1}^{n+1}\mathcal{L}\left(f_{j}\right)$.
Taking the Laplace transform of the $n=2$ example in equation \ref{eq:n2term}
and substituting into \ref{eq:laplace_of_G0} we arrive at
\begin{align}
\mathcal{L}\left[\left\langle \mathrm{...}\right\rangle _{o}\right]= & \sum_{\ell_{i},m_{i}}Y_{l_{1},m_{1}}\left(\vec{\mathrm{u}}_{0}\right)\nonumber\\
& \times \breve{G}_{o}\left(p|\ell_{1}\right)\gamma\int d\vec{\mathrm{u}}_{1}Y_{\ell_{1},m_{1}}\left(\vec{\mathrm{u}}_{1}\right)Y_{2,0}\left(\vec{\mathrm{u}}_{1}\right)Y_{\ell_{2},m_{2}}\left(\vec{\mathrm{u}}_{1}\right)\nonumber\\
 & \times\breve{G}_{o}\left(p|\ell_{2}\right)\gamma\int d\vec{\mathrm{u}}_{2}Y_{\ell_{2},m_{2}}\left(\vec{\mathrm{u}}_{2}\right)Y_{2,0}\left(\vec{\mathrm{u}}_{2}\right)Y_{\ell_{3},m_{3}}\left(\vec{\mathrm{u}}_{2}\right)\nonumber \\
 & \times\breve{G}_{o}\left(p|\ell_{3}\right)Y_{l_{3},m_{2}}\left(\vec{\mathrm{u}}_{f}\right)\label{eq:Ln2term} 
\end{align}
where $\left\langle \mathrm{...}\right\rangle _{o}=\left\langle \gamma^{2}\int_{0}^{N}ds_{2}\int_{0}^{s_{2}}s_{1}Y_{2,0}\left(\vec{\mathrm{u}}_{2}\right)Y_{2,0}\left(\vec{\mathrm{u}}_{1}\right)\right\rangle _{o}$
and $\breve{G}_{o}\left(p|\ell\right)\equiv\frac{1}{p+\ell\left(\ell+1\right)}$
from equation \ref{eq:laplace_of_G0}. The integrals over the spherical
harmonics in the middle can be written in terms of $A_{\ell}^{m}$
and $\beta_{\ell}^{m}$ from appendix \ref{sec:Appendix:-Real-spherical}.
In particular equation \ref{eq:Y20_sum} gives us 
\begin{align}
&\int d\vec{u}_{1}Y_{\ell_{1},m_{1}}\left(\vec{u}_{1}\right)Y_{2,0}\left(\vec{u}_{1}\right)Y_{\ell_{2},m_{2}}\left(\vec{u}_{1}\right)\hspace{0.2cm}=\left(A_{\ell_{1}+2}^{m_{1}}\delta_{\ell_{1}+2,\ell_{2}}+\beta_{\ell_{1}}^{m_{1}}\delta_{\ell_{1},\ell_{2}}+A_{\ell_{1}}^{m_{1}}\delta_{\ell_{1}-2,\ell_{2}}\right)\delta_{m_{1},m_{2}}\label{eq:intYYY}
\end{align}
so that
\begin{align}
\mathcal{L}\left[\left\langle \mathrm{...}\right\rangle _{o}\right]= & \sum_{\ell_{1},m}Y_{l_{1},m}\left(\vec{\mathrm{u}}_{0}\right)\breve{G}_{o}\left(p|\ell_{1}\right)\nonumber \\
 & \times\left(A_{\ell_{1}+2}^{m}\delta_{\ell_{1}+2,\ell_{2}}+\beta_{\ell_{1}}^{m}\delta_{\ell_{1},\ell_{2}}+A_{\ell_{1}}^{m}\delta_{\ell_{1}-2,\ell_{2}}\right)\nonumber \\
 & \times \breve{G}_{o}\left(p|\ell_{2}\right) \nonumber\\
 & \times\left(A_{\ell_{2}+2}^{m}\delta_{\ell_{1}+2,\ell_{2}}+\beta_{\ell_{1}}^{m}\delta_{\ell_{1},\ell_{2}}+A_{\ell_{1}}^{m}\delta_{\ell_{1}-2,\ell_{2}}\right) \nonumber \\
 & \times \breve{G}_{o}\left(p|\ell_{3}\right)Y_{l_{3},m}\left(\vec{\mathrm{u}}_{f}\right)\label{eq:Nine_terms}
\end{align}
which distributes to 9 terms for each $m$ and $\ell_{1}$ (though
some are disallowed by the requirements that $\ell_{1}\geq\left|m\right|$).

We would like to apply the reasoning of equations \ref{eq:Ln2term}
and \ref{eq:intYYY} to all term in the sum \ref{eq:gamma_expansion},
the Laplace transform of $G\left(\vec{\mathrm{u}}_{f},\vec{\mathrm{u}}_{0},N\right)$.
However, multiplying \ref{eq:intYYY} by itself for each of the terms
in \ref{eq:gamma_expansion} will result in a very large number of
terms. The enumeration of these terms is greatly aided by the use
of stone fence diagrams~\cite{yamakawaHelicalWormlikeChains1997, spakowitzExactResultsSemiflexible2004, spakowitzEndtoendDistanceVector2005}. Each diagram represents a
term, with the sum of all diagrams equaling $G\left(\vec{\mathrm{u}}_{f},\vec{\mathrm{u}}_{0},N\right)$.
The simplest such diagram \textendash{} and only type at the $n=0$
level of \ref{eq:gamma_expansion} \textendash{} is the orientation
propagator without an aligning field $\left\langle \ \right\rangle _{o}=\breve{G}_{o}\left(p|\ell_{3}\right)$
and represented by a dot:
\begin{center}
\includegraphics{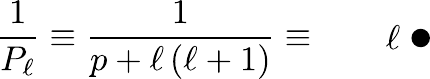}
\par\end{center}

\begin{flushleft}
There is one of these diagrams for each of the values of $\ell$ - indicated
by the $\ell$ to the left of the diagram. In this case, the incoming
$\ell_{0}$ and outgoing $\ell_{f}$ are always the same, as indicated
by the same $\ell$ for the spherical harmonics of $\vec{u}_{f}$
and $\vec{u}_{0}$ in equation \ref{eq:laplace_of_G0}. However, diagrams
for $n>0$, which include factors of $\gamma$, may connect different
$\ell$ values. The form of equation \ref{eq:Ln2term} suggests that
we should connect these dots with products of spherical harmonics.
There are three types of connections in equation \ref{eq:intYYY}
corresponding to keeping the $\ell$ value the same, increasing $\ell$
by 2, or decreasing $\ell$ by 2. These are represented by horizontal,
upwardly slanted, and downwardly slanted connections respectively:
\par\end{flushleft}

\begin{center}
\includegraphics{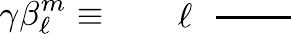}
\par\end{center}

and
\begin{center}
\includegraphics{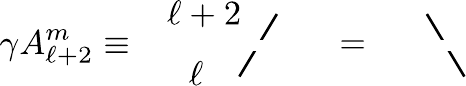}
\par\end{center}

in this way the vertical axis corresponds to $\ell$ as is indicated
by the expressions to the left of the diagrams. The three possible
diagrams starting at $\ell$ and having a single $\gamma$ are
\begin{center}
\includegraphics{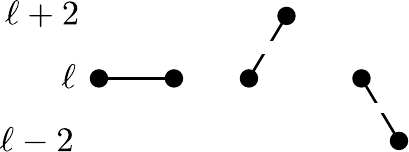}
\par\end{center}

and 4 of the 9 possible diagrams proportional to $\gamma^{2}$ are
\begin{center}
\includegraphics{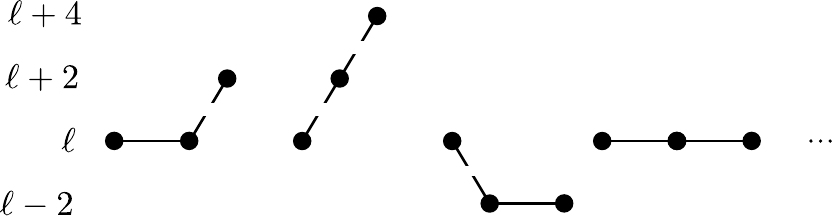}
\par\end{center}

These nine diagrams correspond to the nine terms in equation \ref{eq:Nine_terms}.
For each eigenvalue set $\left\{ m_{0},\ell_{0}\right\} $ describing
the orientation of $\vec{\mathrm{u}}_{0}$, there are $3^{n}$ such
diagrams from the $\gamma^{n}$ term of the expansion \ref{eq:gamma_expansion};
though many of these will be zero due to the requirement that $\ell\geq\left|m\right|$
which is enforced by $A_{\ell}^{\ell}=A_{\ell}^{\ell+1}=0$ and disallows
diagrams that go below $\left|m\right|$ anywhere along their path.
Our task is to sum the allowed diagrams with any number, $n$, of sections.
We define a new symbol to denote the sum of all such diagrams that
don't contain any factors of $A$ so that it starts and ends at $\ell$.
\begin{widetext}
\begin{center}
\includegraphics{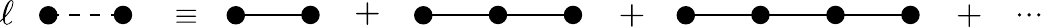}
\par\end{center}
which can be evaluated via a geometric series
\begin{equation}
\sum_{n=0}^{\infty}\left(\frac{1}{P_{\ell}}\right)^{n+1}\left(\gamma\beta_{\ell}^{m}\right)^{n}=\frac{1}{P_{\ell}-\gamma\beta_{\ell}^{m}}
\end{equation}

The critical insight made in~\cite{spakowitzExactResultsSemiflexible2004} is to sum all of the diagrams
that start and end at $\ell$ but never stray below $\ell$ in between.
We define this new quantity as 
\begin{center}
\includegraphics{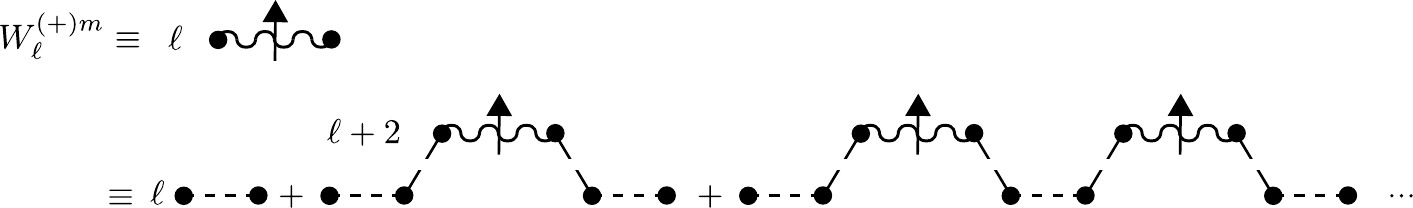}
\par\end{center}
which can be summed with a geometric series
\begin{equation}
W_{\ell}^{\left(+\right)m}=\sum_{n=0}^{\infty}\left(\frac{1}{P_{\ell}-\gamma\beta_{\ell}^{m}}\right)^{n+1}\left(\left(\gamma A_{\ell+2}^{m}\right)^{2}W_{\ell+2}^{\left(+\right)m}\right)^{n}=\frac{1}{P_{\ell}-\gamma\beta_{\ell}^{m}-\left(\gamma A_{\ell+2}^{m}\right)^{2}W_{\ell+2}^{\left(+\right)m}}\label{eq:Wplus}
\end{equation}
which can calculated recursively by truncating at a sufficiently high
$\ell$ value.

Likewise
\begin{center}
\includegraphics{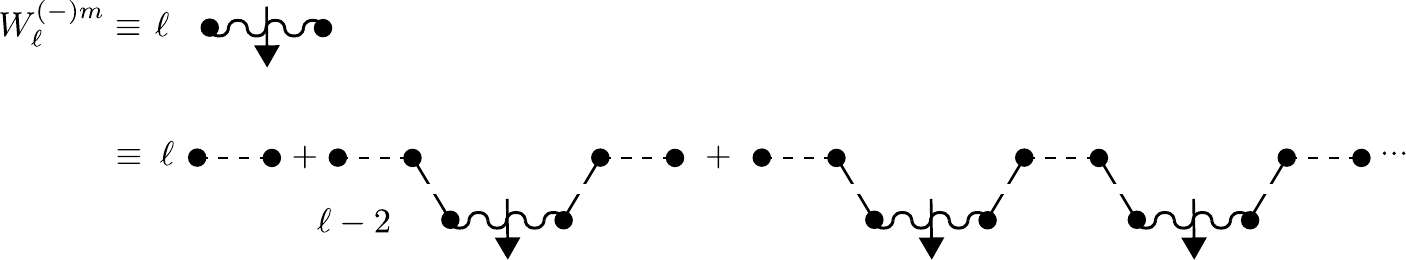}
\par\end{center}
\begin{equation}
W_{\ell}^{\left(-\right)m}=\begin{cases}
\frac{1}{P_{\ell}-\gamma B_{\ell}^{m}-\left(\gamma A_{\ell}^{m}\right)^{2}W_{\ell-2}^{\left(-\right)m}} & \ell\geq\left|m\right|\\
0 & \mathrm{otherwise}
\end{cases}\label{eq:Wminus}
\end{equation}
where we have explicitly cut off the recursion when $\ell$ dips below
$\left|m\right|$.

Also of interest is the diagram similar to $W^{\left(+\right)}$ but
without the left dot and requiring at least one up-down step
\begin{center}
\includegraphics{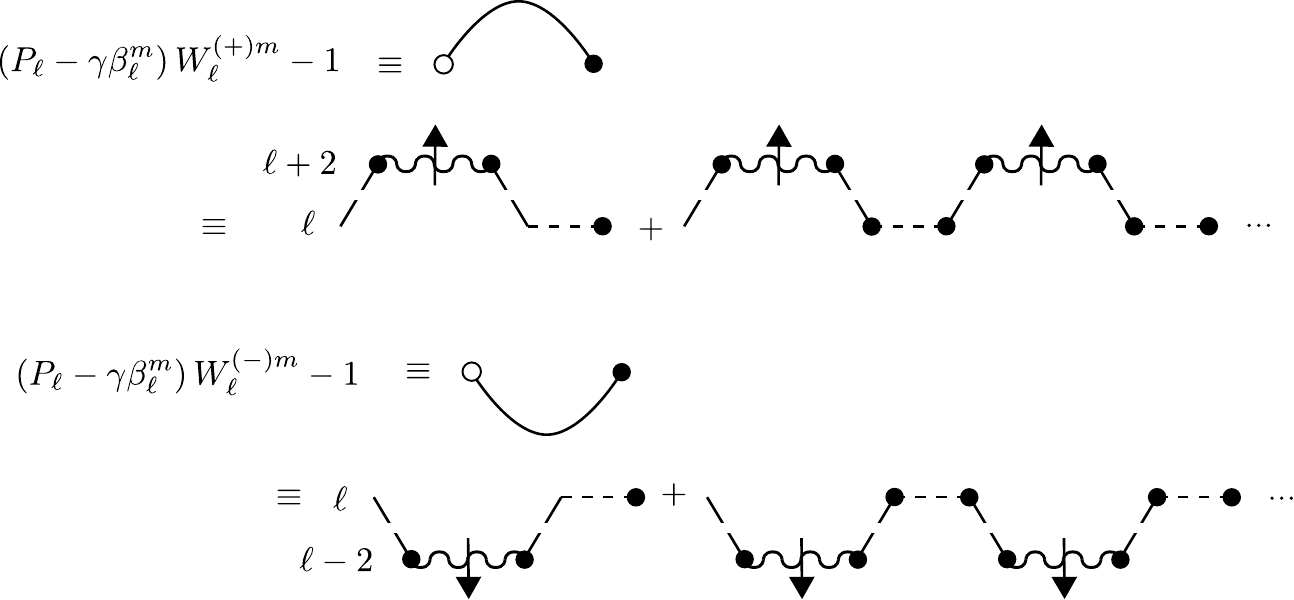}
\par\end{center}

To get all the diagrams that begin and end at $\ell$ we need to include
all possible sets of up steps and down steps. We group these diagrams
into four groups: 1, up first down last; 2, up first up last; 3, down
first up last; and 4, down first down last.
\begin{center}
\includegraphics{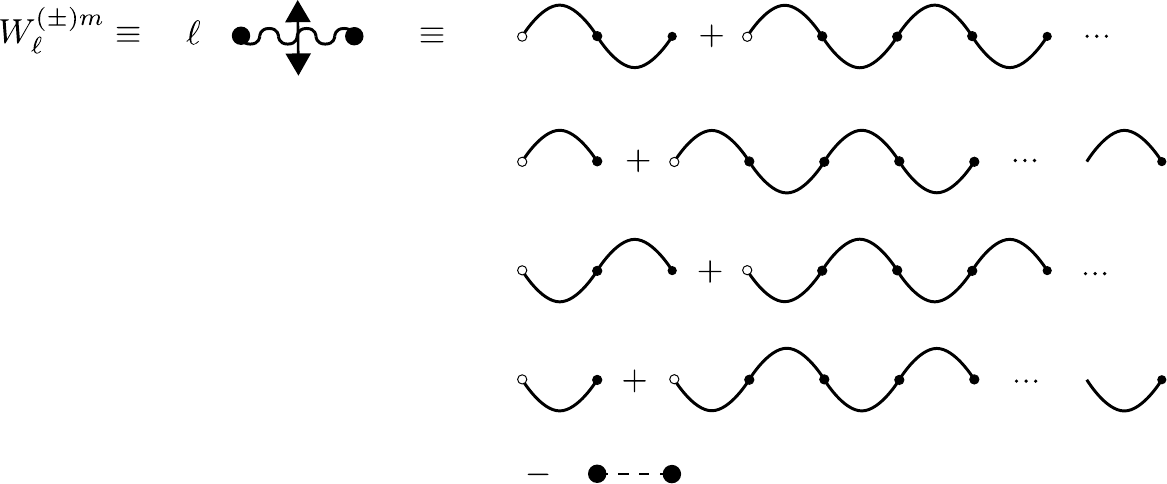}
\par\end{center}

where the final $-1/\left(P_{\ell}-\gamma B_{\ell}^{m}\right)$ is
to remove the double counting of \includegraphics{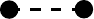}.

We may write the sum

\begin{align}
W_{\ell}^{\left(\pm\right)m} & =\left(\frac{1}{P_{\ell}-\gamma\beta_{\ell}^{m}}\right)\sum_{n=0}^{\infty}\left[\left(\left(P_{\ell}-\gamma\beta_{\ell}^{m}\right)W_{\ell}^{\left(+\right)m}-1\right)\left(\left(P_{\ell}-\gamma\beta_{\ell}^{m}\right)W_{\ell}^{\left(-\right)m}-1\right)\right]^{n}\nonumber \\
 & \ \ \ \ \times\left(2+\left(P_{\ell}-\gamma\beta_{\ell}^{m}\right)W_{\ell}^{\left(+\right)m}-1+\left(P_{\ell}-\gamma\beta_{\ell}^{m}\right)W_{\ell}^{\left(-\right)m}-1\right)-\left(\frac{1}{P_{\ell}-\gamma\beta_{\ell}^{m}}\right)\nonumber \\
 & =\frac{W_{\ell}^{\left(+\right)m}W_{\ell}^{\left(-\right)m}}{W_{\ell}^{\left(-\right)m}-\left(P_{\ell}-\gamma\beta_{\ell}^{m}\right)W_{\ell}^{\left(+\right)m}W_{\ell}^{\left(-\right)m}+W_{\ell}^{\left(+\right)m}}\label{eq:Wpm}
\end{align}

The term $W_{\ell}^{\left(\pm\right)m}$ gives all the diagrams starting
and ending at $m$ and $\ell$. The sum of diagrams that connect different
$\ell$ values 
\begin{center}
\includegraphics{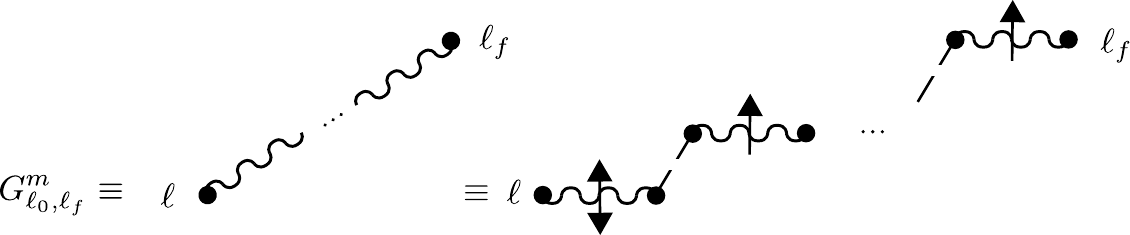}
\par\end{center}
\end{widetext}
can be calculated recursively

\begin{equation}
G_{\ell_{0}\ell_{f}}^{m}=\begin{cases}
0 & \mathrm{if\ }\ell_{f}-\ell_{0}\ \mathrm{is\ odd}\\
W_{\ell_{0}}^{\left(\pm\right)m} & \mathrm{if\ }\ell_{f}=\ell_{0}\\
G_{\ell_{0},\ell_{f}-2}^{m}W_{\ell_{f}}^{\left(+\right)m}A_{\ell_{f}}^{m}\gamma & \mathrm{if\ }\ell_{f}>\ell_{0}\\
G_{\ell_{0},\ell_{f}+2}^{m}W_{\ell_{f}}^{\left(-\right)m}A_{\ell_{f+2}}^{m}\gamma & \mathrm{if\ }\ell_{f}<\ell_{0}
\end{cases}\label{eq:Gmll}
\end{equation}

with the application of the flanking harmonics we have our desired
propagator

\begin{equation}
\breve{G}\left(\vec{\mathrm{u}}|\vec{\mathrm{u}}_{0};p\right)=\sum_{\ell=0}^{\infty}\sum_{m=-\ell}^{\ell}Y_{\ell,m}\left(\vec{\mathrm{u}}_{0}\right)G_{\ell_{0}\ell_{f}}^{m}Y_{\ell,m}\left(\vec{\mathrm{u}}\right)\label{eq:LaplaceProp}
\end{equation}

\section{Evaluating Expectations\label{sec:Evaluating-expectations}}

In this section we will show how the propagator $\breve{G}$ can be
used to calculate expectation values of various quantities (e.x. $\left\langle \phi_{2,0}\right\rangle ^{MF}$
or $\left\langle \phi_{2,1}\phi_{2,1}\right\rangle ^{MF})$ for WLC's
in an aligning mean field.

In the mean field solution, the $n_{p}$ polymers are distributed
over a volume $\mathscr{V}$. Taking the expectation value of equation
\ref{eq:density} we have

\begin{equation}
\left\langle \phi_{2,0}\right\rangle ^{MF}=\left\langle n_{p}\frac{1}{\mathscr{V}}A\sqrt{\frac{4\pi}{5}}2\ell_{p}\int_{0}^{N}dsY_{2,0}\left(\vec{u}\left(s\right)\right)\right\rangle ^{MF}
\end{equation}

or 

\begin{equation}
\left\langle \phi_{2,0}\right\rangle ^{MF}=\sqrt{\frac{4\pi}{5}}\frac{\phi_{0,0}}{N}\int_{0}^{N}ds\left\langle Y_{2,0}\left(\vec{u}\left(s\right)\right)\right\rangle ^{MF}
\end{equation}
The probability the chain will point in direction $\vec{u}$ at path
length $s$ along the chain is weighted by the propagator from the
beginning of the chain to $s$, i.e. $G\left(\vec{u}|\vec{u}_{0};s\right)$,
and the propagator from $s$ to the other end of the chain $G\left(\vec{u}_{f}|\vec{u};L-s\right)$.
The Laplace transform from $N$ to $p$ of $\left\langle \phi_{2,0}\right\rangle ^{MF}$
can be preformed using the convolution rule (see discussion of equation
\ref{eq:Ln2term})

\begin{equation}
\mathcal{L}\left[\int_{0}^{N}dsG\left(\vec{u}_{f}|\vec{u};p\right)G\left(\vec{u}|\vec{u}_{0};s\right)\right]=\breve{G}\left(\vec{u}_{1}|\vec{u}_{0};p\right)\breve{G}\left(\vec{u}_{f}|\vec{u}_{1};p\right)
\end{equation}
To calculate $\left\langle \phi_{2,0}\right\rangle ^{MF}$ we want
to integrate over beginning, middle, and end orientations as in
\begin{widetext}
\begin{equation}
\left\langle \phi_{2,0}\right\rangle ^{MF}=\frac{\phi_{0,0}}{N}\sqrt{\frac{4\pi}{5}}\frac{\mathcal{L}^{-1}\left[\int du_{0}\int du_{1}\int du_{f}\breve{G}\left(\vec{u}_{1}|\vec{u}_{0};p\right)Y_{2,0}\left(\vec{u}_{1}\right)\breve{G}\left(\vec{u}_{f}|\vec{u}_{1};p\right)\right]\left(N\right)}{\mathcal{L}^{-1}\left[\int du_{0}\int du_{f}\breve{G}\left(\vec{u}_{f}|\vec{u}_{0};p\right)\right]\left(N\right)}\label{eq:phi_MF_in_u_basis}
\end{equation}
\end{widetext}
where the denominator provides the appropriate normalization. We can
then substitute equation \ref{eq:LaplaceProp} into \ref{eq:phi_MF_in_u_basis}
to change basis from $\vec{u}$ to $Y_{\ell,m}$. The sums over $\ell$
can be made implicit by writing the result in matrix form where the
rows and columns run over $\ell$ values so that

\begin{equation}
\left(\mathbf{G}^{m}\right)_{\ell_{1},\ell_{2}}=G_{\ell_{0}\ell_{f}}^{m}
\end{equation}
and the product of three real spherical harmonics is defined as

\begin{equation}
\left(\mathbf{J}_{\left(\ell\right)}^{m,m',m''}\right)_{\ell_{1},\ell_{2}}\equiv\int d\vec{u}Y_{\ell_{1},m}\left(\vec{u}\right)Y_{\ell,m'}\left(\vec{u}\right)Y_{\ell_{2}m''}\left(\vec{u}\right)
\end{equation}
The values of $\mathbf{J}_{\left(\ell\right)}^{m,m',m''}$ are derived
in appendix \ref{sec:Appendix:-Real-spherical}. After the change
of basis, equation \ref{eq:phi_MF_in_u_basis} becomes

\begin{equation}
\phi_{2,0}^{MF}=\frac{\phi_{0,0}}{N}\sqrt{\frac{4\pi}{5}}\frac{\mathcal{L}^{-1}\left[\boldsymbol{e}_{0}\cdot\mathbf{G}^{0}\cdot\mathbf{J}_{\left(2\right)}^{0,0,0}\cdot\mathbf{G}^{0}\cdot\boldsymbol{e}_{0}\right]\left(N\right)}{\mathcal{L}^{-1}\left[\boldsymbol{e}_{0}\cdot\mathbf{G}^{0}\cdot\boldsymbol{e}_{0}\right]\left(N\right)}\label{eq:phi_MF_matrix_basis}
\end{equation}
were $\boldsymbol{e}_{0}=\left[1,0,0,0...\right]$ is the unit vector so that ${e}_{0}\cdot\mathbf{G}^{0}\cdot\boldsymbol{e}_{0}=\mathbf{G}^{0}_{1,1}$.
The pinning to $\ell=0$ at the chain ends results from inserting
$Y_{0,0}=\frac{1}{\sqrt{4\pi}}$ and applying $\int d\vec{u}Y_{0,0}Y_{\ell,m}=\delta_{0,\ell}\delta_{0,m}$.
Furthermore, the requirement that $\left|m\right|\leq\ell$ required $m_{1}=m_{2}=0$
in equation \ref{eq:phi_MF_matrix_basis}. The inverse Laplace transforms
can be performed numerically as described in section \ref{sec:Numerical-Laplace-inversion}.

The process of connecting sequential propagators together can be extended
to the expectation value of several spherical harmonics
\begin{widetext}
\begin{eqnarray}
&&\int_{0}^{N}ds_{n}...\int_{0}^{s_{2}}ds_{1}\left\langle \prod_{j=1}^{n}Y_{\ell_{j},m_{j}}\left(s_{j}\right)\right\rangle ^{MF} 
=...\nonumber\\
&&\frac{\mathcal{L}^{-1}\left[\sum_{\mathscr{M},...\mathscr{M}'''}\boldsymbol{e}_{0}\cdot\mathbf{G}^{0}\cdot\mathbf{J}_{\left(\ell_{n}\right)}^{0,m_{n},\mathscr{M}}\cdot\mathbf{G}^{\mathscr{M}}\cdot\mathbf{J}_{\left(\ell_{n-1}\right)}^{\mathscr{M},m_{n-1},\mathscr{M}'}\cdot...\cdot\mathbf{G}^{\mathscr{M}'''}\cdot\mathbf{J}_{\left(\ell_{1}\right)}^{\mathscr{M}''',m_{1},0}\cdot\mathbf{G}^{0}\cdot\boldsymbol{e}_{0}\right]}{\mathcal{L}^{-1}\left[\boldsymbol{e}_{0}\cdot\mathbf{G}^{0}\cdot\boldsymbol{e}_{0}\right]} \label{eq:How_to_take_expectation}
\end{eqnarray}
\end{widetext}
\section{Fluctuations\label{subsec:Fluctuations}}

Having solved for the average field values $\left<\phi\right>^{MF}$ (and by extension
$W_{1}^{MF}=-iV_{12}\left<\phi\right>_{2}^{MF}$) we now investigate local fluctuations
about these values by substituting $\phi=\left<\phi\right>^{MF}+\delta\phi$ and
$W=W^{MF}+\delta W$ into \ref{eq:SimplePartition}. Resulting constant
terms such as $W^{MF}\left<\phi\right>^{MF}$ can be absorbed into the proportionality
constant which only offsets the reference energy. Furthermore, the
mean field solution $W^{MF}$ found above was uniform in space so
the term $W_{1}^{MF}\delta\phi_{1}$ must be zero. Equation \ref{eq:SimplePartition}
can further be simplified by using $iW_{1}^{MF}=V_{12}\left<\phi\right>_{2}^{MF}$
yielding:
\begin{equation}
Z\propto\int\mathscr{D}\delta W\delta\phi\exp\left[-\frac{1}{2}\delta\phi_{1}V_{12}\delta\phi_{2}+i\delta W_{1}\delta\phi_{1}+n_{p}\log\left(z_{p}\right)\right]\label{eq:fluctiationPartition}
\end{equation}
\begin{equation}
z_{p}=\int\mathscr{D}r_{i}^{(1)}\exp\left[-\beta E_{poly}-\hat{\phi}_{1}^{(1)}V_{12}\left<\phi\right>_{2}^{MF}-i\delta W_{1}\hat{\phi}_{1}^{(1)}\right]
\end{equation}

Per assumption \ref{enu:Small_Perterbation}, the field $\phi$ only
makes small deviations from its mean field value $\left<\phi\right>^{MF}$ so $\delta\phi$
is small. Also, under RPA (assumption \ref{enu:RPA}) $W$ only makes
small deviations from its mean value $W^{MF}$ constraining $\delta W$
to be small. Taylor expanding $z_{p}$ about $\delta W=0$ gives
\begin{equation}
z_{p}\approx1+\left\langle -i\delta W_{1}\hat{\phi}_{1}^{(1)}\right\rangle ^{MF}+\frac{1}{2}\left(\left\langle -i\delta W_{1}\hat{\phi}_{1}^{(1)}\right\rangle ^{MF}\right)^{2}
\end{equation}
(up to an additive constant of no consequence). We expand the logarithm
$\log\left(1+x\right)=x-\frac{1}{2}x^{2}$ in equation \ref{eq:fluctiationPartition}
and substituting in $z_{p}$ we have
\begin{align}
Z\propto\int\mathscr{D}W\delta\phi & e^{-\frac{1}{2}\delta\phi_{1}V_{12}\delta\phi_{2}+i\delta W_{1}\delta\phi_{1}
-n_{p}\frac{1}{2}\delta W_{1}S_{12}\delta W_{2}} \label{Zlong}
\end{align}
where
\begin{equation}
S_{12}\equiv\left\langle \hat{\phi}_{1}^{(1)}\hat{\phi}_{2}^{(1)}\right\rangle ^{MF}-\left\langle \hat{\phi}_{1}^{(1)}\right\rangle ^{MF}\left\langle \hat{\phi}_{2}^{(1)}\right\rangle ^{MF}\label{eq:StructureFactor}.
\end{equation}

We then perform a Gaussian integral over $\delta W$ to arrive at
\begin{equation}
Z\propto\int\mathscr{D}\delta\phi\exp\left[-\frac{1}{2}\delta\phi_{1}\left(V_{12}+\frac{1}{n_{p}}S_{12}^{-1}\right)\delta\phi_{2}\right]\label{eq:QuadraticPartision}
\end{equation}
where $S_{12}^{-1}$ is defined via $S_{12}^{-1}S_{23}=\delta_{\ell_{1},\ell_{3}}\delta_{m_{1},m_{3}}\delta\left(\vec{r}_{1}-\vec{r}_{3}\right)$. Appendix \ref{HowToInvertS} discusses how to perform this inverse.

The energy expression in \ref{eq:QuadraticPartision}, which is often
written with $\Gamma_{12}\equiv V_{12}+\frac{1}{n_p}S_{12}^{-1}$, describes
the quadratic fluctuations of $\phi$ around the mean field solution.
The bigger an element of $\Gamma$ is, the smaller the corresponding
fluctuation will be. Given the critical importance of \ref{eq:QuadraticPartision}
we will devote considerable effort to calculating $S^{-1}$.

Because of translational invariance (assumption \ref{enu:TranslationalInvariance}),
the Fourier transform $\vec{r}\to\vec{k}$ of $S_{12}$ is much easier
to invert than the real space $S_{12}$. Following the conventions
in appendix~\ref{sec:Appendix_Fourier}, 
\begin{equation}
Z\propto\int\mathscr{D}\delta\tilde{\phi}\exp\left[-\frac{1}{2}\delta\tilde{\phi}_{1}\left(\tilde{V}_{12}+\frac{1}{n_p}\tilde{S}_{12}^{-1}\right)\delta\tilde{\phi}_{2}\right] 
\end{equation}
where $\tilde{V}_{12}=\left(4\pi\chi\delta_{\ell_{1},0}-\frac{8\pi}{15}a\delta_{\ell_{1},2}\right)\delta_{m_{1},m_{2}}\delta_{\ell_{1},\ell_{2}}\delta\left(\vec{k}_{1}+\vec{k}_{2}\right)$.
In other words, the Fourier modes of fluctuations in $\phi$ are decoupled
from each other (at least to quadratic order in $\phi$). Another
advantage of the Fourier transform is that by assumptions \ref{enu:TranslationalInvariance},
\ref{enu:GoodSolvent}, and \ref{enu:Small_Perterbation} and as discussed
in section \ref{subsec:Mean-Field-Solution} the mean field solution
is uniform in space so the latter term in equation \ref{eq:StructureFactor},
$\left\langle \hat{\phi}_{1}^{(1)}\right\rangle ^{MF}\left\langle \hat{\phi}_{2}^{(1)}\right\rangle ^{MF}$,
contributes only at $\vec{k}=0$. While we will be interested
in the limit as $\vec{k}\to0$, the value at precisely $\vec{k}=0$
will not be used so we can drop this constant term. We will refer
to the first term on the right of equation \ref{eq:StructureFactor}
as $\delta S_{12}\equiv\left\langle \hat{\phi}_{1}^{(1)}\hat{\phi}_{2}^{(1)}\right\rangle ^{MF}$.

\section{Calculating S for Fuzzball and Rigid Rod}

Calculating $\delta S$ for a fuzzball or rigid rod can be accomplished
using $\mathbf{M}$ from equation \ref{eq:Def_of_M}
which amounts to the expectation of a product of two spherical harmonics
in an applied field. Inserting equation \ref{eq:density_FB} into
$\delta S_{12}\equiv\left\langle \hat{\phi}_{1}^{(1)}\hat{\phi}_{2}^{(1)}\right\rangle ^{MF}$
for the fuzzball the spatial and rotational terms, $\rho$ and $v$
respectively, factor because the position and orientation of a
fuzzball are unrelated.
\begin{equation}
\delta\tilde{S}_{12}^{FB}=\rho\left(\vec{r}_{1},\vec{r}_{2}\right)v\left(\ell_{1},m_{1},\ell_{2},m_{2}\right)\label{eq:split_fuzzball}
\end{equation}
\begin{equation}
\rho\left(\vec{r}_{1},\vec{r}_{2}\right)=\frac{1}{\mathscr{V}}\int d\vec{r}_{i}\frac{v_{p}^{2}}{\left(2\pi\sigma^{2}\right)^{3}}\exp\left(-\frac{\left(\vec{r}_{1}-\vec{\mathrm{r}}_{i}\right)^{2}}{2\sigma^{2}}-\frac{\left(\vec{r}_{2}-\vec{\mathrm{r}}_{i}\right)^{2}}{2\sigma^{2}}\right)
\end{equation}
\begin{equation}
v\left(\ell_{1},m_{1},\ell_{2},m_{2}\right)=\frac{4\pi}{\sqrt{\left(2\ell_{1}+1\right)\left(2\ell_{2}+1\right)}}\frac{\mathbf{M}_{\ell_{1},\ell_{2}}^{m_{1}}\delta_{m_{1},m_{2}}}{4\pi\mathbf{M}_{0,0}^{0}}
\end{equation}
where the denominators come from the single fuzzball partition function
$z^{FB}=\mathscr{V}\int d\vec{u}\exp\left(\gamma Y_{2,0}\left(\vec{u}_{i}\right)\right)=\mathscr{V}4\pi M_{0,0}^{0}$.
Taking the Fourier transform of a Gaussian yields a Gaussian, hence
\begin{equation}
\delta\tilde{S}_{12}^{FB}=\frac{v_{p}^{2}}{\mathscr{V}}\delta\left(\vec{k}_{1}+\vec{k}_{2}\right)e^{-\sigma^{2}\vec{k}^{2}}\frac{\mathbf{M}_{\ell_{1},\ell_{2}}^{m_{1}}\delta_{m_{1},m_{2}}}{\sqrt{\left(2\ell_{1}+1\right)\left(2\ell_{2}+1\right)}\mathbf{M}_{0,0}^{0}}\label{eq:S_fuzzball}
\end{equation}
where the $\delta\left(\vec{k}_{1}+\vec{k}_{2}\right)$ arises from
translational invariance.

The position along a rigid rod is inherently coupled to the orientation
of the rod so the factorization into $\rho$ and $v$ is not possible.
In particular, writing out $\delta\tilde{S}=\left\langle \hat{\phi}_{1}^{(1)}\hat{\phi}_{2}^{(1)}\right\rangle ^{MF}$
explicitly from equation \ref{eq:density_RR} we have
\begin{widetext}
\begin{align}
\delta S_{12}^{RR}  =&\frac{1}{z^{RR}}\frac{A^{2}}{\sqrt{\left(2\ell_{1}+1\right)\left(2\ell_{2}+1\right)}}\nonumber \\
&\times\int d\vec{\mathrm{r}}_{j}\int d\vec{\mathrm{u}}_{j}e^{\gamma Y_{2,0}\left(\vec{u}_{j}\right)}\int_{0}^{L}ds_{1}\delta\left(\vec{\mathrm{r}}_{j}+s_{1}\vec{\mathrm{u}}_{j}-\vec{r}_{1}\right)\int_{0}^{L}ds_{2}\delta\left(\vec{\mathrm{r}}_{j}+s_{2}\vec{\mathrm{u}}_{j}-\vec{r}_{2}\right)Y_{\ell_{1,}m_{1}}\left(\vec{u}_{j}\right)Y_{\ell_{2},m_{2}}\left(\vec{u}_{j}\right) \label{S_long}
\end{align}
where the single rod partition function is $z^{RR}=\mathscr{V}\int d\vec{u}\exp\left(\gamma Y_{2,0}\left(\vec{u}_{i}\right)\right)=\mathscr{V}4\pi\mathbf{M}_{0,0}^{0}$.
Applying the Fourier transform from $\vec{r}\to\vec{k}$ takes
\begin{equation}
\delta\left(\vec{\mathrm{r}}_{j}+s_{1}\vec{\mathrm{u}}_{j}-\vec{r}_{1}\right)\delta\left(\vec{\mathrm{r}}_{j}+s_{2}\vec{\mathrm{u}}_{j}-\vec{r}_{2}\right)\to\delta\left(\vec{k}_{1}+\vec{k}_{2}\right)\exp\left(i\vec{k}\cdot\vec{\mathrm{u}}_{j}\left(s_{1}-s_{2}\right)\right).
\end{equation}
The exponential can be expanded to quadratic order and the $s$ integrals
evaluated directly 
\begin{equation}
\int_{0}^{L}ds_{1}\int_{0}^{L}ds_{2}\exp\left(i\vec{k}\cdot\vec{\mathrm{u}}_{j}\left(s_{1}-s_{2}\right)\right)\approx L^{2}-\frac{1}{2}\left(\vec{k}\cdot\vec{\mathrm{u}}_{j}\right)^{2}\frac{L^{4}}{6}.
\end{equation}
Introducing $\kappa$ defined in equation \ref{eq:k_dot_u} and doing some
simplification we arrive at
\begin{align}
\delta\tilde{S}_{12}^{RR}=\frac{A^{2}\delta\left(\vec{k}_{1}+\vec{k}_{2}\right)}{\sqrt{\left(2\ell_{1}+1\right)\left(2\ell_{2}+1\right)}} & \left[\frac{L^{2}}{\mathscr{V}}\frac{\mathbf{M}_{\ell_{1},\ell_{2}}^{m_{1}}}{\mathbf{M}_{0,0}^{0}}\delta_{m_{1},m_{2}} -\frac{4\pi}{z^{RR}}\frac{L^{4}}{12}\sum_{m=-1}^{1}\sum_{m'=-1}^{1}\kappa_{1,m}\kappa_{1,m'}\sum_{\mathscr{M}}\left(\mathbf{J}_{\left(1\right)}^{m_{1},m,\mathscr{M}}\cdot\mathbf{M}^{\mathscr{M}}\cdot\mathbf{J}_{\left(1\right)}^{\mathscr{M},m',m_{2}}\right)_{\ell_{1},\ell_{2}}\right]\label{eq:S_RR_formula}
\end{align}
where the integer $\mathscr{M}$ runs over all values allowed by the selection
rules of $\mathbf{J}$. We now have an explicit formula for the low
$k$ expansion of $\delta\tilde{S}^{RR}$.

\section{Calculating S for WLC}

We now turn to finding $\delta S_{12}$ for the WLC. Writing $\delta S_{12}$
explicitly 
\begin{align}
\left(\delta S_{12}\right)_{\ell_{1}\ell_{2}}^{m_{1}m_{2}}\left(\vec{r}_{1},\vec{r}_{2}\right)= & \frac{4\pi}{\sqrt{\left(2\ell_{1}+1\right)\left(2\ell_{2}+1\right)}}A^{2}\int_{0}^{L}ds_{2}\int_{0}^{L}ds_{1}\left\langle Y_{\ell_{1},m_{1}}\left(\vec{\mathrm{u}}\left(s_{1}\right)\right)\delta\left(\vec{r}_{1}-\mathrm{\vec{r}}\left(s_{1}\right)\right)Y_{\ell_{2},m_{2}}\left(\vec{\mathrm{u}}\left(s_{2}\right)\right)\delta\left(\vec{r}_{2}-\vec{\mathrm{r}}\left(s_{2}\right)\right)\right\rangle ^{MF}  \label{Another_long_S}
\end{align}

Note the distinction between the argument $\vec{r}_1$ and the unitalicized $\vec{\mathrm{r}}(s_1)$ which stands for position of a point $s_1$ along the polymer. We take the Fourier transform $\vec{r}_{1}\to\vec{k}_{1}$ and $\vec{r}_{2}\to\vec{k}_{2}$
via $\frac{1}{\left(2\pi\right)^{3}}\int d\vec{r}_{1}d\vec{r}_{2}\exp\left(i\vec{k}_{1}\vec{r}_{1}+i\vec{k}_{2}\vec{r}_{2}\right)$.
The integrals over $\vec{r}$ are eliminated by delta functions to
give $\exp\left(i\vec{k}_{1}\vec{\mathrm{r}}\left(s_{1}\right)+i\vec{k}_{2}\vec{\mathrm{r}}\left(s_{2}\right)\right)$
which can be factored into $\exp\left[i\vec{k}_{1}\cdot\left(\vec{\mathrm{r}}_{1}-\vec{\mathrm{r}}_{2}\right)\right]\exp\left[i\left(\vec{k}_{1}+\vec{k}_{2}\right)\vec{\mathrm{r}}_{2}\right]$
where we have introduced the notation $\vec{\mathrm{r}}\left(s_{2}\right)=\vec{\mathrm{r}}_{2}$.
By translational invariance (assumption \ref{enu:TranslationalInvariance})
we can average out $\vec{\mathrm{r}}_{2}$ over the system volume
$\mathscr{V}$ by applying $\mathscr{V}^{-1}\int d\vec{\mathrm{r}}_{2}$
and recognizing the delta function $\left(2\pi\right)^{3}\delta\left(\vec{k}_{1}+\vec{k}_{2}\right)=\int d\vec{r}e^{i\vec{r}\left(\vec{k}_{1}+\vec{k}_{2}\right)}$,
giving

\begin{align}
\delta\tilde{S}_{12}= & \frac{4\pi}{\sqrt{\left(2\ell_{1}+1\right)\left(2\ell_{2}+1\right)}}\frac{A^{2}\left(2\ell_{p}\right)^{2}}{\mathscr{V}}\delta\left(\vec{k}_{1}+\vec{k}_{2}\right)
\int_{0}^{N}ds_{2}\int_{0}^{N}ds_{1}\left\langle Y_{\ell_{1},m_{1}}\left(\vec{\mathrm{u}}_{1}\right)\exp\left[i\vec{k}_{1}\cdot\left(\vec{\mathrm{r}}_{1}-\vec{\mathrm{r}}_{2}\right)\right]Y_{\ell_{2},m_{2}}\left(\vec{\mathrm{u}}_{2}\right)\right\rangle ^{MF} \label{eq:Fourier_of_deltaS}
\end{align}
\end{widetext}
where we nondimensionalized $L=\frac{N}{2\ell_{p}}$ so that $\int_{0}^{L}ds_{2}$
became $2\ell_{p}\int_{0}^{N}ds_{2}$. It will be convenient to write
$\vec{k}$ in spherical coordinates. Because $\vec{k}$ is a
vector, it is an $\ell=1$ object. Specifically we will define $\kappa_{\ell,m}$
as
\begin{equation}
\kappa_{1,m}\equiv\sqrt{\frac{4\pi}{3}}\begin{cases}
k_{y} & m=-1\\
k_{z} & m=0\\
k_{x} & m=1
\end{cases}\label{eq:DefinitionOfKappa}
\end{equation}
which has the property
\begin{equation}
\vec{k}\cdot\vec{u}=\sum_{m=-1}^{1}\kappa_{1,m}Y_{1,m}\left(\vec{u}\right)\label{eq:k_dot_u}
\end{equation}
which is useful because $\vec{k}_{1}\cdot\left(\vec{\mathrm{r}}_{1}-\vec{\mathrm{r}}_{2}\right)$
in equation \ref{eq:Fourier_of_deltaS} can be written as $\vec{k}_{1}\cdot\left(2\ell_{p}\right)\int_{s_{1}}^{s_{2}}ds\vec{\mathrm{u}}\left(s\right)=\int_{s_{1}}^{s_{2}}ds\sum_{m=-1}^{1}\kappa'_{1,m}Y_{1,m}\left(\vec{u}\right)$
where $\kappa'=2\ell_{p}\kappa$. It can be shown \textendash{} or
verified numerically \textendash{} that $\int_{0}^{N}ds_{2}\int_{0}^{N}ds_{1}$
may be replaced with $2\int_{0}^{N}ds_{2}\int_{0}^{s_{2}}ds_{1}$
in this expression. To further simplify the expression we define the
nondimensionalized 
\begin{equation}
\tilde{S}'_{12}\left(\vec{k}\right)\equiv\delta\tilde{S}_{12}\left(\vec{k}_{1},-\vec{k}_{1}\right)\frac{\mathscr{V}}{A^{2}\left(2\ell_{p}\right)^{2}N^{2}}
\end{equation}

By assumption \ref{enu:small_k} we are interested in the large distance
(small $\vec{k}$) behavior, so we Taylor expand the exponential in
\ref{eq:Fourier_of_deltaS} which, along with the above modifications,
becomes
\begin{widetext}
\begin{align}
\tilde{S}'_{12} &\approx \frac{4\pi}{\sqrt{\left(2\ell_{1}+1\right)\left(2\ell_{2}+1\right)}}\frac{2}{N^{2}}\nonumber \\
 & \times\left[\int_{0}^{N}ds_{2}\int_{0}^{s_{2}}ds_{1}\left\langle Y_{\ell_{1},m_{1}}\left(\vec{\mathrm{u}}_{1}\right)Y_{\ell_{2},m_{2}}\left(\vec{\mathrm{u}}_{2}\right)\right\rangle ^{MF}\right.\nonumber \\
 & +\sum_{m=-1}^{1}i\kappa'_{1,m}\int_{0}^{N}ds_{3}\int_{0}^{s_{3}}ds_{2}\int_{0}^{s_{2}}ds_{1}\left\langle Y_{\ell_{1},m_{1}}\left(\vec{\mathrm{u}}_{1}\right)Y_{1,m}\left(\vec{\mathrm{u}}_{2}\right)Y_{\ell_{2},m_{2}}\left(\vec{\mathrm{u}}_{3}\right)\right\rangle ^{MF}\nonumber \\
 & \left.-\sum_{m,m'=-1}^{1}\kappa'_{1,m}\kappa'_{1,m'}\int_{0}^{N}ds_{4}\int_{0}^{s_{4}}ds_{3}\int_{0}^{s_{3}}ds_{2}\int_{0}^{s_{2}}ds_{1}\left\langle Y_{\ell_{1},m_{1}}\left(\vec{\mathrm{u}}_{1}\right)Y_{1,m}\left(\vec{\mathrm{u}}_{2}\right)Y_{1,m'}\left(\vec{\mathrm{u}}_{3}\right)Y_{\ell_{2},m_{2}}\left(\vec{\mathrm{u}}_{4}\right)\right\rangle ^{MF}\right]\label{eq:Sprime_expand}
\end{align}

The expectation values in \ref{eq:Sprime_expand} can be evaluated
as described in equation \ref{eq:How_to_take_expectation}. The first
term in the series becomes
\begin{equation}
\int_{0}^{N}ds_{2}\int_{0}^{s_{2}}ds_{1}\left\langle Y_{\ell_{1},m_{1}}\left(\vec{\mathrm{u}}_{1}\right)Y_{\ell_{2},m_{2}}\left(\vec{\mathrm{u}}_{2}\right)\right\rangle ^{MF}=\frac{\mathscr{\mathcal{L}}^{-1}\left[\boldsymbol{e}_{0}\cdot\mathbf{G}^{0}\cdot\mathbf{J}_{\left(\ell_{1}\right)}^{0,m_{1},m_{1}}\cdot\mathbf{G}^{m_{1}}\cdot\mathbf{J}_{\left(\ell_{2}\right)}^{m_{1},m_{1},0}\cdot\mathbf{G}^{0}\cdot\boldsymbol{e}_{0}\right]\delta_{m_{1},m_{2}}}{\mathscr{\mathcal{L}}^{-1}\left[\boldsymbol{e}_{0}\cdot\mathbf{G}^{0}\cdot\boldsymbol{e}_{0}\right]}
\end{equation}
the linear term is killed by selection rules for $\mathbf{J}$ and
the quadratic term is given by
\begin{align}
 & \int_{0}^{N}ds_{4}\int_{0}^{s_{4}}ds_{3}\int_{0}^{s_{3}}ds_{2}\int_{0}^{s_{2}}ds_{1}\left\langle Y_{\ell_{1},m_{1}}\left(\vec{\mathrm{u}}_{1}\right)Y_{1,m}\left(\vec{\mathrm{u}}_{2}\right)Y_{1,m'}\left(\vec{\mathrm{u}}_{3}\right)Y_{\ell_{2},m_{2}}\left(\vec{\mathrm{u}}_{4}\right)\right\rangle ^{MF}\nonumber \\
 & =\frac{\mathscr{\mathcal{L}}^{-1}\left[\sum_{\mathscr{M}}\boldsymbol{e}_{0}\cdot\mathbf{G}^{0}\cdot\mathbf{J}_{\left(\ell_{1}\right)}^{0,m_{1},m_{1}}\cdot\mathbf{G}^{m_{1}}\cdot\mathbf{J}_{\left(1\right)}^{m_{1},m,\mathscr{M}}\cdot\mathbf{G}^{\mathscr{M}}\cdot\mathbf{J}_{\left(1\right)}^{\mathscr{M},m',m_{2}}\cdot\mathbf{G}^{m_{2}}\cdot\mathbf{J}_{\left(\ell_{2}\right)}^{m_{2},m_{2},0}\cdot\mathbf{G}^{0}\cdot\boldsymbol{e}_{0}\right]}{\mathscr{\mathcal{L}}^{-1}\left[\boldsymbol{e}_{0}\cdot\mathbf{G}^{0}\cdot\boldsymbol{e}_{0}\right]}. 
\end{align}
\end{widetext}

\section{Cancellation}

Having gone though the effort to calculate the low $k$ behavior of
$\delta\tilde{S}$ it is worth taking some time to consider the form
of the result. In this paper we are primarily interested in the
density $\ell=0$ and quadrupole $\ell=2$ interactions. It is convenient
two write $\delta\tilde{S}$ in the basis 
\begin{equation}
\left(\ell_{i},m_{i}\right)=\left[\left(0,0\right),\left(2,0\right),\left(2,-1\right),\left(2,1\right),\left(2,-2\right),\left(2,2\right)\right].
\end{equation}

In the long wavelength limit there are only correlations between fluctuations
of the same $m$ values, i.e. $\lim_{k\to0}\delta\tilde{S}\propto\delta_{m_{1},m_{2}}$.
As it turns out, in this basis we have
\begin{equation}
\lim_{\vec{\kappa}\to0}\delta\tilde{S}_{\ell_{1}\ell_{2}}^{m_{1},m_{2}}=\begin{bmatrix}t & h & 0 & 0 & 0 & 0\\
h & x & 0 & 0 & 0 & 0\\
0 & 0 & y & 0 & 0 & 0\\
0 & 0 & 0 & y & 0 & 0\\
0 & 0 & 0 & 0 & z & 0\\
0 & 0 & 0 & 0 & 0 & z
\end{bmatrix}\label{eq:thxyzMtrx}
\end{equation}
where the expressions for $t,\ h,\ x,\ y,\ \mathrm{and\ z}$ depend
on the whether we are talking about a WLC, rigid rod, or fuzzball.
In section \ref{subsec:Frank-Elastic-Energies1} we will discuss the
meaning of the various fluctuations (see equation \ref{eq:Interp_of_delta_phi} and figure \ref{fig:delta_realY}).
For now, we will point out that $t$ governs density-density interactions,
$x$ governs variations in alignment strength, $y$ governs variations
in alignment direction, and $z$ governs fluctuations in a secondary
direction of alignment. As the Frank elastic constants are of most
interest to us we will focus on $y$. Here we provide plots of $\lim_{\vec{\kappa}\to0}\delta\tilde{S}_{\ell_{1}\ell_{2}}^{m_{1},m_{2}}$
for a nearly rigid WLC.
\begin{figure}
\begin{centering}
\includegraphics[width=0.5\linewidth]{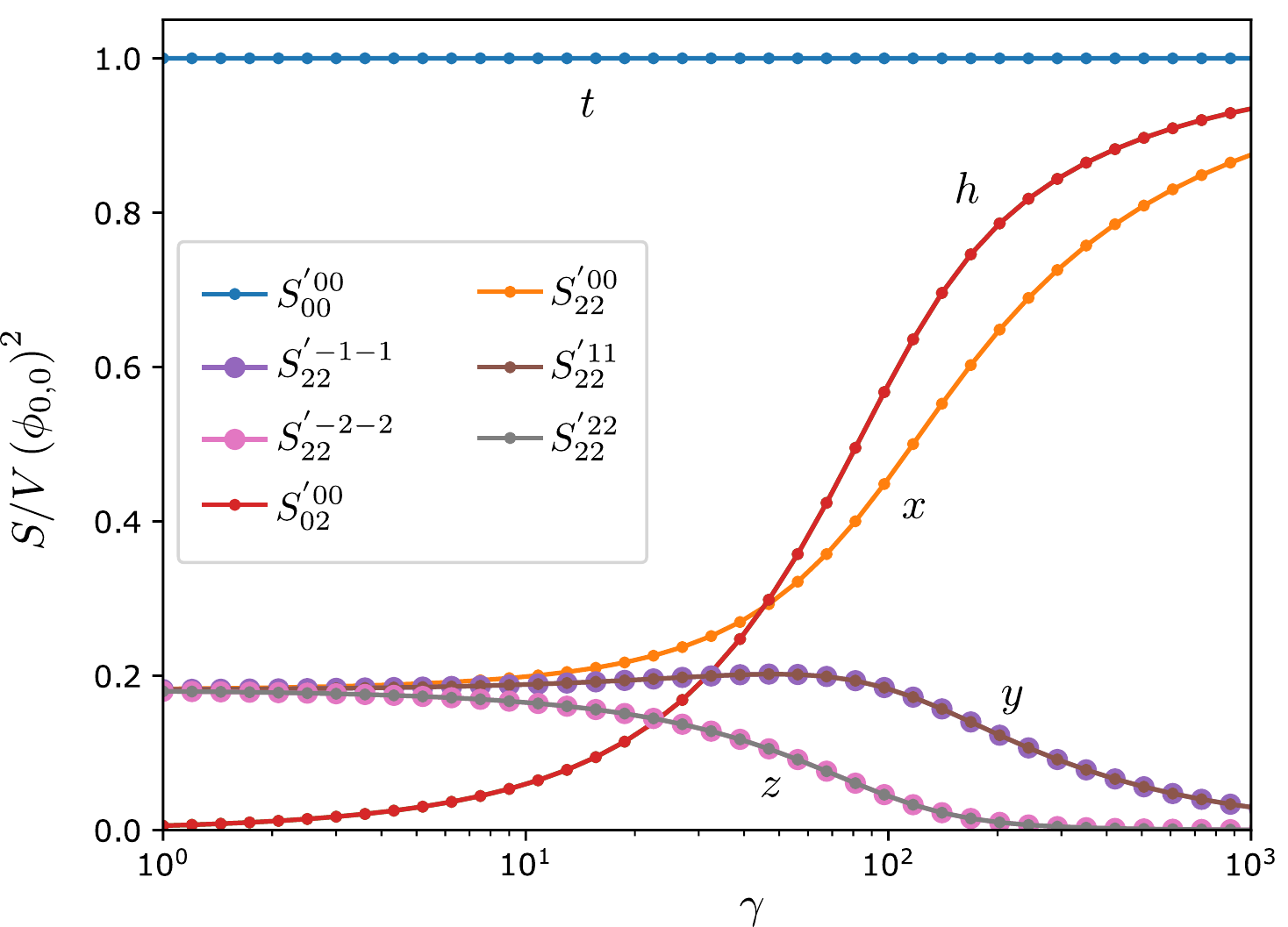}
\par\end{centering}
\caption[Density and orientation correlations]{\label{fig:S_plot} Plot of selected terms of $\lim_{k\to 0}\delta\tilde{S}$ as a function of field strength. Because $\lim_{k\to 0}\delta\tilde{S}^{1,1}_{2,2}=\lim_{k\to 0}\delta\tilde{S}^{-1,-1}_{2,2}$ and $\lim_{k\to 0}\delta\tilde{S}^{2,2}_{2,2}=\lim_{k\to 0}\delta\tilde{S}^{-2,-2}_{2,2}$ some curves are the same.}
\end{figure}

In the isotropic limit, $\gamma\to0$, all directions $Y_{2,m}$ are
identical. As alignment in the $\hat{z}$ direction (i.e. $Y_{2,0}$)
grows, it does so at the expense of the alignment in the other directions. 

When $\delta\tilde{S}$ is inverted and substituted into equation
\ref{eq:QuadraticPartision} we get the energy coefficients 
\begin{equation}
\tilde{\Gamma}_{12}=\left(\chi\delta_{\ell_{1},0}-\frac{2}{3}a\delta_{\ell_{1},2}\right)\delta_{m_{1},m_{2}}\delta_{\ell_{1},\ell_{2}}\delta\left(\vec{k}_{1}+\vec{k}_{2}\right)+\frac{1}{n_{p}}\delta\tilde{S}_{12}^{-1}\label{eq:Gamma_and_cancel}
\end{equation}
for the various modes where the inverse is discussed in~\ref{HowToInvertS}. Critically, we have verified numerically, that
in the limit that $\vec{\kappa}\to0$ that $y$ precisely cancels
with $\frac{2}{3}a$ term from the original interaction potential.
This cancellation occurs at all field strengths $\gamma$. This cancellation
comes about because of the invariance of the energy to a rotation
of the entire system. Hence, the system will exhibit long distance
deformations in the direction of alignment. It is only the rate of
change of these directions that contribute to the energy, hence the
derivatives, $\nabla$, in the Frank elastic expressions.

\section{Frank Elastic Energies\label{subsec:Frank-Elastic-Energies1}}

In this section we will define the nematic director and Frank elastic
constants and their relation to the orientation field $\hat{\phi}$.
The Frank elastic energy density depends on the orientations $\vec{u}$
in the vicinity $\Delta V$. The orientation distribution can be described
by the 3x3 matrix 
\begin{equation}
M=\frac{\int^{\Delta V}d\vec{r}\sum_{j=1}^{n_{p}}A\int_{0}^{L}ds\delta\left(\vec{r}-\vec{\mathrm{r}}_{j}\left(s\right)\right)\vec{\mathrm{u}}_{j}\left(s\right)\otimes\vec{\mathrm{u}}_{j}\left(s\right)}{\int^{\Delta V}d\vec{r}\sum_{j=1}^{n_{p}}A\int_{0}^{L}ds\delta\left(\vec{r}-\vec{\mathrm{r}}_{j}\left(s\right)\right)}\label{eq:DefinitionOfM}
\end{equation}
the denominator of which normalizes out the local polymer density,
$\phi_{0,0}$, leaving only the directional information. The $\Delta V$
is the coarse-graining volume such that details of the structure smaller
than this volume will not be treated. Going forward we will drop the
$\int^{\Delta V}$ . The eigen-decomposition of the $M$ matrix is~\cite{turziDistortioninducedEffectsNematic2007}
\begin{equation}
M=\lambda_{1}\vec{n}\otimes\vec{n}+\lambda_{2}\vec{e}_{2}\otimes\vec{e}_{2}+\lambda_{3}\vec{e}_{3}\otimes\vec{e}_{3}\label{eq:DefinitionOfn}
\end{equation}
where $\lambda_{1}\geq\lambda_{2}\geq\lambda_{3}$  This defines the unit
length nematic director $\vec{n}$ which indicates the general direction
of alignment within $\Delta V$. As $\vec{n}$ is an eigenvector,
its sign is arbitrarily chosen. The eigenvalues of $M$, which obey
$\lambda_{1}+\lambda_{2}+\lambda_{3}=1$ and $\lambda_{i}\geq0$,
designate the degree of alignment~\cite{turziDistortioninducedEffectsNematic2007}. The interpretation
of $\lambda_{i}$ can be understood in terms of the traceless alignment
matrix 
\begin{equation}
Q\equiv M-\frac{1}{3}Ic_{1}=c_{1}\left(\vec{n}\otimes\vec{n}-\frac{1}{3}I\right)+c_{2}\left(\vec{e}_{2}\otimes\vec{e}_{2}-\vec{e}_{3}\otimes\vec{e}_{3}\right)\label{eq:definition_of_Q}
\end{equation}
where 
\begin{equation}
c_{1}=\frac{3\lambda_{1}-1}{2},\ \ \ \ c_{2}=\lambda_{2}-\lambda_{1}
\end{equation}
with $c_{1}\in\left[0,1\right]$ describing the degree of alignment
in the primary direction and $c_{2}$ describing the degree to which
a secondary direction of alignment exists. 

The Frank elastic energy density is the sum of bend, twist, and splay
energies and is defined in terms of the nematic director
\begin{align}
E_{Frank}=&\frac{1}{2}K_{bend}\left\langle \left(\vec{n}\times\nabla\times\vec{n}\right)^{2}\right\rangle +\frac{1}{2}K_{twist}\left\langle \left(\vec{n}\cdot\nabla\times\vec{n}\right)^{2}\right\rangle  +\frac{1}{2}K_{splay}\left\langle \left(\nabla\cdot\vec{n}\right)^{2}\right\rangle \tag{\ref{eq:FrankElasticEnergy} revisited}\nonumber
\end{align}

\begin{figure}[htpb]
\begin{centering}
\includegraphics[width=0.5\linewidth]{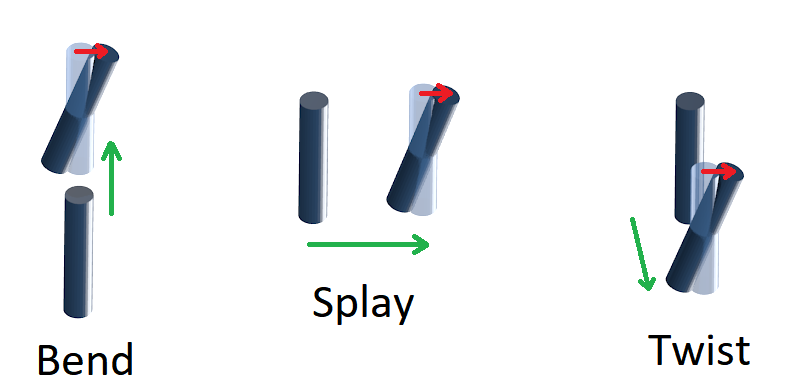}
\par\end{centering}
\caption[Bend, twist, and splay definitions]{\label{fig:Frank_Elastic_Pic_2}Pictorial representation clarifying the chief difference between bend, twist, and splay. The axes of the cylinders represent the direction of nematic ordering, $\vec{n}$. Each green arrow represents a displacement to a different position in a liquid crystal. Red arrows represent the change in orientation associated with the change in position.  A semitransparent cylinder represents the new position but with the old orientation as a guide to the eye.}
\end{figure}

To see why there are exactly three deformation modes, consider figure~\ref{fig:Frank_Elastic_Pic_2}.  Each mode prescribes a change in orientation associated with a displacement.  For each mode there is a direction $\vec{n}$ of alignment (represented by the cylinder), a displacement (green arrow), and a change in the direction of alignment (red arrow).  The displacement must be perpendicular to $\vec{n}$ as it is a differential change.  The displacement, therefore, can be in one of three direction: 1) parallel to $\vec{n}$, in which case the deformation is bend; 2) parallel to the change in direction, in which case the deformation is splay; and 3) perpendicular to both, in which case the deformation is twist.

We will devote the reminder of this derivation to finding the Frank elastic
constants $K_{bend}$, $K_{twist}$, and $K_{splay}$ in terms of
polymer properties. Our first goal will be to solve for the associated
expectation values in terms of $\hat{\phi}_{\ell,m}$. We begin by
exploring the relation between $\hat{\phi}$ and $\vec{\mathrm{u}}_{j}\left(s\right)\otimes\vec{\mathrm{u}}_{j}\left(s\right)$.
Using the real spherical harmonic for $\ell=2$
\begin{equation}
Y_{2,m}\left(\vec{u}\right)=\sqrt{\frac{5}{4\pi}}\begin{cases}
\sqrt{3}u_{x}u_{y} & m=-2\\
\sqrt{3}u_{y}u_{z} & m=-1\\
\frac{1}{2}\left(3u_{z}^{2}-1\right) & m=0\\
\sqrt{3}u_{z}u_{x} & m=1\\
\frac{\sqrt{3}}{2}\left(u_{x}^{2}-u_{y}^{2}\right) & m=2
\end{cases}
\end{equation}
and definition \ref{eq:density} we can write 
\begin{align}
Q & =\frac{1}{\phi_{0,0}}\sum_{j=1}^{n_{p}}A\int_{0}^{L}ds\delta\left(\vec{r}-\vec{\mathrm{r}}_{j}\left(s\right)\right)\left(\vec{\mathrm{u}}_{j}\left(s\right)\otimes\vec{\mathrm{u}}_{j}\left(s\right)-\frac{1}{3}\mathbb{I}\right)\label{eq:RestateQ}\\
 & =\frac{1}{\phi_{0,0}\sqrt{3}}\begin{bmatrix}\hat{\phi}_{2,2}-\frac{1}{\sqrt{3}}\hat{\phi}_{2,0} & \hat{\phi}_{2,-2} & \hat{\phi}_{2,1}\\
\hat{\phi}_{2,-2} & -\hat{\phi}_{2,2}-\frac{1}{\sqrt{3}}\hat{\phi}_{2,0} & \hat{\phi}_{2,-1}\\
\hat{\phi}_{2,1} & \hat{\phi}_{2,-1} & \frac{2}{\sqrt{3}}\hat{\phi}_{2,0}
\end{bmatrix}\label{eq:phi_2asMatrix}
\end{align}

The magnitude of $\hat{\phi}_{2,m}$ integrated over a coarse-graining
volume $\Delta V$ depends on the direction and degree of local alignment
of $\vec{u}$ vectors and grows with amount of polymer within $\Delta V$. 

Explicitly writing the relation between $\hat{n}$ and $\phi_{2,m}$
is complicated as it involves the cubic equation for the eigenvalues
of $M$. This relation can be much simplified by first making some assumptions. By assumption \ref{eq:SimplePartition},
there is a primary direction of alignment requiring $\lambda_{1}>\lambda_{2}$
implying $c_{1}>0$. Furthermore, assumption \ref{enu:Small_Perterbation}
requires the direction of alignment to be approximately in the $\hat{z}$
direction. Excluding terms that are quadratic in the deviation from
$\hat{z}$ (i.e. of order $n_{x/y}^{2}$ or higher) we write 
\begin{equation}
\vec{n}\approx\hat{z}+n_{x}\hat{x}+n_{y}\hat{y}
\end{equation}
\begin{equation}
\vec{e}_{2}\approx\left(\hat{x}-n_{x}\hat{z}\right)\cos\left(\theta\right)+\left(\hat{y}-n_{y}\hat{z}\right)\sin\left(\theta\right)
\end{equation}
\begin{equation}
\vec{e}_{3}\approx\left(\hat{y}-n_{y}\hat{z}\right)\cos\left(\theta\right)-\left(\hat{x}-n_{x}\hat{z}\right)\sin\left(\theta\right)
\end{equation}
where $\theta$ specifies the orientation of any azimuthal asymmetry.
Substituting into \ref{eq:definition_of_Q}
\begin{equation}
\vec{e}_{2}\approx\cos\left(\theta\right)\hat{x}+\sin\left(\theta\right)\hat{y}-\left(n_{y}\sin\left(\theta\right)+n_{x}\cos\left(\theta\right)\right)\hat{z}
\end{equation}
\begin{equation}
\vec{e}_{3}\approx-\sin\left(\theta\right)\hat{x}+\cos\left(\theta\right)\hat{y}+\left(-n_{y}\cos\left(\theta\right)+n_{x}\sin\left(\theta\right)\right)\hat{z}
\end{equation}
and
\begin{widetext}
\begin{equation}
Q=\begin{bmatrix}-\frac{1}{3}c_{1}+c_{2}\cos2\theta & c_{2}\sin2\theta & c_{1}n_{x}-c_{2}\left(\sin2\theta n_{y}+\cos2\theta n_{x}\right)\\
- & -\frac{1}{3}c_{1}-c_{2}\cos2\theta & c_{1}n_{y}+c_{2}\left(\cos2\theta n_{y}-\sin2\theta n_{x}\right)\\
- & - & \frac{2}{3}c_{1}
\end{bmatrix}
\end{equation}
where the dashes indicate a symmetric matrix.

By assumption \ref{enu:AzimuthallySymmetric} we will work with systems
where $c_{2}\approx0$ and therefore $\lambda_{2}\approx\lambda_{3}\approx\left(1-\lambda_{1}\right)/2$.
We write $c_{1}$ and $c_{2}$ in terms of their deviations so that
$c_{1}\to c_{1}+\delta c_{1}$ and $c_{2}\to0+\delta c_{2}$. Keeping
only first order terms we have
\begin{equation}
Q=\begin{bmatrix}-\frac{1}{3}\left(c_{1}+\delta c_{1}\right)+\delta c_{2}\cos\left(2\theta\right) & \delta c_{2}\sin\left(2\theta\right) & c_{1}n_{x}\\
- & -\frac{1}{3}\left(c_{1}+\delta c_{1}\right)-\delta c_{2}\cos\left(2\theta\right) & c_{1}n_{y}\\
- & - & \frac{2}{3}\left(c_{1}+\delta c_{1}\right)
\end{bmatrix}
\end{equation}
\end{widetext}

\begin{figure}
\begin{centering}
\includegraphics[width=0.5\linewidth]{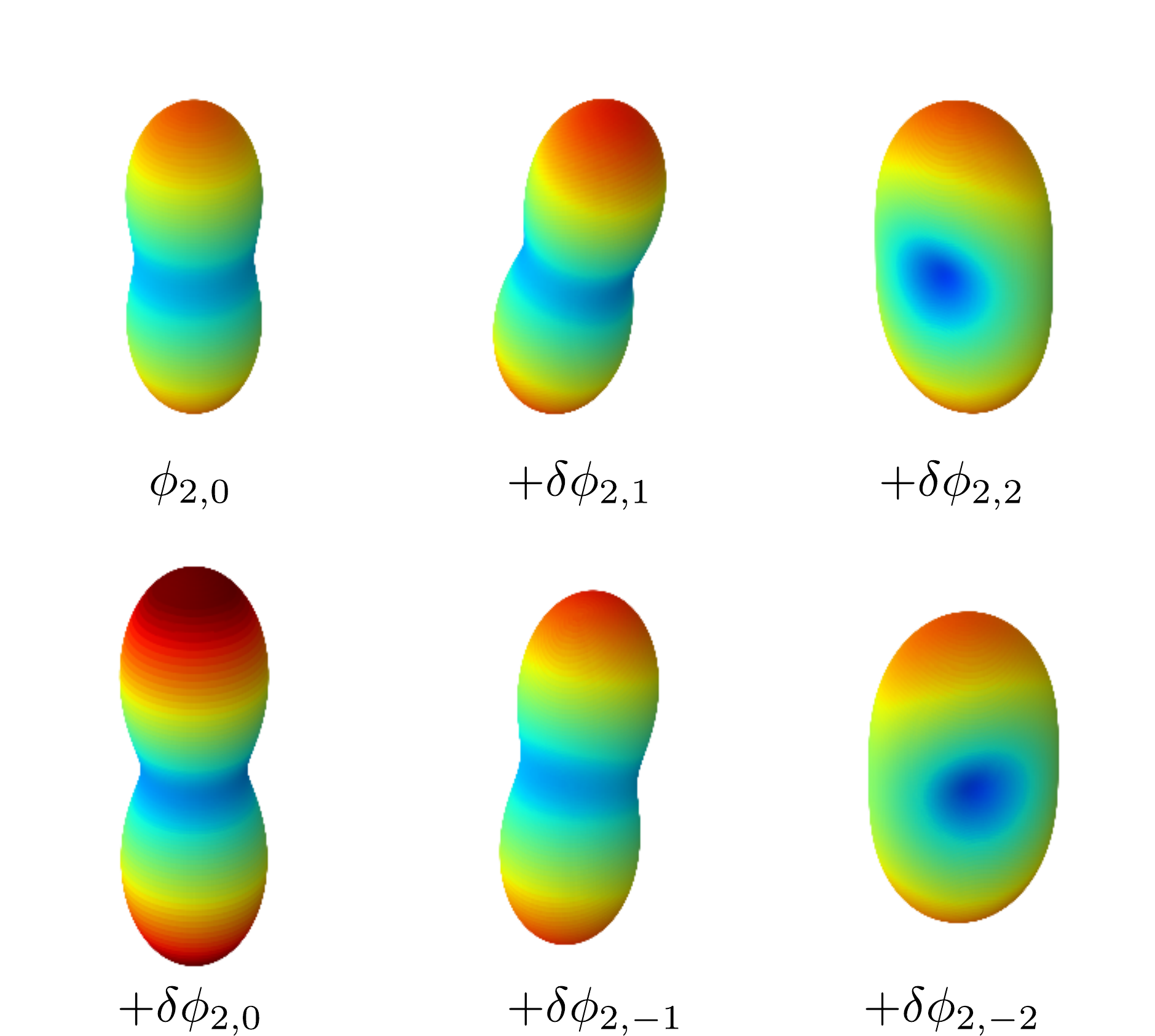}
\par\end{centering}
\caption[Orientation distributions]{\label{fig:delta_realY} Distortions of the local orientation probability due to changes in the field strength at various $m$ values.  If a small spinner (e.g small molecule) is placed inside the field $\phi$ then the probability distribution of the orientation of the spinner will go as $\propto\exp\left(a'\sum_{m}\phi_{2,m}Y_{2,m}(\vec{u})\right)$ where $a'$ depends on the magnitude of the spinner. In the above plots the radius in each direction is proportional to the probability that a molecule will point in that direction when it is subject to a field of the form $\phi_{0,0}$ and for various deviations away from $\phi_{0,0}$.}
\end{figure}

Comparing to equation \ref{eq:phi_2asMatrix} we have
\begin{equation}
\hat{\phi}_{2,m}=\phi_{0,0}\begin{cases}
\sqrt{3}\delta c_{2}\sin\left(2\theta\right) & m=-2\\
\sqrt{3}c_{1}n_{y} & m=-1\\
c_{1}+\delta c_{1} & m=0\\
\sqrt{3}c_{1}n_{x} & m=1\\
\sqrt{3}\delta c_{2}\cos\left(2\theta\right) & m=2
\end{cases}\label{eq:Interp_of_delta_phi}
\end{equation}
This gives us an interpretation of $\phi_{2,m}$ under the assumption
that alignment only slightly differs from the $\hat{z}$ direction as shown in figure~\ref{fig:delta_realY}.
The value of $\phi_{2,0}$ dictates the magnitude of alignment, $\phi_{2,1}$
and $\phi_{2,-1}$ denote fluctuations that rotate the direction of
the alignment into the $\hat{x}$ and $\hat{y}$ directions respectively.
Meanwhile $\phi_{2,-2}$ and $\phi_{2,2}$ denote fluctuations that
create a secondary direction of alignment in the $\hat{x}$ and $\hat{y}$
directions respectively.

Solving for $\hat{n}$ we have
\begin{equation}
\vec{n}=\frac{1}{\phi_{0,0}\sqrt{3}c_{1}}\phi_{2,1}\hat{x}+\frac{1}{\phi_{0,0}\sqrt{3}c_{1}}\phi_{2,-1}\hat{y}+\hat{z}
\end{equation}
from which we can write the bend, twist and splay
\begin{equation}
\left(\hat{n}\times\nabla\times\hat{n}\right)^{2}=\left(\frac{1}{\phi_{0,0}\sqrt{3}c_{1}}\right)^{2}\left(\left(\frac{\partial}{\partial z}\phi_{2,1}\right)^{2}+\left(\frac{\partial}{\partial z}\phi_{2,-1}\right)^{2}\right)
\end{equation}
\begin{equation}
\left(\hat{n}\cdot\nabla\times\hat{n}\right)^{2}=\left(\frac{1}{\phi_{0,0}\sqrt{3}c_{1}}\right)^{2}\left(\frac{\partial}{\partial x}\phi_{2,-1}-\frac{\partial}{\partial y}\phi_{2,1}\right)^{2}
\end{equation}
\begin{equation}
\left(\nabla\cdot\hat{n}\right)^{2}=\left(\frac{1}{\phi_{0,0}\sqrt{3}c_{1}}\right)^{2}\left(\frac{\partial}{\partial x}\phi_{2,1}+\frac{\partial}{\partial y}\phi_{2,-1}\right)^{2}
\end{equation}

In appendix~\ref{sec:Appendix_Fourier} we show that
\begin{equation}
\int dx\left(\frac{\partial}{\partial x}\phi\right)^{2}=\int k^{2}\tilde{f}\left(k\right)\tilde{f}\left(-k\right)dk \label{eq:Fourer_2nd_derivative}
\end{equation}
allowing us to write the Fourier representation of bend, twist, and splay as
\begin{equation}
    \left(\hat{n}\cdot\nabla\times\hat{n}\right)^{2}=\left(\frac{k_{x}\tilde{\phi}_{2,-1}-k_{y}\tilde{\phi}_{2,1}}{\phi_{0,0}\sqrt{3}c_{1}}\right)^{2}
\end{equation}

\begin{equation}
    \left(\hat{n}\times\nabla\times\hat{n}\right)^{2}=\frac{\left(k_{z}\tilde{\phi}_{2,1}\right)^{2}+\left(k_{z}\tilde{\phi}_{2,-1}\right)^{2}}{\left(\phi_{0,0}\sqrt{3}c_{1}\right)^{2}}
\end{equation}

\begin{equation}
    \left(\nabla\cdot\hat{n}\right)^{2}=\left(\frac{k_{x}\tilde{\phi}_{2,1}+k_{y}\tilde{\phi}_{2,-1}}{\phi_{0,0}\sqrt{3}c_{1}}\right)^{2}
\end{equation}

Using 
\begin{equation}
\kappa_{1,m}=\sqrt{\frac{4\pi}{3}}\begin{cases}
k_{y} & m=-1\\
k_{z} & m=0\\
k_{x} & m=1
\end{cases}
\end{equation}
and  $\phi_{0,0}c_{1}=\left\langle \phi_{2,0}\right\rangle ^{MF}$
we get
\begin{equation}
    \left(\hat{n}\cdot\nabla\times\hat{n}\right)^{2} =\frac{\left(\kappa_{1,1}\tilde{\phi}_{2,-1}-\kappa_{1,-1}\tilde{\phi}_{2,1}\right)^{2}}{4\pi\left(\left\langle \phi_{2,0}\right\rangle ^{MF}\right)^{2}} \label{eq:Frank_in_phiphi_A}
\end{equation}

\begin{equation}
    \left(\hat{n}\times\nabla\times\hat{n}\right)^{2}  =\frac{\left(\kappa_{1,0}\tilde{\phi}_{2,1}\right)^{2}+\left(\kappa_{1,0}\tilde{\phi}_{2,-1}\right)^{2}}{4\pi\left(\left\langle \phi_{2,0}\right\rangle ^{MF}\right)^{2}}\label{eq:Frank_in_phiphi_B}
\end{equation}

\begin{equation}
    \left(\nabla\cdot\hat{n}\right)^{2}  =\frac{\left(\kappa_{1,1}\tilde{\phi}_{2,1}+\kappa_{1,-1}\tilde{\phi}_{2,-1}\right)^{2}}{4\pi\left(\left\langle \phi_{2,0}\right\rangle ^{MF}\right)^{2}}\label{eq:Frank_in_phiphi_C}
\end{equation}

Equation \ref{eq:FrankElasticEnergy} can be written in matrix form
as
\begin{widetext}
\begin{equation}
E_{Frank}=\frac{1}{2}\left(\left\langle \phi_{2,0}\right\rangle ^{MF}\right)^{-2}\frac{1}{4\pi}\vec{\left(\kappa\phi\right)}\cdot\begin{bmatrix}K_{splay} & 0 & 0 & 0 & 0 & B\\
0 & K_{twist} & 0 & 0 & A & 0\\
0 & 0 & K_{bend} & 0 & 0 & 0\\
0 & 0 & 0 & K_{bend} & 0 & 0\\
0 & A & 0 & 0 & K_{twist} & 0\\
B & 0 & 0 & 0 & 0 & K_{splay}
\end{bmatrix}\cdot\vec{\left(\kappa\phi\right)}\label{eq:K_matrix}
\end{equation}

where $A+B=K_{splay}-K_{twist}$ and

\begin{equation}
\vec{\left(\kappa\phi\right)}=\begin{bmatrix}\kappa_{1,-1}\tilde{\phi}_{2,-1} & \kappa_{1,-1}\tilde{\phi}_{2,1} & \kappa_{1,0}\tilde{\phi}_{2,-1} & \kappa_{1,0}\tilde{\phi}_{2,1} & \kappa_{1,1}\tilde{\phi}_{2,-1} & \kappa_{1,1}\tilde{\phi}_{2,1}\end{bmatrix}
\end{equation}
\end{widetext}
The Frank elastic constants are related to the deformation energy
of the system given in equation~\ref{eq:Gamma_and_cancel}, in that they comprise the
portion of the energy that comes from $\tilde{\phi}_{2,\pm1}$ deformations
that are quadratic in $\kappa$. The comparison between equations
\ref{eq:Frank_in_phiphi_A}-\ref{eq:Frank_in_phiphi_B} and $\delta\tilde{S}^{-1}$ in equation
\ref{eq:Gamma_and_cancel} will allow us to write the Frank elastic
constants in terms of the microscopic details of the system (e.g.
$L$, $A$, $\phi_{0,0}$, $a$, $v_{p}$, etc.).

\section{Frank Elastic Constants for Fuzzball and Rigid Rod}

Comparing equations \ref{eq:Frank_in_phiphi_A}-\ref{eq:Frank_in_phiphi_B} and \ref{eq:K_matrix}
to equation \ref{eq:S_fuzzball} we find that the Frank elastic constants
for a fuzzball are identical and equal
\begin{equation}
K_{bend}^{FB}=K_{twist}^{FB}=K_{splay}^{FB}=3\phi_{0,0}\frac{\sigma^{2}}{v_{p}}\frac{\left(M_{0,2}^{0}\right)^{2}}{M_{2,2}^{1}M_{0,0}^{0}}
\end{equation}

Comparing equations \ref{eq:Frank_in_phiphi_A}-\ref{eq:Frank_in_phiphi_B} and \ref{eq:K_matrix}
to equation \ref{eq:S_RR_formula} we get the Frank elastic constants
for a rigid rod
\begin{equation}
K_{bend}^{RR}=\phi_{0,0}\frac{L}{A}\frac{2\pi}{3}\frac{\left(M_{0,2}^{0}\right)^{2}}{M_{2,2}^{1}M_{2,2}^{1}M_{0,0}^{0}}\left(J_{\left(1\right)}^{1,0,1}\cdot M^{1}\cdot J_{\left(1\right)}^{1,0,1}\right)_{2,2}
\end{equation}
\begin{eqnarray}
K_{twist}^{RR}=&&\phi_{0,0}\frac{L}{A}\frac{2\pi}{3}\frac{\left(M_{0,2}^{0}\right)^{2}}{M_{2,2}^{-1}M_{2,2}^{-1}M_{0,0}^{0}}\nonumber\\
&&\times\sum_{\mathscr{M}\in\left\{ -2,0\right\} }\left(J_{\left(1\right)}^{-1,1,\mathscr{M}}\cdot M^{\mathscr{M}}\cdot J_{\left(1\right)}^{\mathscr{M},1,-1}\right)_{2,2}
\end{eqnarray}
\begin{eqnarray}
K_{splay}^{RR}=&&\phi_{0,0}\frac{L}{A}\frac{2\pi}{3}\frac{\left(M_{0,2}^{0}\right)^{2}}{M_{2,2}^{1}M_{2,2}^{1}M_{0,0}^{0}}\nonumber\\
&&\times\sum_{\mathscr{M}\in\left\{ 0,2\right\} }\left(J_{\left(1\right)}^{1,1,\mathscr{M}}\cdot M^{\mathscr{M}}\cdot J_{\left(1\right)}^{\mathscr{M},1,1}\right)_{2,2}
\end{eqnarray}
\section{Frank Elastic Constants for WLC \label{sec:Calculate_FEC}}

Comparing equations \ref{eq:Frank_in_phiphi_A}-\ref{eq:Frank_in_phiphi_B} and \ref{eq:K_matrix}
to equation \ref{eq:Sprime_expand} we find the dimensionless elastic constants
\begin{equation}
K'_{splay}=\frac{4\pi\left(B_{1}\right)^{2}B_{4}\left(1,1,1,1\right)}{N^{2}\left(B_{2}\left(1\right)\right)^{2}B_{0}}
\end{equation}
\begin{equation}
K'_{twist}=\frac{4\pi\left(B_{1}\right)^{2}B_{4}\left(-1,-1,1,1\right)}{N^{2}\left(B_{2}\left(-1\right)\right)^{2}B_{0}}
\end{equation}
\begin{equation}
K'_{bend}=\frac{4\pi\left(B_{1}\right)^{2}B_{4}\left(1,1,0,0\right)}{N^{2}\left(B_{2}\left(1\right)\right)^{2}B_{0}}
\end{equation}
where 
\begin{equation}
K'=K\frac{A}{\phi_{0,0}\left(2\ell_{p}\right)N}
\end{equation}
\begin{equation}
B_{0}=\mathscr{\mathcal{L}}^{-1}\left[\boldsymbol{e}_{0}\cdot\mathbf{G}^{0}\cdot\boldsymbol{e}_{0}\right]
\end{equation}
\begin{equation}
B_{1}=\mathcal{L}^{-1}\left[\boldsymbol{e}_{0}\cdot\mathbf{G}^{0}\cdot\mathbf{J}_{\left(2\right)}^{0,0,0}\cdot\mathbf{G}^{0}\cdot\boldsymbol{e}_{0}\right]
\end{equation}
\begin{equation}
B_{2}\left(m_{1}\right)=\mathscr{\mathcal{L}}^{-1}\left[\boldsymbol{e}_{0}\cdot\mathbf{G}^{0}\cdot\mathbf{J}_{\left(2\right)}^{0,m_{1},m_{1}}\cdot\mathbf{G}^{m_{1}}\cdot\mathbf{J}_{\left(2\right)}^{m_{1},m_{1},0}\cdot\mathbf{G}^{0}\cdot\boldsymbol{e}_{0}\right]
\end{equation}
\begin{widetext}
\begin{align}
B_{4}\left(m_{1},m_{2},m,m'\right) =\mathscr{\mathcal{L}}^{-1}\left[\sum_{\mathscr{M}}\boldsymbol{e}_{0}\cdot\mathbf{G}^{0}\cdot\mathbf{J}_{\left(2\right)}^{0,m_{1},m_{1}}\cdot\mathbf{G}^{m_{1}}\cdot\mathbf{J}_{\left(1\right)}^{m_{1},m,\mathscr{M}}\cdot\mathbf{G}^{\mathscr{M}}\cdot\mathbf{J}_{\left(1\right)}^{\mathscr{M},m',m_{2}}\cdot\mathbf{G}^{m_{2}}\cdot\mathbf{J}_{\left(2\right)}^{m_{2},m_{2},0}\cdot\mathbf{G}^{0}\cdot\boldsymbol{e}_{0}\right]
\end{align}
\end{widetext}

\section{Numerical Laplace Inversion\label{sec:Numerical-Laplace-inversion}}

Laplace inversion is accomplished by the numerical integration \\
\begin{equation}
f\left(N\right)=\frac{1}{2\pi i}\int_{\lambda-i\infty}^{\lambda+i\infty}e^{pN}F\left(p\right)dp.
\end{equation}
\begin{figure}
\begin{centering}
\includegraphics[width=0.4\linewidth]{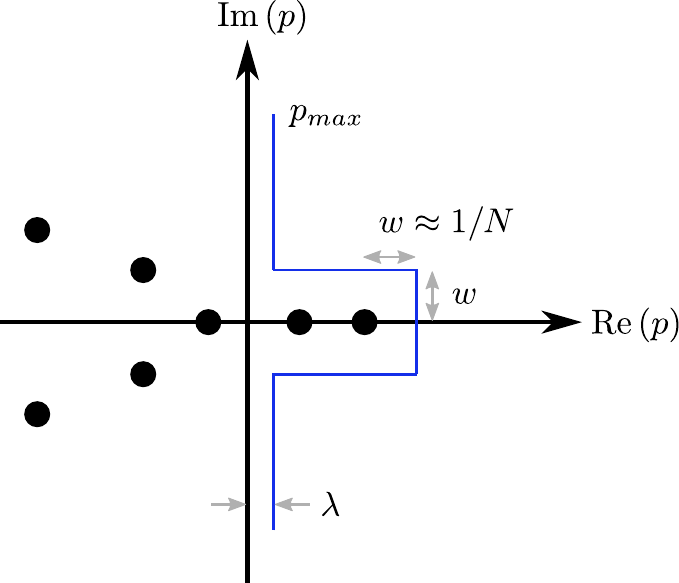}
\par\end{centering}
\caption[Numerical inverse Laplace transform]{\label{fig:p_path} Diagram of typical path of integration on the complex plane of Laplace variable $p$.  The path goes from bottom to top along the blue lines.  Black dots represent typical poll locations. Not drawn to scale, in practice $p_{max}\gg w$.}
\end{figure}
The complex integration is done along the path in figure~\ref{fig:p_path}
where the dots are poles. The position of the first pole is found
with binary search using the facts that $\boldsymbol{G}_{0,0}^{0}$
changes sign as you go horizontally through the poles and that the
first pole is always on the real axis. As the pole moves to the right
the $\exp\left(pN\right)$ in the integral becomes extremely large.
To avoid overflow we can multiply by $\exp\left(-p_{max}N\right)$
which will cancel when we take the ratio of two inverse Laplace transforms,
as always occurs in this derivation.

Python code to do the inverse Laplace transform and generate the plots in this paper can be found at \url{https://github.com/SpakowitzLab/wlcsim/tree/master/wlcsim/FrankElastic}.  The documentation can be found at \url{https://wlcsim.readthedocs.io/en/latest/FrankExample.html}.

\section{Results\label{sec:Frank_Results}}

\begin{figure}
\begin{centering}
\includegraphics[width=0.7\linewidth]{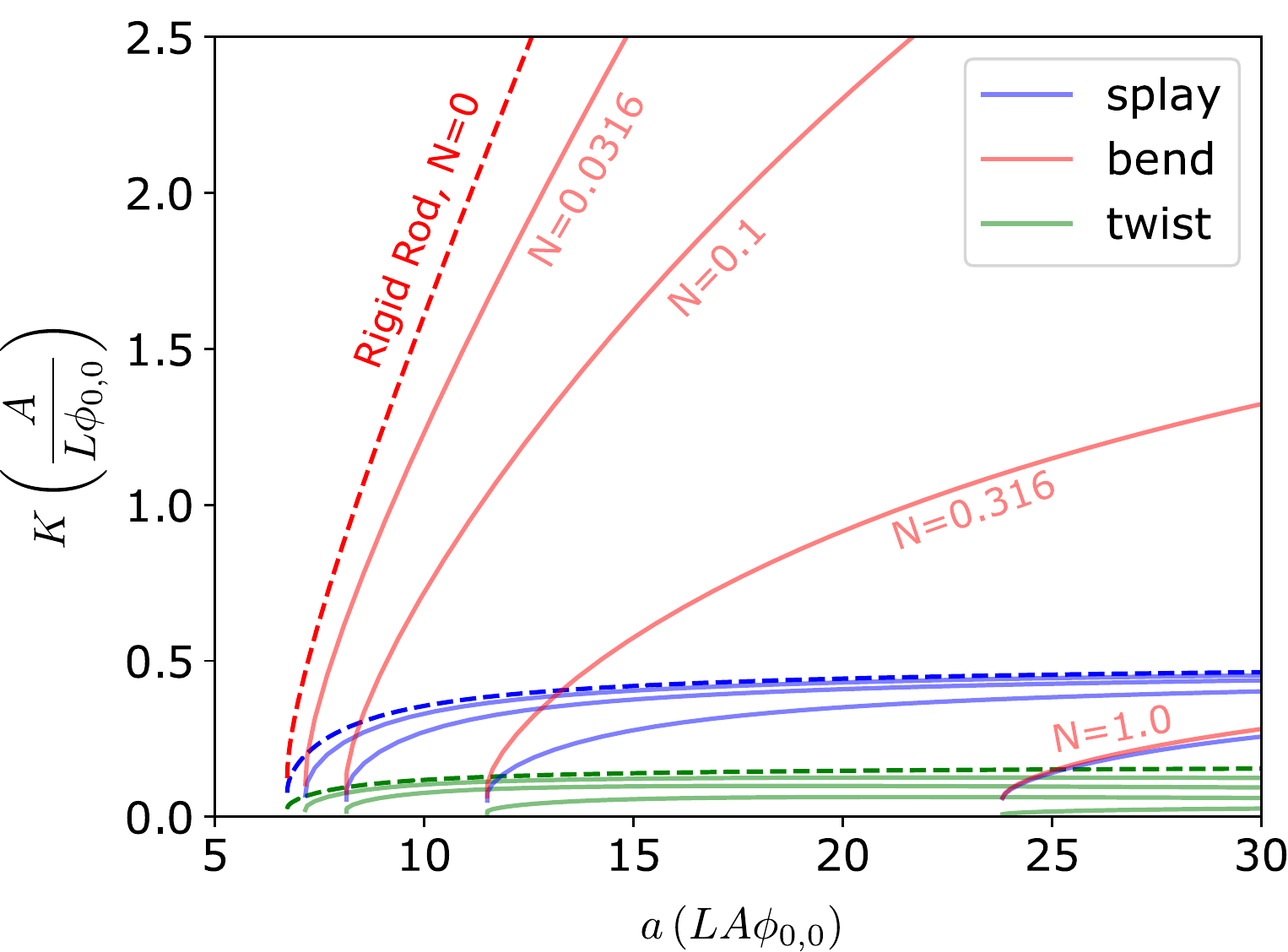}
\par\end{centering}
\caption[Frank elastic constants for stiff polymer solutions]{\label{fig:Low_N}Frank elastic constants for stiff polymer solutions for stiff chains that are $N$ Kuhn lengths long.
Dashed curves are for the rigid rod. For the rigid rod, $K_{splay}=3K_{twist}$.}
\end{figure}

The units of the Frank elastic constants are energy times length because
they denote energy per unit volume per rate of change of direction
squared. To compare the magnitudes of the Frank elastic constants
we first nondimensionalize them. Throughout this paper we have dropped
the implied energy units of $k_b T$'s, so energy is already dimensionless.
For relatively stiff polymers, it makes sense to nondimensionalize
$K$ by multiplying it by $A/L$ because $L$ and $A$ are the only
relevant length scales for the rigid rod. When nondimensionalizing
we also divide the Frank elastic constants by the volume fraction
of polymer $\phi_{0,0}$ because denser solutions experience proportionally
more aligning field. Figure \ref{fig:Low_N} presents the Frank elastic
constants nondemensinalized via $K\left(\frac{A}{L\phi_{0,0}}\right)$.
The Maier-Saupe parameter $a$, which captures the energetic coupling strength of aligning molecules, has units of energy (also in $k_b T$'s) per
volume, therefore we nondimensionalize it via $a\left(LA\phi_{00}\right)$.
This nondimensionalization is convenient because a rigid rod that is twice as long will
experience twice the aligning potential.

\begin{figure}
\begin{centering}
\includegraphics[width=0.5\linewidth]{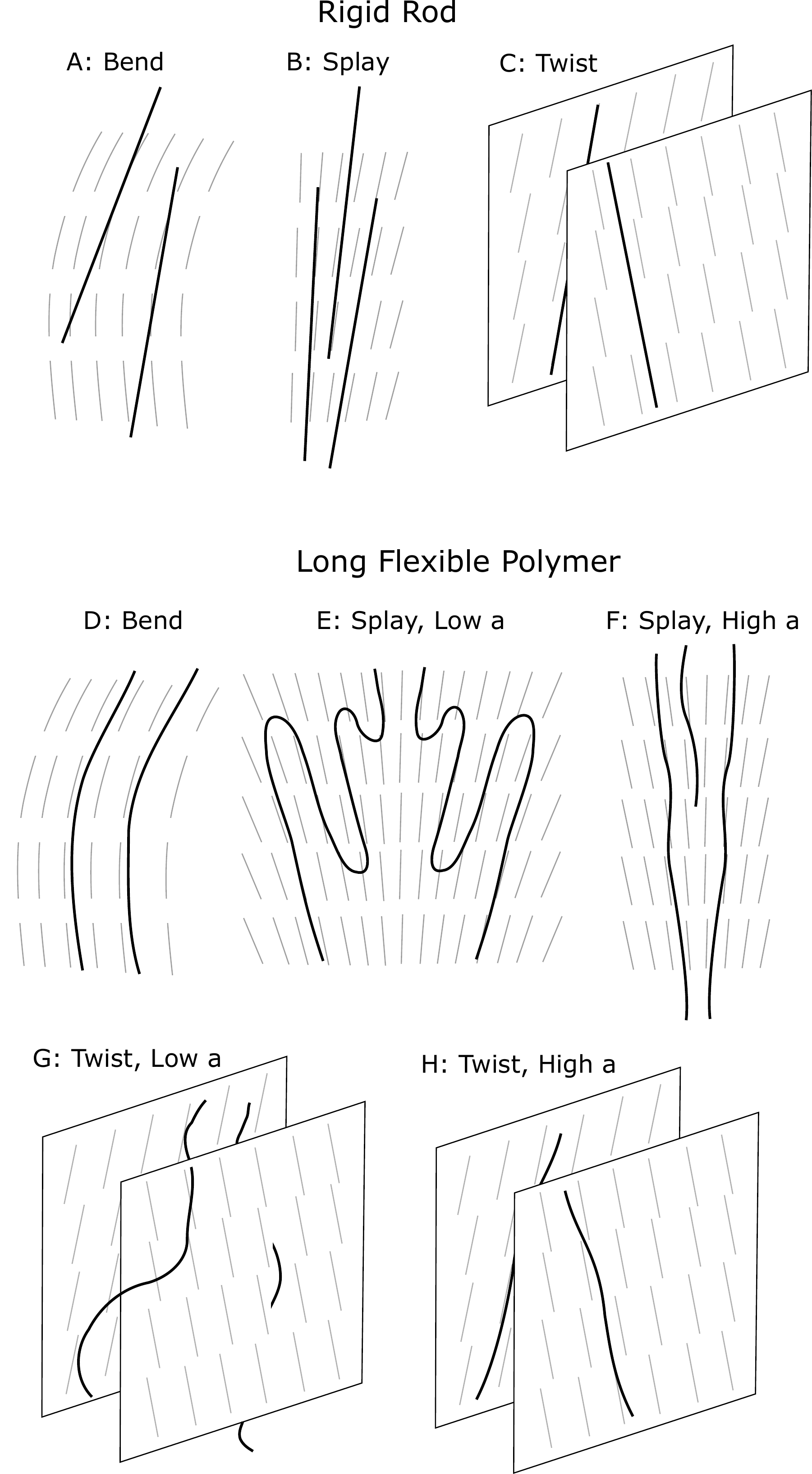}
\par\end{centering}
\caption[Schematic explanation of Frank elastic constants]{\label{fig:discussionSketch} Schematics providing qualitative arguments for the observed trends in the Frank elastic constants.  A-C are for rigid rods or very stiff ($N\ll1$) polymers.  D-H are for long semiflexible polymers ($N\gg1$).}
\end{figure}

In Figure~\ref{fig:Low_N} the value of the elastic constants $K$ generally grow with increased  Maier-Saupe nematic coupling strength $a$. The coupling strength drives the formation of a liquid crystal state, so it is natural that higher $a$ could induce higher $K$ and thereby reduce deviations from the perfectly aligned state. Figure~\ref{fig:Low_N} shows that liquid crystal solutions of rigid rods and other stiff polymers ($N\ll 1$) have a relatively high bend modulus $K_{bend}$.  Figure~\ref{fig:discussionSketch}A shows rigid rods fit poorly into a bending field with their elongated nature inevitably spanning regions of different orientation.  In contrast, rods neatly connect together regions of similar orientation making it easy for a solution of rods to twist and splay (see figure~\ref{fig:discussionSketch}B and C).  

The fundamental difference that leads to $K_{bend}$ being greater than the other $K$'s is that bend is a change in orientation associated with a displacement in the direction of alignment (see figure~\ref{fig:Frank_Elastic_Pic_2}).  This is precisely the direction in which the rod is elongated.  To emphasise this point, compare the rod to the fuzzball which is equally elongated in all directions and has $K^{FB}_{twist}=K^{FB}_{bend}=K^{FB}_{Splay}$.

In figure \ref{fig:Low_N} we see that making a rod-like molecule
more flexible (i.e. increasing $N$) reduces all three elastic constants.
The dramatic reduction in $K_{bend}$ reflects the ability of
the polymers to bend with a bending field. Increased flexibility
also reduces splay and twist constants, but to a lesser extent. 

Note that at the lower left corner of figure~\ref{fig:Low_N} the elastic constants turn downward
and end. They do this because the lower left end of the curve represents
the limit of metastability corresponding to the minimum $a$ value in figure~\ref{fig:Mean-field-strength}. At values of $a$ below this cutoff
the solution will revert to an isotropic state.

\begin{figure}
\begin{centering}
\includegraphics[width=0.6\linewidth]{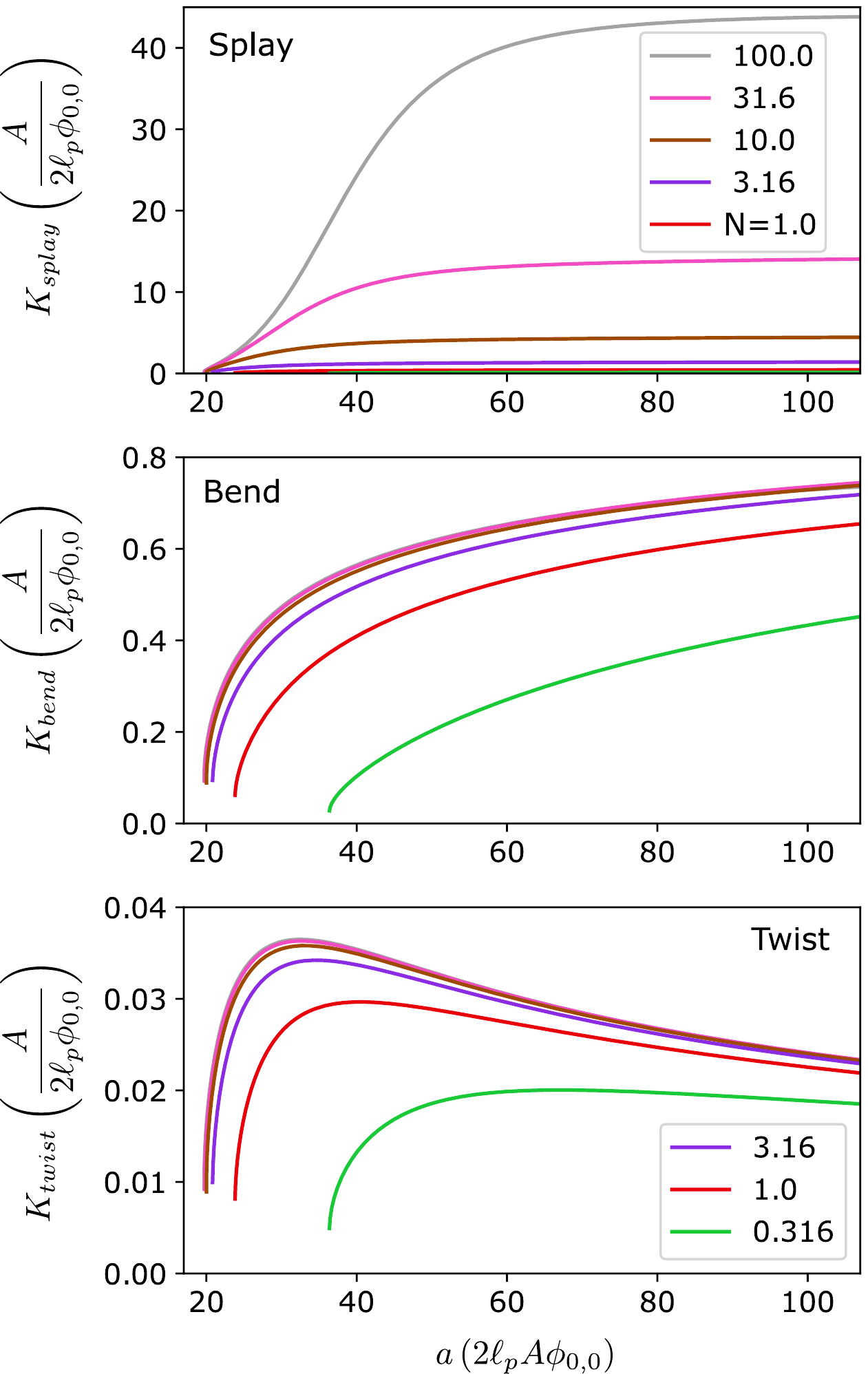}
\par\end{centering}
\caption[Frank Elastic constants for semiflexible polymers]{\label{fig:High_N}Frank elastic constants for polymers of different lengths, $N$, in Kuhn lengths.}
\end{figure}

For relatively flexible polymers it makes sense to nondimensionalize
by the Kuhn length, ${2\ell_{p}=L/N}$, rather than the length of the entire
polymer. In figure~\ref{fig:Low_N} the Frank elastic constants decrease with increasing $N$, in contrast to figure~\ref{fig:High_N} where they increase.  The difference between the two is their different nondimensionalization of $K$.  In~\ref{fig:Low_N} the polymer length ${L=2\ell_p N}$ is held constant so that increasing $N$ makes the chain more flexible and decreases $K$.  In figure~\ref{fig:High_N} the Kuhn length $2\ell_p$ is held constant so that increasing $N$ makes the chain longer and increases $K$.

Nondimensionalizing by the Kuhn length in figure~\ref{fig:High_N} accentuates the universal behavior of polymers at large $N$. When comparing a solution of polymers of length $N=100$ and $N=200$ but equivalent volume fractions of polymer $\phi_{0,0}$ the solutions look quite similar other than the $N=200$ solution will have half as many polymer ends.
Figure \ref{fig:High_N} shows that above about $N=10$ the bend and twist no longer depend on $N$.
Indeed, for relatively flexible ($N\gg1$) polymers, physical properties (such as the bend and twist moduli) depend much more on the Kuhn length $2\ell_p$ than the total length $L$.
This is consistent with the intuition from figure~\ref{fig:discussionSketch}D.

The splay modulus $K_{splay}$ for long polymers is larger and more complex.  At low coupling strength splay can be accommodated by hairpins as shown in figure~\ref{fig:discussionSketch}E.  However, increasing $a$ makes hairpins increasingly energetically unfavorable leading to a exponential growth in the spacing between hairpins and the resulting splay modulus~\cite{petschekMolecularstatisticalTheoryCurvature1992}.  At sufficiently high $a$, all hairpins are removed from the chain and the splay is accommodated by chain ends (figure~\ref{fig:discussionSketch}F).  As the number of chain ends depends on the chain length (at a fixed concentration) rather then $a$, $K_{splay}$ will plateau at a value that increases with $N$.

Surprisingly, after initially increasing, the twist modulus $K_{twist}$ moderately decreases with increasing $a$ as shown in figure~\ref{fig:High_N}.  We justify this by arguing that at low $a$ polymers will mix between twist planes as shown in figure~\ref{fig:discussionSketch}G.  Each time a polymer mixes between twist planes it introduces stress pulling the orientations of the planes into alignment.  As $a$ increases the amount of stress introduced when a polymer mixes between plains increases but the polymers are much less likely to stray from their plane as depicted in~\ref{fig:discussionSketch}H.  The latter effect apparently overwhelms the former at high $a$.

\begin{figure}
\begin{centering}
\includegraphics[width=0.5\linewidth]{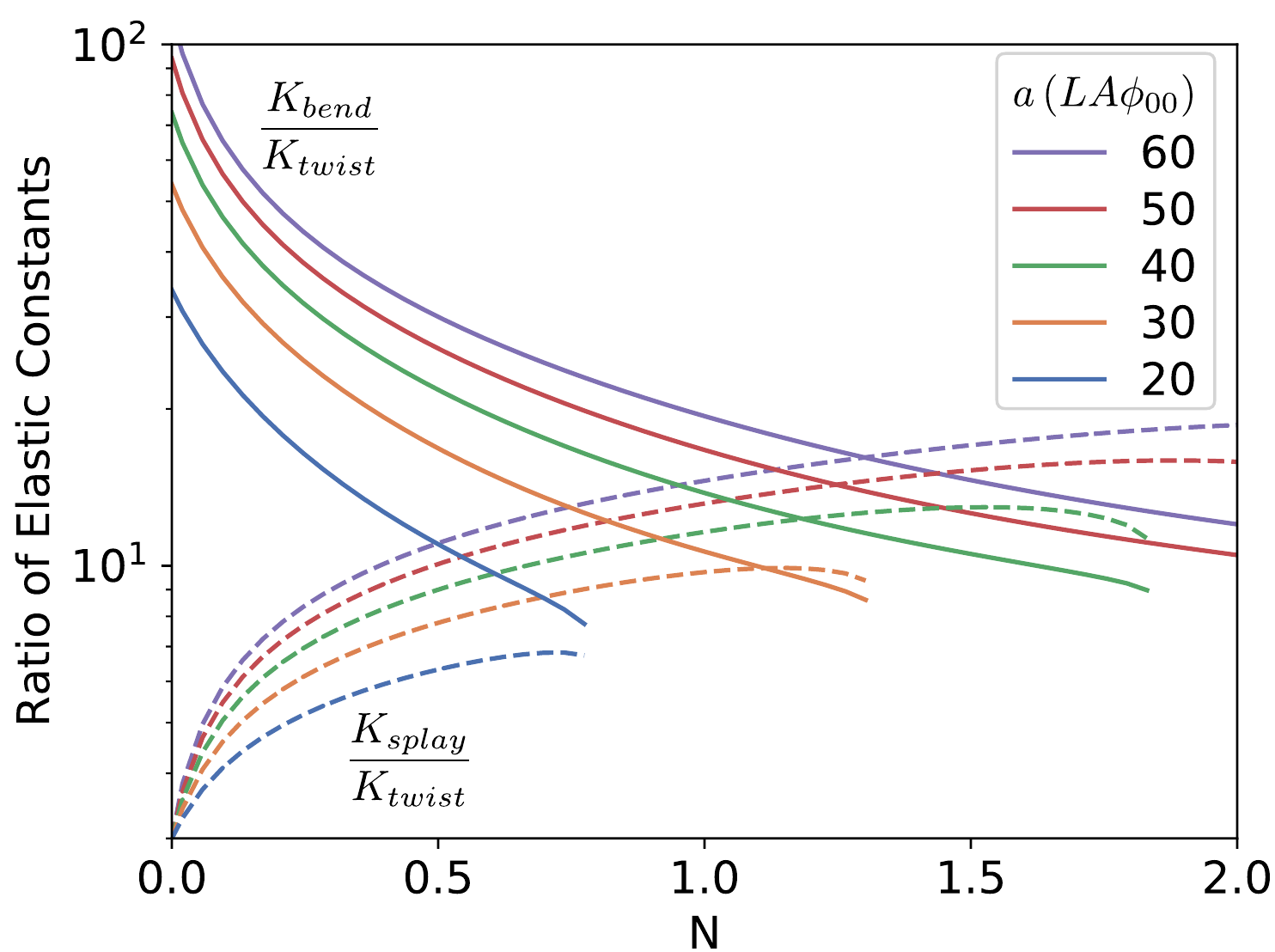}
\par\end{centering}
\caption[Ratio of Frank elastic constants]{\label{fig:Nsweep} Ratio of Frank elastic constants for stiff polymers.}
\end{figure}

The primary trend presented above is that for rigid polymers $N\ll 1$ the $K_{bend}>K_{splay}$ and for long flexible polymers $K_{splay}>K_{bend}$.  Naturally, we'd like to know when the crossover between bend and splay occurs.  In figure~\ref{fig:Nsweep} we plot the ratios of the Frank elastic constants.  The crossover where $K_{bend}=K_{splay}$ depends modestly on $a$ and occurs a lengths a little above one Kuhn length.  The curves in this figure end when the polymers become flexible to the point where the aligned phase is no longer metastable and spontaneously transforms to the isotropic state.  Note that we have nondimensionalized $a$ in this plot by the total polymer length.  If we instead had nondimensionalized $a$ by the Kuhn length, the curves would still cross at $N$ values just above one, but the shape of the curves would be different.

\section{Summary}

We have presented a method for evaluating the size of orientation deformations for nematic solutions of semiflexible polymers.  These deformations, described by the Frank elastic constants, are accurately found for a wide range of chain flexibilities and alignment interaction strengths (i.e. persistence length, $\ell_p$ and Maier-Saupe parameter, $a$).  Our method shows the convenience of combining a previously-derived, stone-fence-diagram-based, exact statistics for the worm-like chain in a quadrupole field~\cite{spakowitzExactResultsSemiflexible2004, spakowitzEndtoendDistanceVector2005} with a real-spherical-harmonic orientational density distribution.  This method allows us provide both quantitative plots and qualitative explanations for the Frank elastic constants over this range. Here we summarize the features of the Frank elastic constants, some of which have been previously reported.

Below a minimum alignment interaction strength the nematic state will spontaneously degenerate into the isotropic one.  That is, the aligned state is not even metastable.  This minimum interaction is associated with finite - if small - Frank elastic constants and a large but finite amount of deformation.  The following features are for solutions in a stable (or at least metastable) nematic state.

We will divide our results into those for relatively rigid polymers ($L \lessapprox \ell_p$) and those for relatively flexible polymers. First, we will make three observations that hold over both ranges.  1) The twist modulus is always smaller than bend and splay moduli.  For most polymer solutions $K_{twist}$ is order(s) of magnitude smaller than $K_{bend}$ and $K_{splay}$.  This means that the primary deformation mode of polymer solutions will be twist. 2) At the level of the present theory (no coupling between deformation modes) the Flory Huggins $\chi$ parameter does not affect the Frank elastic constants.  This means that using a better solvent will not directly\footnote{A better solvent could effect behavior on shorter length scales which could renormalize into a different Maier-Saupe parameter on longer length scales and thereby indirectly effect the Frank elastic constants.  We have also assumed that all solvents are good enough to dissolve the polymers.} affect the Frank elastic constants. 3) Frank elastic constants are inversely proportional to polymer concentration.  Higher concentrations lead to more alignment.

For nematic solutions of rigid polymers the Frank elastic constants increase as $~L\phi_{0,0}$ where $L$ is the polymer length.  That is, increasing polymer length and concentration decreases the expected amount of bend, twist, and splay.  Over the range of Maier-Saupe coupling strengths $a$ where liquid crystal alignment can be found, we find $K_{bend} > K_{splay} > K_{twist}$. When coupling strength $a$ is just above the limit of metastability of the nematic phase, then $K_{bend}\approx K_{splay}$ as predicted by \cite{shimadaCoefficiensGradientTerms1988}.  However, with increasing $a$ away from the metastability limit, we find that $K_{bend}$ increases rapidly so that $K_{bend}\gg K_{splay}$.  In other words, solutions of long rigid molecules exhibit little bend deformation. Interestingly, for perfectly rigid rods ($N=0$) we have $K_{splay}=3K_{twist}$ independent of coupling strength.  Furthermore, away from the metastability limit, $K_{splay}$ and $3K_{twist}$ are nearly independent of $a$, leaving the amount of splay and twist primarily determined by polymer length and concentration.

For nematic solutions of semiflexible and flexible polymers the qualitative elastic behavior differs significantly from that of rigid polymers.  For flexible polymers $K_{splay} > K_{bend}$.  The crossover where $K_{bend} = K_{splay}$ occurs at $N\approx1.25$ with the crossover value increasing modestly with aligning strength $a$.  Unlike solutions of rigid polymers where the elastic constants depended primarily on the polymer length, the elastic constants of semiflexible polymers depend more heavily on the persistence length. In particular, the twist and bend constants, $K_{twist}$ and $K_{bend}$, grow as $~\ell_p \phi_{0,0}$ where $\ell_p$ is the polymer persistence length.  The splay constant $K_{splay}$ grows as $~\ell_p \phi_{0,0}$ at low coupling strengths but as $L\phi_{0,0}$ at high coupling strengths.
    
The dependence of the elastic deformations of flexible polymer solutions have an interesting dependence on coupling strength $a$. The value of $K_{splay}$ increases quickly with $a$ as hairpins are removed from chains as shown by~\cite{petschekMolecularstatisticalTheoryCurvature1992}.  The increase flattens off when all the hairpins are removed and splay is accommodated by chain ends. In contrast, after initially increasing, the twist modulus actually \emph{decreases} with increasing coupling strength.  Over this range, an increase in the level of alignment of the polymers is associated with an increase in the amount of twist deformation.

Taken together, the elastic deformations depend on the the polymer length, rigidity, and local alignment strength in a varied but predicable fashion. Armed with the knowledge of these dependencies, the reader can both predict and rationally design the liquid crystalline behavior of polymer solutions.

\begin{acknowledgments}
I thank Andrew Spakowitz for is advising during the research and writing of this paper.  I thank Ashesh Ghosh for his comments during the editing or this document.
\end{acknowledgments}
\appendix
\section{Real Spherical Harmonics\label{sec:Appendix:-Real-spherical}}

The real spherical harmonics, $Y_{\ell,m}$, are defined in terms
the more widely know complex spherical harmonics, $Y_{l}^{m}$, by
\begin{equation}
Y_{\ell,m}=\begin{cases}
\frac{i}{\sqrt{2}}\left(Y_{\ell}^{m}-Y_{\ell}^{m*}\right) & m<0\\
Y_{\ell}^{0} & m=0\\
\frac{1}{\sqrt{2}}\left(Y_{\ell}^{-m}+Y_{\ell}^{-m*}\right) & m>0
\end{cases}
\end{equation}
and conversely
\begin{equation}
Y_{\ell}^{m}=\begin{cases}
\frac{1}{\sqrt{2}}\left(Y_{\ell,-m}-iY_{\ell,m}\right) & m<0\\
Y_{l,m} & m=0\\
\frac{\left(-1\right)^{m}}{\sqrt{2}}\left(Y_{\ell,m}+iY_{\ell,-m}\right) & m>0
\end{cases}
\end{equation}
where we note that $Y_{\ell}^{m*}=\left(-1\right)^{m}Y_{\ell}^{-m}$.
In particular
\begin{equation}
Y_{0,0}=\frac{1}{\sqrt{4\pi}}
\end{equation}
\parbox{\linewidth}{
\begin{equation}
Y_{1,m}=\sqrt{\frac{3}{4\pi}}\begin{cases}
u_{y} & m=-1\\
u_{z} & m=0\\
u_{x} & m=1
\end{cases}
\end{equation}
}
\vspace{0.5cm}
\parbox{\linewidth}{
\begin{equation}
Y_{2,m}\left(\vec{u}\right)=\sqrt{\frac{5}{4\pi}}\begin{cases}
\sqrt{3}u_{x}u_{y} & m=-2\\
\sqrt{3}u_{y}u_{z} & m=-1\\
\frac{1}{2}\left(3u_{z}^{2}-1\right) & m=0\\
\sqrt{3}u_{z}u_{x} & m=1\\
\frac{\sqrt{3}}{2}\left(u_{x}^{2}-u_{y}^{2}\right) & m=2
\end{cases}
\end{equation}
}

The orthogonality of real spherical harmonics is
\begin{equation}
\int d\vec{u}_{i}Y_{\ell_{1},m_{1}}\left(\vec{u}_{i}\right)Y_{\ell_{2},m_{2}}\left(\vec{u}_{i}\right)=\delta_{\ell_{1},\ell_{2}}\delta_{m_{1},m_{2}}
\end{equation}

A useful property of products of real spherical harmonics is their
relation the the Legendre polynomial
\begin{equation}
P_{\ell}\left(\vec{u}_{1}\cdot\vec{u}_{2}\right)=\frac{4\pi}{2\ell+1}\sum_{m=-\ell}^{\ell}Y_{\ell,m}\left(\vec{u}_{1}\right)Y_{\ell,m}\left(\vec{u}_{2}\right)
\end{equation}

Furthermore, products of harmonics can be factored into sums of single
harmonics
\begin{equation}
Y_{1,0}Y_{\ell,m}=\alpha_{\ell+1}^{m}Y_{\ell+1,m}+\alpha_{\ell}^{m}Y_{\ell-1,m}
\end{equation}
\begin{align}
Y_{2,0}Y_{\ell,m}=&\frac{\alpha_{\ell+1}^{m}\alpha_{\ell+2}^{m}}{\alpha_{2}^{0}}Y_{\ell+2,m}+\frac{\alpha_{\ell+1}^{m}a_{\ell+1}^{m}-\alpha_{1}^{0}Y_{0}^{0}+\alpha_{\ell}^{m}\alpha_{\ell}^{m}}{\alpha_{2}^{0}}Y_{\ell,m}+\frac{\alpha_{\ell}^{m}\alpha_{\ell-1}^{m}}{\alpha_{2}^{0}}Y_{\ell-2,m}
\end{align}
where
\begin{equation}
\alpha_{\ell}^{m}\equiv\sqrt{\frac{3\left(\ell-m\right)\left(\ell+m\right)}{4\pi\left(2\ell-1\right)\left(2\ell+1\right)}}\label{eq:alpha_def}
\end{equation}

Because it will be used frequently, we will rewrite 
\begin{equation}
Y_{2,0}Y_{\ell,m}=A_{\ell+2,}^{m}Y_{\ell+2,m}+\beta_{\ell}^{m}Y_{\ell,m}+A_{\ell}^{m}Y_{\ell-2,m}\label{eq:Y20_sum}
\end{equation}
where
\begin{equation}
A_{\ell}^{m}\equiv\frac{\alpha_{\ell}^{m}\alpha_{\ell-1}^{m}}{\alpha_{2}^{0}},\ \beta_{\ell}^{m}\equiv\frac{\alpha_{\ell+1}^{m}a_{\ell+1}^{m}-\alpha_{1}^{0}Y_{0,0}+\alpha_{\ell}^{m}\alpha_{\ell}^{m}}{\alpha_{2}^{0}}\label{eq:AlmBlmdef}
\end{equation}
Furthermore,
\begin{widetext}
\begin{align}
Y_{1,1}Y_{\ell,m}= & \alpha_{\ell+1}^{\left(+\right)m}Y_{\ell+1,m+1}f\left(m+1\right)-\alpha_{\ell+1}^{\left(+\right)-m}Y_{\ell+1,m-1}f\left(m\right)\nonumber \\
 & +\alpha_{\ell}^{\left(+\right)m-1}Y_{\ell-1,m-1}f\left(m\right)-\alpha_{\ell}^{\left(+\right)-m-1}Y_{\ell-1,m+1}f\left(m+1\right)
\end{align}
\begin{align}
Y_{1,-1}Y_{\ell,m}= & \alpha_{\ell+1}^{\left(+\right)-m}\frac{1}{\sqrt{2}}Y_{\ell+1,-m+1}g\left(m\right)-\alpha_{\ell}^{\left(+\right)m-1}\frac{1}{\sqrt{2}}Y_{\ell-1,-m+1}g\left(m\right)\nonumber \\
 & +\alpha_{\ell+1}^{\left(+\right)m}\frac{1}{\sqrt{2}}Y_{\ell+1,-m-1}h\left(m\right)-\alpha_{\ell}^{\left(+\right)-m-1}\frac{1}{\sqrt{2}}Y_{\ell-1,-m-1}h\left(m\right)
\end{align}
where
\begin{equation}
\alpha_{\ell}^{\left(\pm\right)m}\equiv\sqrt{\frac{3\left(\ell+m\right)\left(\ell+m\pm1\right)}{8\pi\left(2\ell-1\right)\left(2\ell+1\right)}}\label{eq:alpha_pm_def}
\end{equation}
\begin{equation}
f\left(m\right)=\frac{1}{\sqrt{2}}\cdot\begin{cases}
-1 & m<-1\\
-1 & m=-1\\
0 & m=0\\
\sqrt{2} & m=1\\
1 & m>1
\end{cases}\ \ \ \ \mathrm{and}\ \ \ g\left(m\right)=\begin{cases}
-1 & m<-1\\
-1 & m=-1\\
0 & m=0\\
0 & m=1\\
1 & m>1
\end{cases}\ \ \ \ \mathrm{and}\ \ \ h\left(m\right)=\begin{cases}
-1 & m<-1\\
-\sqrt{2} & m=-1\\
\sqrt{2} & m=0\\
1 & m=1\\
1 & m>1
\end{cases}
\end{equation}
Furthermore
\begin{equation}
Y_{2,\pm1}Y_{\ell,0}=\frac{\alpha_{\ell+1}^{\left(+\right)0}\alpha_{\ell+2}^{1}}{\alpha_{2}^{1}}Y_{\ell+2,\pm1}+\frac{\alpha_{\ell+1}^{\left(+\right)0}\alpha_{\ell+1}^{1}-\alpha_{\ell}^{(-)0}\alpha_{\ell}^{1}}{\alpha_{2}^{1}}Y_{\ell,\pm1}-\frac{\alpha_{\ell}^{(-)0}\alpha_{\ell-1}^{1}}{\alpha_{2}^{1}}Y_{\ell-2,\pm1}
\end{equation}
\begin{equation}
Y_{2,\pm2}Y_{\ell,0}=\frac{\alpha_{\ell+1}^{\left(+\right)0}\alpha_{\ell+2}^{\left(+\right)1}}{\alpha_{2}^{\left(+\right)1}}Y_{\ell+2,\pm2}-\frac{\alpha_{\ell+1}^{\left(+\right)0}\alpha_{\ell+1}^{\left(-\right)-1}+\alpha_{\ell}^{\left(-\right)0}\alpha_{\ell}^{\left(+\right)1}}{\alpha_{2}^{\left(+\right)1}}Y_{\ell,\pm2}+\frac{\alpha_{\ell}^{\left(-\right)0}\alpha_{\ell-1}^{\left(-\right)-1}}{\alpha_{2}^{\left(+\right)1}}Y_{\ell-2,\pm2}
\end{equation}
\end{widetext}
At several points throughout this derivation we use the product of
three real spherical harmonics combined in matrix form
\begin{equation}
\left(\mathbf{J}_{\left(\ell\right)}^{m,m',m''}\right)_{\ell_{1},\ell_{2}}\equiv\int d\vec{u}Y_{\ell_{1},m}\left(\vec{u}\right)Y_{\ell,m'}\left(\vec{u}\right)Y_{\ell_{2}m''}\left(\vec{u}\right)\label{eq:Definition_of_J}
\end{equation}
The elements of the matrix are mostly zero except a few diagonals
which follow from the products above and the orthogonality property.
The selection rules for real spherical harmonics differ from that
of the complex version. The selection rules dictate that $\mathbf{J}_{\left(\ell\right)}^{m,m',m''}=0$
unless $m+m'-m''=0$ or $m-m'+m''=0$ or $-m+m'+m''=0$.

\section{Fourier Transform Conventions \label{sec:Appendix_Fourier}}

We use the Fourier transform convention
\begin{equation}
\tilde{f}\left(\vec{k}\right)=\frac{1}{\left(2\pi\right)^{3/2}}\int d\vec{r}\exp\left(i\vec{k}\cdot\vec{r}\right)f(\vec{r})\label{eq:FourierTransform}
\end{equation}
and
\begin{equation}
f\left(\vec{r}\right)=\frac{1}{\left(2\pi\right)^{3/2}}\int d\vec{k}\exp\left(-i\vec{k}\cdot\vec{r}\right)\tilde{f}(\vec{k})\label{eq:InverseFourierTransform}
\end{equation}

When we extend the summation notation to integrate over $\vec{k}$
rather than $\vec{r}$ it has the effect of reversing the latter $\vec{k}$.
For real functions $A$ and $B$
\begin{align}
\tilde{A}_{1}\tilde{B}_{1} & \equiv\sum_{\ell=0}^{\infty}\sum_{m=-\ell}^{\ell}\int d\vec{k}A_{l}^{m}\left(\vec{k}\right)B_{l}^{m*}\left(\vec{k}\right)\nonumber \\
 & =\sum_{\ell=0}^{\infty}\sum_{m=-\ell}^{\ell}\int d\vec{k}A_{l}^{m}\left(\vec{k}\right)B_{\ell}^{m}\left(-\vec{k}\right) 
\end{align}
which, along with $\tilde{S}_{12}\tilde{\mathbb{I}}_{23}=\tilde{S}_{13}$,
implies that the identity is $\tilde{\mathbb{I}}_{12}=\delta_{\ell_{1},\ell_{2}}\delta_{m_{1,}m_{2}}\delta\left(\vec{k}_{1}+\vec{k}_{2}\right)$.
{} By plugging \ref{eq:InverseFourierTransform} into $A_{1}B_{1}$
we find that
\begin{align}
A_{1}B_{1} & =\frac{1}{\left(2\pi\right)^{3}}\int d\vec{k}\int d\vec{k}'\tilde{A}(\vec{k})\tilde{B}^{*}(\vec{k}')\int d\vec{r} e^{-i\left(\vec{k}-\vec{k}'\right)\cdot\vec{r}}\nonumber \\
 & = \int d\vec{k}\int d\vec{k}'\tilde{A}(\vec{k})\tilde{B}^{*}(\vec{k}')\delta\left(\vec{k}-\vec{k}'\right)\nonumber \\
 & =\tilde{A}_{1}\tilde{B}_{1}\label{eq:conversion_of_summation}
\end{align}

Of particular interest to this paper is the forier transform of a derivative squared.  To this end we prove equation~\ref{eq:Fourer_2nd_derivative}
\begin{align}
\int dx\left(\frac{\partial}{\partial x}\phi\right)^{2} & =\int dx\left(\frac{1}{\sqrt{2\pi}}\frac{\partial}{\partial x}\int e^{-ikx}\tilde{f}\left(k\right)dk\right)^{2}\\
 & =\int dx\left(\frac{1}{\sqrt{2\pi}}\int\left(-ik\right)e^{-ikx}\tilde{f}\left(k\right)dk\right)^{2}\\
 & =\int dk_{1}\int dk_{2}\left(-ik_{1}\right)\left(-ik_{2}\right)\tilde{f}\left(k_{1}\right)\tilde{f}\left(k_{2}\right)\times\frac{1}{2\pi}\int dxe^{-i\left(k_{1}+k_{2}\right)x}\\
 & =-\int dk_{1}\int dk_{2} k_{1}k_{2}\tilde{f}\left(k_{1}\right)\tilde{f}\left(k_{2}\right)\delta\left(k_{1}+k_{2}\right)\\
 & =\int k^{2}\tilde{f}\left(k\right)\tilde{f}\left(-k\right)dk
\end{align}

All interactions discussed in this paper are assumed to be translationally
invariant (assumption \ref{enu:TranslationalInvariance}) meaning
that for some interaction $S\left(\vec{x}_{1},\vec{x}_{2}\right)=S\left(\vec{x}_{1}+\vec{x}_{0},\vec{x}_{2}+\vec{x}_{0}\right)$
for all $\vec{x}_{0}$. Taking the Fourier transform
\begin{equation}
\tilde{S}\left(\vec{k}_{1},\vec{k}_{2}\right)=\frac{1}{\left(2\pi\right)^{3}}\int_{-\infty}^{\infty}d\vec{x}_{2}d\vec{x}_{1}e^{i\vec{k}_{1}\cdot\vec{x}_{1}+i\vec{k}_{2}\cdot\vec{x}_{2}}S\left(\vec{x}_{1},\vec{x}_{2}\right)
\end{equation}
because of the infinite extent of the system we can do a change of
integration variables $\vec{x}_{3}=\vec{x}_{1}+\vec{x}_{0}$ 
\begin{align}
\tilde{S}\left(\vec{k}_{1},\vec{k}_{2}\right)=&\frac{1}{\left(2\pi\right)^{3}}\int_{-\infty}^{\infty}d\vec{x}_{2}d\vec{x}_{3}e^{i\vec{k}_{1}\cdot\left(\vec{x}_{3}-\vec{x}_{0}\right)+i\vec{k}_{2}\cdot\vec{x}_{2}}\nonumber\\
&\times S\left(\vec{x}_{3}-\vec{x}_{0},\vec{x}_{2}\right)
\end{align}
using the definition of translation invariance $S\left(\vec{x}_{3}-\vec{x}_{0},\vec{x}_{2}\right)=S\left(\vec{x}_{3},\vec{x}_{2}+\vec{x}_{0}\right)$
and setting $\vec{x}_{0}=-\vec{x}_{2}$ yields
\begin{align}
\tilde{S}\left(\vec{k}_{1},\vec{k}_{2}\right)=&\frac{1}{\left(2\pi\right)^{3}}\int_{-\infty}^{\infty}d\vec{x}_{2}d\vec{x}_{3}e^{i\vec{k}_{1}\cdot\left(\vec{x}_{3}\right)+i\left(\vec{k}_{1}+\vec{k}_{2}\right)\cdot\vec{x}_{2}}\nonumber\\
&\times S\left(\vec{x}_{3},0\right)
\end{align}

Recognizing the delta function $\left(2\pi\right)^{3}\delta\left(\vec{k}_{1}+\vec{k}_{2}\right)=\int_{-\infty}^{\infty}d\vec{x}_{2}\exp\left(i\left(\vec{k}_{1}+\vec{k}_{2}\right)\cdot\vec{x}_{2}\right)$
and defining \\ $\tilde{S}\left(\vec{k}\right)=\int_{-\infty}^{\infty}d\vec{x}\exp\left(i\vec{k}\cdot\vec{x}\right)S\left(\vec{x},0\right)$
we have
\begin{equation}
\tilde{S}\left(\vec{k}_{1},\vec{k}_{2}\right)=\tilde{S}\left(\vec{k}_{1}\right)\delta\left(\vec{k}_{1}+\vec{k}_{2}\right)
\end{equation}

Being diagonal with respect to $\vec{k}$ makes $\tilde{S}$ easy
to invert. We define the inverse by $\tilde{S}\cdot\tilde{S}^{-1}\equiv\mathbb{I}$
which written out implies 
\begin{equation}
\tilde{S}^{-1}(k_{1},k_{3})=\frac{\delta\left(\vec{k}_{1}+\vec{k}_{2}\right)}{\tilde{S}\left(\vec{k}_{1}\right)}\label{eq:HowToTakeS_inverse}
\end{equation}

Note that if $\tilde{S}$ contains $\ell$ and $m$ indices, it still
needs to be inverted with respect them in the normal way. 

\section{Inverting S\label{HowToInvertS}}

As the deformation energy \ref{eq:Gamma_and_cancel} requires the
inverse of $S$, we will take a moment to discuss how to take the
inverse of a power series of matrices. The derivative of an inverse
can be calculated by differentiating the identity $\partial I=0=\partial\left(SS^{-1}\right)=S\partial\left(S^{-1}\right)+\left(\partial S\right)S^{-1}$
which gives
\begin{equation}
\partial\left(S^{-1}\right)=-S^{-1}\left(\partial S\right)S^{-1}
\end{equation}

\begin{align}
\frac{\partial}{\partial\kappa_{1}}\frac{\partial}{\partial\kappa_{2}}\left(S^{-1}\right)= & S^{-1}\left(\frac{\partial}{\partial\kappa_{1}}S\right)S^{-1}\left(\frac{\partial}{\partial\kappa_{2}}S\right)S^{-1}\nonumber \\
& +S^{-1}\left(\frac{\partial}{\partial\kappa_{2}}S\right)S^{-1}\left(\frac{\partial}{\partial\kappa_{1}}S\right)S^{-1}\nonumber \\
& -S^{-1}\left(\frac{\partial}{\partial\kappa_{2}}\frac{\partial}{\partial\kappa_{1}}S\right)S^{-1}
\end{align}
\begin{equation}
S^{-1}\left(k\right)=\left.S^{-1}\right|_{\kappa=0}+\left.\frac{\partial}{\partial\kappa}S^{-1}\left(\kappa\right)\right|_{\kappa=0}\kappa+\frac{1}{2}\left.\frac{\partial}{\partial\kappa}S^{-2}\left(\kappa\right)\right|_{\kappa=0}\kappa^{2}
\end{equation}

In other words, if we have a Taylor expansion
\begin{equation}
S\left(\kappa\right)=S+S_{\kappa\kappa}\left(\kappa^{2}\right)
\end{equation}
then
\begin{equation}
S^{-1}\left(\kappa\right)=S^{-1}-S^{-1}S_{\kappa\kappa}\left(\kappa^{2}\right)S^{-1}.
\end{equation}

\section{List of Variables and Notation}

\textbf{Variables}

$A$ \textendash{} polymer cross sectional area (also matrix element in equation~\ref{eq:K_matrix})

$A_{\ell}^{m}$ \textendash{} defined in equations \ref{eq:AlmBlmdef} 

$a$ \textendash{} Maier-Saupe parameter

$B$ \textendash{} a particular matrix element in equation~\ref{eq:K_matrix}

$B_{\left(n\right)}$ \textendash{} products of propagators defined
and used in section \ref{sec:Calculate_FEC}

$c_{i}$ \textendash{} degree of alignment, expressed as $s_{i}$
in~\cite{turziDistortioninducedEffectsNematic2007}, see equation \ref{eq:definition_of_Q}

$\boldsymbol{e}_{0}$ \textendash{} unit vector for $\ell=0$, i.e.
$[1,0,0,....0]$

$\vec{e}_{i}$ \textendash{} eigenvector of alignment matrix, see
equation \ref{eq:DefinitionOfn}

$E_{FH}$\textendash{} Flory-Huggins Separation energy

$E_{Frank}$\textendash{} Frank elastic energy

$E_{MS}$\textendash{} Maier-Saupe alignment energy

$E_{poly}$\textendash{} polymer bending energy

$f$ \textendash{} used to denote arbitrary function

$\ ^{FB}$ \textendash{} super script indicating fuzzball system

$G$ \textendash{} propagator for wormlike chain in aligning field
(not normalized!), see equation \ref{eq:orientation_partition}

$G_{o}$ \textendash{} propagator for wormlike chain given in equation
\ref{eq:FreeChainPropagator}

$G_{\ell_{0}\ell_{f}}^{m}$ \textendash{} $\breve{G}$ coefficient,
see equation \ref{eq:Gmll}

$\text{\ensuremath{\mathbf{G}}}^{m}$ \textendash{} $\breve{G}$ coefficient
matrix representation $\left(\text{\ensuremath{\mathbf{G}}}^{m}\right)_{\ell_{1},\ell_{2}}\equiv G_{\ell_{1},\ell_{2}}^{m}$

$h$ \textendash{} short for $\lim_{k\to 0} S'_{02}{00}$, see equation~\ref{eq:thxyzMtrx}

$I$ \textendash{} 3x3 identity matrix

$\mathbb{I}_{12}$ \textendash{} the identity in the space defined
by our summation notation

$i$ \textendash{} $\sqrt{-1}$ or index depending on context

$j$ \textendash{} polymer or monomer index

$\mathbf{J}$ \textendash{} product of three spherical harmonics defined
in \ref{eq:Definition_of_J}

$\vec{k}$ \textendash{} Fourier conjugate of position

$K$ \textendash{} Frank elastic content.

$K'$ \textendash{} Nondimensionalized Frank elastic content.

$L$ \textendash{} path length of polymer

$\mathcal{L}\left[...\right]$ \textendash{} Laplace transform $N\to p$

$\mathcal{L}^{-1}\left[...\right]\left(N\right)$ \textendash{} Inverse
Laplace transform $N\leftarrow p$

$\ell$ \textendash{} angular momentum eigenvalue

$\ell_{p}$ \textendash{} persistence length of the polymer when placed in isotropic solution

$\mathbf{M}_{\ell_{1},\ell_{2}}^{m}$ \textendash{} rigid
rod/fuzzball matrix, see equation \ref{eq:Def_of_M}

$M$ \textendash{} 3x3 alignment matrix defined by equation \ref{eq:DefinitionOfM}

$m$ \textendash{} z-component of angular momentum eigenvalue

$\mathscr{M}$ \textendash{} m value in particular summations.

$N$ \textendash{} polymer length in Kuhn lengths, $N\equiv L/\left(2\ell_{p}\right)$

$n$ \textendash{} order in $\gamma$ expansion, see equation \ref{eq:gamma_expansion}

$\vec{n}$ \textendash{} nematic director defined in equation \ref{eq:DefinitionOfn}

$n_{p}$ \textendash{} number of polymers

$p$ \textendash{} Laplace transform variable of $N$ 

$P$ \textendash{} Legendre polynomial

$P$ \textendash{} probability propagator, see equation \ref{eq:rewightedPropagator}

$P_{\ell}$ \textendash{} shorthand for $p+\ell\left(\ell+1\right)$ 

$Q$ \textendash{} traceless alignment matrix $M-\frac{1}{3}I$

$\ ^{RR}$ \textendash{} super script indicating rigid rod

$\vec{\mathrm{r}}_{j}\left(s\right)$ \textendash{} position of point
$s$ along polymer $j$

$\vec{\mathrm{r}}_{1}$ \textendash{} shorthand for $\vec{\mathrm{r}}\left(s_{1}\right)$

$\vec{r}$ \textendash{} arbitrary position is space

$s$ \textendash{} path length along polymer

$S_{12}$ \textendash{} two point correlation (single polymer structure
factor), see equation \ref{eq:StructureFactor}

$t$ \textendash{} short for $\lim_{k\to 0} S'_{00}{00}$, see equation~\ref{eq:thxyzMtrx}

$\vec{\mathrm{u}}_{j}\left(s\right)$ orientation unit vector $\frac{\partial\vec{\mathrm{r}}_{j}\left(s\right)}{\partial s}$
for polymer $j$ at path length $s$

$\vec{\mathrm{u}}_{1}$ \textendash{} shorthand for $\vec{\mathrm{u}}\left(s_{1}\right)$ 

$\vec{u}_{1}$ \textendash{} arbitrary unit vector

$\mathscr{V}$ \textendash{} volume of space (assumed to be large)

$V_{12}$ \textendash{} interaction potential defined in equation \ref{eq:V_potential}

$v$ \textendash{} rotational term in equation~\ref{eq:split_fuzzball}

$v_{p}$ \textendash{} volume of one polymer, rod, or fuzzball

$W$ \textendash{} field conjugate to $\phi$, similar to chemical
potential, see equation \ref{eq:functional_delta_function}

$W_{\ell}^{\left(\pm\right)m}$ \textendash{} see equation \ref{eq:Wpm}

$W_{\ell}^{\left(+\right)m}$ \textendash{} see equation \ref{eq:Wplus}

$W_{\ell}^{\left(-\right)m}$ \textendash{} see equation \ref{eq:Wminus}

$w$ \textendash{} distance to avoid pole by inverse Laplace transform,
see section \ref{sec:Numerical-Laplace-inversion}

$x$ \textendash{} short for $\lim_{k\to 0} S'_{22}{00}$, see equation~\ref{eq:thxyzMtrx}

$\hat{x}$ \textendash{} Cartesian unit vector

$y$ \textendash{} short for $\lim_{k\to 0} S'_{22}{\pm1\pm1}$, see equation~\ref{eq:thxyzMtrx}

$\hat{y}$ \textendash{} Cartesian unit vector

$Y_{l}^{m}$ \textendash{} real spherical harmonic. We use the normalization
$\int d\vec{u}Y_{l}^{m}\left(\vec{u}\right)Y_{l'}^{*m'}\left(\vec{u}\right)=\delta_{\ell,\ell'}\delta_{m,m'}$

$\hat{z}$ \textendash{} Cartesian unit vector

$Z$ \textendash{} multi-polymer partition function

$z$ \textendash{} short for $\lim_{k\to 0} S'_{22}{\pm2\pm2}$, see equation~\ref{eq:thxyzMtrx}

$z_{p}$\textendash{} single polymer partition function, see equation
\ref{eq:SimplePartition}

$\alpha_{\ell}^{m}$ \textendash{} raising symbol, defined in equation
\ref{eq:alpha_def}

$\alpha_{\ell}^{\left(\pm\right)m}$\textendash{} raising symbol with
change in $m$, defined in equation \ref{eq:alpha_pm_def}

$\beta_{\ell}^{m}$ \textendash{} defined in equations \ref{eq:AlmBlmdef}

$\Gamma_{12}$ \textendash{} the quadratic order fluctuation coefficient,
see \ref{eq:QuadraticPartision}

$\gamma$ \textendash{} aligning field strength 

$\nabla$ \textendash{} derivative vector $\hat{x}\partial_{x}+\hat{y}\partial_{y}+\hat{z}\partial_{x}$

$\Delta V$ \textendash{} coarse-graining volume element

$\delta\left(...\right)$ \textendash{} delta function or product
of delta function at every point is space as in equation \ref{eq:functional_delta_function}

$\delta\phi$ \textendash{} fluctuations in $\phi$ from it's mean
field value

$\delta W$ \textendash{} fluctuations in $W$ from it's mean field
value

$\delta S_{12}$ \textendash{} first term in \ref{eq:StructureFactor}

$\kappa_{1}^{m}$ \textendash{} spherical representation of $\vec{k}$
as defined in \ref{eq:DefinitionOfKappa}

$\lambda$ \textendash{} real offset in numerical inverse Laplace
transform see section \ref{sec:Numerical-Laplace-inversion}

$\rho$ \textendash{} spatial term in equation~\ref{eq:split_fuzzball}

$\sigma$ \textendash{} standard deviation of a fuzzball

$\phi$ \textendash{} smooth field approximately equal to $\hat{\phi}$

$\hat{\phi}_{l}^{m}\left(\vec{r}\right)$ \textendash{} polymer position
and orientation field defined in equation \ref{eq:density}

$\chi$\textendash{} Flory-Huggins interaction parameter

\textbf{Notation}

$\int\mathscr{D}\left[\vec{r}_{j}\left(s\right)\right]$ \textendash{}
functional integral of positions of polymers defined in equation \ref{eq:functional_intigral}

$\int\mathscr{D}W=\prod_{\vec{r}}\left(\int dW\left(\vec{r}\right)\right)$
\textendash{} functional integral of $W$ over all space

$\int d\vec{u}$ \textendash{} Integral over surface of ball $\int\sin\left(\theta\right)d\theta d\phi$

$\propto$ \textendash{} proportional to

$\equiv$ \textendash{} definition

$\vec{u}\otimes\vec{v}$ \textendash{} outer product matrix $\left(\vec{u}\otimes\vec{v}\right)_{i,j}=u_{i}v_{j}$

$A_{1}B_{1}\equiv\sum_{\ell=0}^{\infty}\sum_{m=-\ell}^{\ell}\int d\vec{r}A_{l}^{m}\left(\vec{r}\right)B_{l}^{m*}\left(\vec{r}\right)$
\textendash{} extended summation notation

$f*g$ \textendash{} convolution $f*g=\int_{0}^{t}d\tau f\left(\tau\right)g\left(t-\tau\right)$

$G\left(\vec{u}|\vec{u}_{0};L\right)$ \textendash{} distribution
$G\left(\vec{u}\right)$ given $\vec{u}_{0}$ for polymer of length
$L$

$...^{\left(1\right)}$ \textendash{} superscript denoting single
polymer

$...^{MF}$ \textendash{} superscript denoting mean field

$\left\langle ...\right\rangle ^{MF}$ \textendash{} mean field expectation
value as in equation \ref{eq:SelfConsistantEquation}

$\left\langle ...\right\rangle _{o}$ \textendash{} expectation value
for free chain (no applied field) with fixed end orientations, see
example \ref{eq:IntermediateOrientation}

$\breve{...}$ \textendash{} indicates Laplace transform $N\to p$ 

$\tilde{...}$ \textendash{} indicates Fourier transform $\vec{r}\to\vec{k}$

$\vec{...}$ \textendash{} Euclidean vector in 3 dimensions

\textbf{Stone Fence symbols}

All symbols are assumed to start from $\ell$ on the left.

\includegraphics{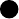} $=1/P_{\ell}$

\includegraphics{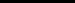} $=\gamma\beta_{\ell}^{m}$

\includegraphics{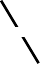} $=\gamma A_{\ell}^{m}$

\includegraphics{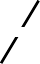} $=\gamma A_{\ell+2}^{m}$

\includegraphics{Figures/diagrams/dashed.pdf} $=1/\left(P_{\ell}-\gamma\beta_{\ell}^{m}\right)$

\includegraphics{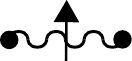} $=W_{\ell}^{\left(+\right)m}$

\includegraphics{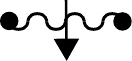} $=W_{\ell}^{\left(-\right)m}$

\includegraphics{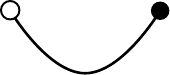}
$=\left(P_{\ell}-\gamma\beta_{\ell}^{m}\right)W_{\ell}^{\left(+\right)m}-1$

\includegraphics{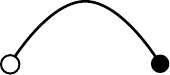}
$=\left(P_{\ell}-\gamma\beta_{\ell}^{m}\right)W_{\ell}^{\left(-\right)m}-1$

\includegraphics{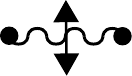} $=W_{\ell}^{\left(\pm\right)m}$

\bibliography{mylibrarybibtex}% Produces the bibliography via BibTeX.

\begin{thebibliography}{30}%
\makeatletter
\providecommand \@ifxundefined [1]{%
 \@ifx{#1\undefined}
}%
\providecommand \@ifnum [1]{%
 \ifnum #1\expandafter \@firstoftwo
 \else \expandafter \@secondoftwo
 \fi
}%
\providecommand \@ifx [1]{%
 \ifx #1\expandafter \@firstoftwo
 \else \expandafter \@secondoftwo
 \fi
}%
\providecommand \natexlab [1]{#1}%
\providecommand \enquote  [1]{``#1''}%
\providecommand \bibnamefont  [1]{#1}%
\providecommand \bibfnamefont [1]{#1}%
\providecommand \citenamefont [1]{#1}%
\providecommand \href@noop [0]{\@secondoftwo}%
\providecommand \href [0]{\begingroup \@sanitize@url \@href}%
\providecommand \@href[1]{\@@startlink{#1}\@@href}%
\providecommand \@@href[1]{\endgroup#1\@@endlink}%
\providecommand \@sanitize@url [0]{\catcode `\\12\catcode `\$12\catcode
  `\&12\catcode `\#12\catcode `\^12\catcode `\_12\catcode `\%12\relax}%
\providecommand \@@startlink[1]{}%
\providecommand \@@endlink[0]{}%
\providecommand \url  [0]{\begingroup\@sanitize@url \@url }%
\providecommand \@url [1]{\endgroup\@href {#1}{\urlprefix }}%
\providecommand \urlprefix  [0]{URL }%
\providecommand \Eprint [0]{\href }%
\providecommand \doibase [0]{https://doi.org/}%
\providecommand \selectlanguage [0]{\@gobble}%
\providecommand \bibinfo  [0]{\@secondoftwo}%
\providecommand \bibfield  [0]{\@secondoftwo}%
\providecommand \translation [1]{[#1]}%
\providecommand \BibitemOpen [0]{}%
\providecommand \bibitemStop [0]{}%
\providecommand \bibitemNoStop [0]{.\EOS\space}%
\providecommand \EOS [0]{\spacefactor3000\relax}%
\providecommand \BibitemShut  [1]{\csname bibitem#1\endcsname}%
\let\auto@bib@innerbib\@empty
%</preamble>
\bibitem [{\citenamefont
  {Selinger}(2018)}]{selingerInterpretationSaddlesplayOseenFrank2018}%
  \BibitemOpen
  \bibfield  {author} {\bibinfo {author} {\bibfnamefont {J.~V.}\ \bibnamefont
  {Selinger}},\ }\bibfield  {title} {\bibinfo {title} {Interpretation of
  saddle-splay and the {Oseen}-{Frank} free energy in liquid crystals},\ }\href
  {https://doi.org/10.1080/21680396.2019.1581103} {\bibfield  {journal}
  {\bibinfo  {journal} {Liquid Crystals Reviews}\ }\textbf {\bibinfo {volume}
  {6}},\ \bibinfo {pages} {129} (\bibinfo {year} {2018})},\ \bibinfo {note}
  {arXiv: 1901.06306}\BibitemShut {NoStop}%
\bibitem [{\citenamefont {Rudnicki}\ \emph {et~al.}(2019)\citenamefont
  {Rudnicki}, \citenamefont {MacPherson}, \citenamefont {Balhorn},
  \citenamefont {Feng}, \citenamefont {Qin}, \citenamefont {Salleo},\ and\
  \citenamefont {Spakowitz}}]{rudnickiImpactLiquidCrystallineChain2019}%
  \BibitemOpen
  \bibfield  {author} {\bibinfo {author} {\bibfnamefont {P.~E.}\ \bibnamefont
  {Rudnicki}}, \bibinfo {author} {\bibfnamefont {Q.}~\bibnamefont
  {MacPherson}}, \bibinfo {author} {\bibfnamefont {L.}~\bibnamefont {Balhorn}},
  \bibinfo {author} {\bibfnamefont {B.}~\bibnamefont {Feng}}, \bibinfo {author}
  {\bibfnamefont {J.}~\bibnamefont {Qin}}, \bibinfo {author} {\bibfnamefont
  {A.}~\bibnamefont {Salleo}},\ and\ \bibinfo {author} {\bibfnamefont {A.~J.}\
  \bibnamefont {Spakowitz}},\ }\bibfield  {title} {\bibinfo {title} {Impact of
  {Liquid}-{Crystalline} {Chain} {Alignment} on {Charge} {Transport} in
  {Conducting} {Polymers}},\ }\bibfield  {journal} {\bibinfo  {journal}
  {Macromolecules}\ }\href {https://doi.org/10.1021/acs.macromol.9b01729}
  {10.1021/acs.macromol.9b01729} (\bibinfo {year} {2019})\BibitemShut {NoStop}%
\bibitem [{\citenamefont {Svenšek}\ \emph {et~al.}(2010)\citenamefont
  {Svenšek}, \citenamefont {Veble},\ and\ \citenamefont
  {Podgornik}}]{svensekConfinedNematicPolymers2010}%
  \BibitemOpen
  \bibfield  {author} {\bibinfo {author} {\bibfnamefont {D.}~\bibnamefont
  {Svenšek}}, \bibinfo {author} {\bibfnamefont {G.}~\bibnamefont {Veble}},\
  and\ \bibinfo {author} {\bibfnamefont {R.}~\bibnamefont {Podgornik}},\
  }\bibfield  {title} {\bibinfo {title} {Confined nematic polymers: {Order} and
  packing in a nematic drop},\ }\href
  {https://doi.org/10.1103/PhysRevE.82.011708} {\bibfield  {journal} {\bibinfo
  {journal} {Physical Review E}\ }\textbf {\bibinfo {volume} {82}},\ \bibinfo
  {pages} {011708} (\bibinfo {year} {2010})},\ \bibinfo {note} {publisher:
  American Physical Society}\BibitemShut {NoStop}%
\bibitem [{\citenamefont {Khokhlov}\ and\ \citenamefont
  {Semenov}(1981)}]{khokhlovLiquidcrystallineOrderingSolution1981}%
  \BibitemOpen
  \bibfield  {author} {\bibinfo {author} {\bibfnamefont {A.~R.}\ \bibnamefont
  {Khokhlov}}\ and\ \bibinfo {author} {\bibfnamefont {A.~N.}\ \bibnamefont
  {Semenov}},\ }\bibfield  {title} {\bibinfo {title} {Liquid-crystalline
  ordering in the solution of long persistent chains},\ }\href
  {https://doi.org/10.1016/0378-4371(81)90148-5} {\bibfield  {journal}
  {\bibinfo  {journal} {Physica A: Statistical Mechanics and its Applications}\
  }\textbf {\bibinfo {volume} {108}},\ \bibinfo {pages} {546} (\bibinfo {year}
  {1981})}\BibitemShut {NoStop}%
\bibitem [{\citenamefont {Khokhlov}\ and\ \citenamefont
  {Semenov}(1982{\natexlab{a}})}]{khokhlovLiquidcrystallineOrderingSolution1982}%
  \BibitemOpen
  \bibfield  {author} {\bibinfo {author} {\bibfnamefont {A.~R.}\ \bibnamefont
  {Khokhlov}}\ and\ \bibinfo {author} {\bibfnamefont {A.~N.}\ \bibnamefont
  {Semenov}},\ }\bibfield  {title} {\bibinfo {title} {Liquid-crystalline
  ordering in the solution of partially flexible macromolecules},\ }\href
  {https://doi.org/10.1016/0378-4371(82)90199-6} {\bibfield  {journal}
  {\bibinfo  {journal} {Physica A: Statistical Mechanics and its Applications}\
  }\textbf {\bibinfo {volume} {112}},\ \bibinfo {pages} {605} (\bibinfo {year}
  {1982}{\natexlab{a}})}\BibitemShut {NoStop}%
\bibitem [{\citenamefont {Khokhlov}\ and\ \citenamefont
  {Semenov}(1982{\natexlab{b}})}]{khokhlovInfluenceExternalField1982}%
  \BibitemOpen
  \bibfield  {author} {\bibinfo {author} {\bibfnamefont {A.~R.}\ \bibnamefont
  {Khokhlov}}\ and\ \bibinfo {author} {\bibfnamefont {A.~N.}\ \bibnamefont
  {Semenov}},\ }\bibfield  {title} {\bibinfo {title} {Influence of external
  field on the liquid-crystalline ordering in the solutions of stiff-chain
  macromolecules},\ }\href {https://doi.org/10.1021/ma00233a012} {\bibfield
  {journal} {\bibinfo  {journal} {Macromolecules}\ }\textbf {\bibinfo {volume}
  {15}},\ \bibinfo {pages} {1272} (\bibinfo {year} {1982}{\natexlab{b}})},\
  \bibinfo {note} {publisher: American Chemical Society}\BibitemShut {NoStop}%
\bibitem [{\citenamefont {Khokhlov}\ and\ \citenamefont
  {Semenov}(1985)}]{khokhlovTheoryLiquidcrystallineOrdering1985}%
  \BibitemOpen
  \bibfield  {author} {\bibinfo {author} {\bibfnamefont {A.~R.}\ \bibnamefont
  {Khokhlov}}\ and\ \bibinfo {author} {\bibfnamefont {A.~N.}\ \bibnamefont
  {Semenov}},\ }\bibfield  {title} {\bibinfo {title} {On the theory of
  liquid-crystalline ordering of polymer chains with limited flexibility},\
  }\href {https://doi.org/10.1007/BF01017855} {\bibfield  {journal} {\bibinfo
  {journal} {Journal of Statistical Physics}\ }\textbf {\bibinfo {volume}
  {38}},\ \bibinfo {pages} {161} (\bibinfo {year} {1985})}\BibitemShut
  {NoStop}%
\bibitem [{\citenamefont {Khokhlov}\ and\ \citenamefont
  {Semenov}(1986)}]{khokhlovTheoryNematicOrdering1986}%
  \BibitemOpen
  \bibfield  {author} {\bibinfo {author} {\bibfnamefont {A.~R.}\ \bibnamefont
  {Khokhlov}}\ and\ \bibinfo {author} {\bibfnamefont {A.~N.}\ \bibnamefont
  {Semenov}},\ }\bibfield  {title} {\bibinfo {title} {Theory of nematic
  ordering in the melts of macromolecules with different flexibility
  mechanisms},\ }\href {https://doi.org/10.1021/ma00156a025} {\bibfield
  {journal} {\bibinfo  {journal} {Macromolecules}\ }\textbf {\bibinfo {volume}
  {19}},\ \bibinfo {pages} {373} (\bibinfo {year} {1986})},\ \bibinfo {note}
  {publisher: American Chemical Society}\BibitemShut {NoStop}%
\bibitem [{\citenamefont {Semenov}\ and\ \citenamefont
  {Khokhlov}(1988)}]{semenovStatisticalPhysicsLiquidcrystalline1988}%
  \BibitemOpen
  \bibfield  {author} {\bibinfo {author} {\bibfnamefont {A.~N.}\ \bibnamefont
  {Semenov}}\ and\ \bibinfo {author} {\bibfnamefont {A.~R.}\ \bibnamefont
  {Khokhlov}},\ }\bibfield  {title} {\bibinfo {title} {Statistical physics of
  liquid-crystalline polymers},\ }\href
  {https://doi.org/10.1070/PU1988v031n11ABEH005643} {\bibfield  {journal}
  {\bibinfo  {journal} {Soviet Physics Uspekhi}\ }\textbf {\bibinfo {volume}
  {31}},\ \bibinfo {pages} {988} (\bibinfo {year} {1988})},\ \bibinfo {note}
  {publisher: IOP Publishing}\BibitemShut {NoStop}%
\bibitem [{\citenamefont {Liu}\ and\ \citenamefont
  {Fredrickson}(1993)}]{liuFreeEnergyFunctionals1993}%
  \BibitemOpen
  \bibfield  {author} {\bibinfo {author} {\bibfnamefont {A.~J.}\ \bibnamefont
  {Liu}}\ and\ \bibinfo {author} {\bibfnamefont {G.~H.}\ \bibnamefont
  {Fredrickson}},\ }\bibfield  {title} {\bibinfo {title} {Free energy
  functionals for semiflexible polymer solutions and blends},\ }\href
  {https://doi.org/10.1021/ma00063a028} {\bibfield  {journal} {\bibinfo
  {journal} {Macromolecules}\ }\textbf {\bibinfo {volume} {26}},\ \bibinfo
  {pages} {2817} (\bibinfo {year} {1993})}\BibitemShut {NoStop}%
\bibitem [{\citenamefont {Spakowitz}\ and\ \citenamefont
  {Wang}(2003)}]{spakowitzSemiflexiblePolymerSolutions2003}%
  \BibitemOpen
  \bibfield  {author} {\bibinfo {author} {\bibfnamefont {A.~J.}\ \bibnamefont
  {Spakowitz}}\ and\ \bibinfo {author} {\bibfnamefont {Z.-G.}\ \bibnamefont
  {Wang}},\ }\bibfield  {title} {\bibinfo {title} {Semiflexible polymer
  solutions. {I}. {Phase} behavior and single-chain statistics},\ }\href
  {https://doi.org/10.1063/1.1628669} {\bibfield  {journal} {\bibinfo
  {journal} {The Journal of Chemical Physics}\ }\textbf {\bibinfo {volume}
  {119}},\ \bibinfo {pages} {13113} (\bibinfo {year} {2003})}\BibitemShut
  {NoStop}%
\bibitem [{\citenamefont {Straley}(1973)}]{straleyFrankElasticConstants1973}%
  \BibitemOpen
  \bibfield  {author} {\bibinfo {author} {\bibfnamefont {J.~P.}\ \bibnamefont
  {Straley}},\ }\bibfield  {title} {\bibinfo {title} {Frank {Elastic}
  {Constants} of the {Hard}-{Rod} {Liquid} {Crystal}},\ }\href
  {https://doi.org/10.1103/PhysRevA.8.2181} {\bibfield  {journal} {\bibinfo
  {journal} {Physical Review A}\ }\textbf {\bibinfo {volume} {8}},\ \bibinfo
  {pages} {2181} (\bibinfo {year} {1973})},\ \bibinfo {note} {publisher:
  American Physical Society}\BibitemShut {NoStop}%
\bibitem [{\citenamefont {Lee}\ and\ \citenamefont
  {Meyer}(1986)}]{leeComputationsPhaseEquilibrium1986}%
  \BibitemOpen
  \bibfield  {author} {\bibinfo {author} {\bibfnamefont {S.}~\bibnamefont
  {Lee}}\ and\ \bibinfo {author} {\bibfnamefont {R.~B.}\ \bibnamefont
  {Meyer}},\ }\bibfield  {title} {\bibinfo {title} {Computations of the phase
  equilibrium, elastic constants, and viscosities of a hard‐rod nematic
  liquid crystal},\ }\href {https://doi.org/10.1063/1.450228} {\bibfield
  {journal} {\bibinfo  {journal} {The Journal of Chemical Physics}\ }\textbf
  {\bibinfo {volume} {84}},\ \bibinfo {pages} {3443} (\bibinfo {year}
  {1986})},\ \bibinfo {note} {publisher: American Institute of
  Physics}\BibitemShut {NoStop}%
\bibitem [{\citenamefont {Marrucci}\ and\ \citenamefont
  {Greco}(1991)}]{marrucciElasticConstantsMaierSaupe1991}%
  \BibitemOpen
  \bibfield  {author} {\bibinfo {author} {\bibfnamefont {G.}~\bibnamefont
  {Marrucci}}\ and\ \bibinfo {author} {\bibfnamefont {F.}~\bibnamefont
  {Greco}},\ }\bibfield  {title} {\bibinfo {title} {The {Elastic} {Constants}
  of {Maier}-{Saupe} {Rodlike} {Molecule} {Nematics}},\ }\href
  {https://doi.org/10.1080/00268949108037714} {\bibfield  {journal} {\bibinfo
  {journal} {Molecular Crystals and Liquid Crystals}\ }\textbf {\bibinfo
  {volume} {206}},\ \bibinfo {pages} {17} (\bibinfo {year} {1991})},\ \bibinfo
  {note} {publisher: Taylor \& Francis \_eprint:
  https://doi.org/10.1080/00268949108037714}\BibitemShut {NoStop}%
\bibitem [{\citenamefont {Odijk}(1986)}]{odijkElasticConstantsNematic1986}%
  \BibitemOpen
  \bibfield  {author} {\bibinfo {author} {\bibfnamefont {T.}~\bibnamefont
  {Odijk}},\ }\bibfield  {title} {\bibinfo {title} {Elastic constants of
  nematic solutions of rod-like and semi-flexible polymers},\ }\href
  {https://doi.org/10.1080/02678298608086279} {\bibfield  {journal} {\bibinfo
  {journal} {Liquid Crystals}\ }\textbf {\bibinfo {volume} {1}},\ \bibinfo
  {pages} {553} (\bibinfo {year} {1986})},\ \bibinfo {note} {publisher: Taylor
  \& Francis \_eprint: https://doi.org/10.1080/02678298608086279}\BibitemShut
  {NoStop}%
\bibitem [{\citenamefont {Shimada}\ \emph {et~al.}(1988)\citenamefont
  {Shimada}, \citenamefont {Doi},\ and\ \citenamefont
  {Okano}}]{shimadaCoefficiensGradientTerms1988}%
  \BibitemOpen
  \bibfield  {author} {\bibinfo {author} {\bibfnamefont {T.}~\bibnamefont
  {Shimada}}, \bibinfo {author} {\bibfnamefont {M.}~\bibnamefont {Doi}},\ and\
  \bibinfo {author} {\bibfnamefont {K.}~\bibnamefont {Okano}},\ }\bibfield
  {title} {\bibinfo {title} {Coefficiens of {Gradient} {Terms} in {Landau}-de
  {Gennes} {Free} {Energy} {Expansion} for {Polymeric} {Liquid} {Crystals}},\
  }\href {https://doi.org/10.1143/JPSJ.57.2432} {\bibfield  {journal} {\bibinfo
   {journal} {Journal of the Physical Society of Japan}\ }\textbf {\bibinfo
  {volume} {57}},\ \bibinfo {pages} {2432} (\bibinfo {year} {1988})},\ \bibinfo
  {note} {publisher: The Physical Society of Japan}\BibitemShut {NoStop}%
\bibitem [{\citenamefont {Doussal}\ and\ \citenamefont
  {Nelson}(1991)}]{doussalStatisticalMechanicsDirected1991}%
  \BibitemOpen
  \bibfield  {author} {\bibinfo {author} {\bibfnamefont {P.~L.}\ \bibnamefont
  {Doussal}}\ and\ \bibinfo {author} {\bibfnamefont {D.~R.}\ \bibnamefont
  {Nelson}},\ }\bibfield  {title} {\bibinfo {title} {Statistical {Mechanics} of
  {Directed} {Polymer} {Melts}},\ }\href
  {https://doi.org/10.1209/0295-5075/15/2/009} {\bibfield  {journal} {\bibinfo
  {journal} {Europhysics Letters (EPL)}\ }\textbf {\bibinfo {volume} {15}},\
  \bibinfo {pages} {161} (\bibinfo {year} {1991})},\ \bibinfo {note}
  {publisher: IOP Publishing}\BibitemShut {NoStop}%
\bibitem [{\citenamefont {Petschek}\ and\ \citenamefont
  {Terentjev}(1992)}]{petschekMolecularstatisticalTheoryCurvature1992}%
  \BibitemOpen
  \bibfield  {author} {\bibinfo {author} {\bibfnamefont {R.~G.}\ \bibnamefont
  {Petschek}}\ and\ \bibinfo {author} {\bibfnamefont {E.~M.}\ \bibnamefont
  {Terentjev}},\ }\bibfield  {title} {\bibinfo {title} {Molecular-statistical
  theory for curvature elasticity of thermotropic main-chain-polymer liquid
  crystals},\ }\href {https://doi.org/10.1103/PhysRevA.45.930} {\bibfield
  {journal} {\bibinfo  {journal} {Physical Review A}\ }\textbf {\bibinfo
  {volume} {45}},\ \bibinfo {pages} {930} (\bibinfo {year} {1992})},\ \bibinfo
  {note} {publisher: American Physical Society}\BibitemShut {NoStop}%
\bibitem [{\citenamefont {Sato}\ and\ \citenamefont
  {Teramoto}(1996)}]{satoFrankElasticConstants1996}%
  \BibitemOpen
  \bibfield  {author} {\bibinfo {author} {\bibfnamefont {T.}~\bibnamefont
  {Sato}}\ and\ \bibinfo {author} {\bibfnamefont {A.}~\bibnamefont
  {Teramoto}},\ }\bibfield  {title} {\bibinfo {title} {On the {Frank} {Elastic}
  {Constants} of {Lyotropic} {Polymer} {Liquid} {Crystals}},\ }\href
  {https://doi.org/10.1021/ma950986a} {\bibfield  {journal} {\bibinfo
  {journal} {Macromolecules}\ }\textbf {\bibinfo {volume} {29}},\ \bibinfo
  {pages} {4107} (\bibinfo {year} {1996})},\ \bibinfo {note} {publisher:
  American Chemical Society}\BibitemShut {NoStop}%
\bibitem [{Note1()}]{Note1}%
  \BibitemOpen
  \bibinfo {note} {It appears that this approximation roughly corresponds to
  including only the leading pole in the complex integral we perform in
  section~\ref {sec:Numerical-Laplace-inversion}}\BibitemShut {NoStop}%
\bibitem [{\citenamefont {Spakowitz}\ and\ \citenamefont
  {Wang}(2004)}]{spakowitzExactResultsSemiflexible2004}%
  \BibitemOpen
  \bibfield  {author} {\bibinfo {author} {\bibfnamefont {A.~J.}\ \bibnamefont
  {Spakowitz}}\ and\ \bibinfo {author} {\bibfnamefont {Z.-G.}\ \bibnamefont
  {Wang}},\ }\bibfield  {title} {\bibinfo {title} {Exact {Results} for a
  {Semiflexible} {Polymer} {Chain} in an {Aligning} {Field}},\ }\href
  {https://doi.org/10.1021/ma049958v} {\bibfield  {journal} {\bibinfo
  {journal} {Macromolecules}\ }\textbf {\bibinfo {volume} {37}},\ \bibinfo
  {pages} {5814} (\bibinfo {year} {2004})}\BibitemShut {NoStop}%
\bibitem [{\citenamefont {Spakowitz}\ and\ \citenamefont
  {Wang}(2005)}]{spakowitzEndtoendDistanceVector2005}%
  \BibitemOpen
  \bibfield  {author} {\bibinfo {author} {\bibfnamefont {A.~J.}\ \bibnamefont
  {Spakowitz}}\ and\ \bibinfo {author} {\bibfnamefont {Z.-G.}\ \bibnamefont
  {Wang}},\ }\bibfield  {title} {\bibinfo {title} {End-to-end distance vector
  distribution with fixed end orientations for the wormlike chain model},\
  }\href {https://doi.org/10.1103/PhysRevE.72.041802} {\bibfield  {journal}
  {\bibinfo  {journal} {Physical Review. E, Statistical, Nonlinear, and Soft
  Matter Physics}\ }\textbf {\bibinfo {volume} {72}},\ \bibinfo {pages}
  {041802} (\bibinfo {year} {2005})}\BibitemShut {NoStop}%
\bibitem [{\citenamefont {Priest}(1973)}]{priestTheoryFrankElastic1973}%
  \BibitemOpen
  \bibfield  {author} {\bibinfo {author} {\bibfnamefont {R.~G.}\ \bibnamefont
  {Priest}},\ }\bibfield  {title} {\bibinfo {title} {Theory of the {Frank}
  {Elastic} {Constants} of {Nematic} {Liquid} {Crystals}},\ }\href
  {https://doi.org/10.1103/PhysRevA.7.720} {\bibfield  {journal} {\bibinfo
  {journal} {Physical Review A}\ }\textbf {\bibinfo {volume} {7}},\ \bibinfo
  {pages} {720} (\bibinfo {year} {1973})}\BibitemShut {NoStop}%
\bibitem [{Note2()}]{Note2}%
  \BibitemOpen
  \bibinfo {note} {Without the factor of $1/2$ the $E_{MS}$ would have a $2/3$
  rather than a $1/3$ per the customary definition of $a$}\BibitemShut
  {NoStop}%
\bibitem [{\citenamefont {Yamakawa}(1997)}]{yamakawaHelicalWormlikeChains1997}%
  \BibitemOpen
  \bibfield  {author} {\bibinfo {author} {\bibfnamefont {H.}~\bibnamefont
  {Yamakawa}},\ }\href {https://doi.org/10.1007/978-3-642-60817-9} {\emph
  {\bibinfo {title} {Helical {Wormlike} {Chains} in {Polymer} {Solutions}}}}\
  (\bibinfo  {publisher} {Springer-Verlag},\ \bibinfo {address} {Berlin
  Heidelberg},\ \bibinfo {year} {1997})\BibitemShut {NoStop}%
\bibitem [{\citenamefont
  {Leibler}(1980)}]{leiblerTheoryMicrophaseSeparation1980}%
  \BibitemOpen
  \bibfield  {author} {\bibinfo {author} {\bibfnamefont {L.}~\bibnamefont
  {Leibler}},\ }\bibfield  {title} {\bibinfo {title} {Theory of microphase
  separation in block copolymers},\ }\href
  {https://doi.org/10.1021/ma60078a047} {\bibfield  {journal} {\bibinfo
  {journal} {Macromolecules}\ }\textbf {\bibinfo {volume} {13}},\ \bibinfo
  {pages} {1602} (\bibinfo {year} {1980})}\BibitemShut {NoStop}%
\bibitem [{\citenamefont {Mao}\ \emph {et~al.}(2016)\citenamefont {Mao},
  \citenamefont {Macpherson}, \citenamefont {He}, \citenamefont {Coletta},\
  and\ \citenamefont {Spakowitz}}]{maoImpactConformationalChemical2016}%
  \BibitemOpen
  \bibfield  {author} {\bibinfo {author} {\bibfnamefont {S.}~\bibnamefont
  {Mao}}, \bibinfo {author} {\bibfnamefont {Q.}~\bibnamefont {Macpherson}},
  \bibinfo {author} {\bibfnamefont {S.}~\bibnamefont {He}}, \bibinfo {author}
  {\bibfnamefont {E.}~\bibnamefont {Coletta}},\ and\ \bibinfo {author}
  {\bibfnamefont {A.}~\bibnamefont {Spakowitz}},\ }\bibfield  {title} {\bibinfo
  {title} {Impact of conformational and chemical correlations on microphase
  segregation in random copolymers},\ }\bibfield  {journal} {\bibinfo
  {journal} {Macromolecules}\ }\textbf {\bibinfo {volume} {49}},\ \href
  {https://doi.org/10.1021/acs.macromol.5b02639} {10.1021/acs.macromol.5b02639}
  (\bibinfo {year} {2016})\BibitemShut {NoStop}%
\bibitem [{Note3()}]{Note3}%
  \BibitemOpen
  \bibinfo {note} {For WLCs the Kuhn length is twice the persistence
  length.}\BibitemShut {Stop}%
\bibitem [{\citenamefont
  {Turzi}(2007)}]{turziDistortioninducedEffectsNematic2007}%
  \BibitemOpen
  \bibfield  {author} {\bibinfo {author} {\bibfnamefont {S.}~\bibnamefont
  {Turzi}},\ }\emph {\bibinfo {title} {Distortion-induced eﬀects in nematic
  liquid crystals}},\ \href@noop {} {Ph.D. thesis},\ \bibinfo  {school}
  {Politecnico di Milano}, \bibinfo {address} {Milan, Italy} (\bibinfo {year}
  {2007})\BibitemShut {NoStop}%
\bibitem [{Note4()}]{Note4}%
  \BibitemOpen
  \bibinfo {note} {A better solvent could effect behavior on shorter length
  scales which could renormalize into a different Maier-Saupe parameter on
  longer length scales and thereby indirectly effect the Frank elastic
  constants. We have also assumed that all solvents are good enough to dissolve
  the polymers.}\BibitemShut {Stop}%
\end{thebibliography}%

\end{document}